\documentclass[twocolumn]{aastex631}

\usepackage[super]{nth}
\usepackage{hyperref}

\usepackage{color}
\usepackage{threeparttable}
\usepackage{url}

\newcommand\kms{\ifmmode{\rm km\thinspace s^{-1}}\else km\thinspace s$^{-1}$\fi}

\usepackage{xcolor}

\begin{document}

\title{The Discovery of Two Quadruple Star Systems with the Second and Third Shortest Outer Periods}

\correspondingauthor{Brian P. Powell}
\email{brian.p.powell@nasa.gov}

\author[0000-0003-0501-2636]{Brian P. Powell}
\affiliation{NASA Goddard Space Flight Center, 8800 Greenbelt Road, Greenbelt, MD 20771, USA}
%
\author[0000-0002-5286-0251]{Guillermo Torres}
\affiliation{Center for Astrophysics $\vert$ Harvard \& Smithsonian, 60 Garden Street, Cambridge, MA 02138, USA}
%
\author[0000-0001-9786-1031]{Veselin~B.~Kostov}
\affiliation{NASA Goddard Space Flight Center, 8800 Greenbelt Road, Greenbelt, MD 20771, USA}
\affiliation{SETI Institute, 189 Bernardo Ave, Suite 200, Mountain View, CA 94043, USA}
%
\author[0000-0002-8806-496X]{Tam\'as Borkovits}
\affiliation{Baja Astronomical Observatory of University of Szeged, H-6500 Baja, Szegedi út, Kt. 766, Hungary}
\affiliation{HUN-REN -- SZTE Stellar Astrophysics Research Group,  H-6500 Baja, Szegedi út, Kt. 766, Hungary}
\affiliation{Konkoly Observatory, Research Centre for Astronomy and Earth Sciences, H-1121 Budapest, Konkoly Thege Miklós út 15-17, Hungary}

%
\author[0000-0003-3182-5569]{Saul A. Rappaport}
\affiliation{Department of Physics, Kavli Institute for Astrophysics and Space Research, M.I.T., Cambridge, MA 02139, USA}

\author[0000-0002-0870-6388]{Maxwell Moe}
\affiliation{Department of Physics \& Astronomy, University of Wyoming, Laramie, WY 82072, USA}

\author[0000-0001-9911-7388]{David W. Latham}
\affiliation{Center for Astrophysics $\vert$ Harvard \& Smithsonian, 60 Garden Street, Cambridge, MA 02138, USA}

\author[0000-0003-3988-3245]{Thomas L. Jacobs}
\affiliation{Amateur Astronomer, Missouri City, TX 77459}
%

\author[0000-0002-5665-1879]{Robert Gagliano}
\affiliation{Amateur Astronomer, Glendale, AZ 85308}
%

\author[0000-0002-2607-138X]{Martti~H.~K.~Kristiansen}
\affil{Brorfelde Observatory, Observator Gyldenkernes Vej 7, DK-4340 T\o{}ll\o{}se, Denmark}

\author{Mark Omohundro}
\affiliation{Citizen Scientist, c/o Zooniverse, Department of Physics, University of Oxford, Denys Wilkinson Building, Keble Road, Oxford, OX13RH, UK}
%

\author[0000-0002-1637-2189]{Hans M. Schwengeler}
\affiliation{Citizen Scientist, Planet Hunter, Bottmingen, Switzerland}
%

\author[0000-0002-8527-2114]{Daryll M. LaCourse}
\affiliation{Amateur Astronomer, 7507 52nd Place NE Marysville, WA 98270}

\author[0000-0002-0654-4442]{Ivan A. Terentev}
\affiliation{Citizen Scientist, Planet Hunter, Petrozavodsk, Russia}
%
\author[0000-0002-5034-0949]{Allan R. Schmitt}
\affiliation{Citizen Scientist, 616 W. 53rd. St., Apt. 101, Minneapolis, MN 55419, USA}

\received{March 3, 2025}
\revised{April 12, 2025}
\accepted{April 16, 2025}

\begin{abstract}

We present the discovery of two quadruple star systems -- TIC 285853156 and TIC 392229331 -- each consisting of two bound eclipsing binary stars. Among the most compact quadruples known, TIC 392229331 and TIC 285853156 have the second and third shortest outer orbital periods (145 days and 152 days, respectively) after BU Canis Minoris (122 days, \citealt{2023MNRAS.524.4220P}). We demonstrate that both systems are long-term dynamically stable despite substantial outer orbital eccentricities (0.33 for TIC 285853156 and 0.56 for TIC 392229331). We previously reported these systems in \citet{2022ApJS..259...66K} and \citet{2024MNRAS.527.3995K} as 2+2 hierarchical quadruple candidates producing two sets of primary and secondary eclipses in {\em TESS} data, as well as prominent eclipse timing variations on both binary components. We combine all available {\em TESS} data and new spectroscopic observations into a comprehensive photodynamical model, proving that the component binary stars are gravitationally bound in both systems and finding accurate stellar and orbital parameters for both systems, including very precise determinations of the outer periods. TIC 285853156 and TIC 392229331 represent the latest addition to the small population of well-characterized proven quadruple systems dynamically interacting on detectable timescales. 

\end{abstract}



\section{Introduction} \label{sec:intro}

Substantial progress has recently been made in the exploration of eclipsing multiple stellar systems, with a large number of new discoveries in the last few years alone \citep[e.g.][]{2022MNRAS.513.4341R,2022MNRAS.510.1352B,2022A&A...664A..96Z,2022ApJS..259...66K,2021AJ....161..162P,2022ApJ...938..133P,2025AJ....169..124T}
. Most of these discoveries are based on data from NASA's Transiting Exoplanet Survey Satellite ({\em TESS}) mission \citep{Ricker14}, and include a wide variety of triply-eclipsing triple systems \citep{2022MNRAS.513.4341R,2022MNRAS.510.1352B,2024A&A...686A..27R}, eclipsing sextuple systems \citep{2021AJ....161..162P,2022AcA....72..103Z,2023MNRAS.520.3127Z}, as well as hundreds of hierarchical 2+2 eclipsing quadruple candidates \citep{2022ApJS..259...66K,2024MNRAS.527.3995K,2022A&A...664A..96Z,2024A&A...682A.164V}. 

The latter set of targets, in particular, has recently broken the record for the most compact orbital configuration twice, first in June 2023 with the discovery of TIC 219006972 (outer period 168 days with eccentricity of 0.25, \citealt{2023MNRAS.522...90K}), and then in September 2023 with the announcement of BU Canis Minoris  (outer period 122 days with eccentricity of 0.27, \citealt{2023MNRAS.524.4220P}). The outer periods of both targets are more than a factor of two shorter than the previous record holder, VW LMi, which stood uncontested for nearly 15 years \citep{2008MNRAS.390..798P,2020MNRAS.494..178P}. In Table \ref{tbl:topten}, we show, to the best of our knowledge, the current top ten shortest known 2+2 quadruple outer orbital periods for context.  We populated this table from literature searches as well as A. Tokovinin's Multiple Star Catalog \citep{2018ApJS..235....6T}.  It is clear from this table that the four 2+2 quadruples with outer orbits less than 200 days (including the two identified in this work) are in a class by themselves.

These discoveries provide important new insights into the mechanisms regulating the formation and evolution of multiple stellar systems \citep{2021Univ....7..352T}, and raise a number of interesting questions. For example, given the still woefully incomplete sample of quadruple star systems, it is currently unclear if the systems listed in Table \ref{tbl:topten} are representative of the general population of such systems or notable outliers. If the latter, one cannot help but wonder what would be the most extreme outlier and how did it achieve its present configuration. 

In seeking to augment the small sample of tight, compact quadruples, as part of our ongoing search for candidates for multiple stellar systems based on {\em TESS} photometry, we continue to monitor sources listed in the \citet{2022ApJS..259...66K} and \citet{2024MNRAS.527.3995K} catalogs (hereafter, K22 and K24, respectively) as new data become available. In particular, we are paying especially close attention to systems with prominent eclipse timing variations, {\em Gaia}'s \texttt{non\_single\_star} parameter greater than zero, and targets with measured astrometric orbital solutions in {\em Gaia} DR3 \citep{2023A&A...674A...1G}. 

\citet{2024A&A...687A...6Z} emphasized that, although 2+2 eclipsing quadruple candidates continue to be discovered and their number is growing rapidly, only a small fraction of these have been proven to be bound quadruples.  For example, in K22 and K24, we take great care to confirm through photocenter analysis that both sets of eclipses originate from the same source of light in {\em TESS}.\footnote{Vetting algorithms developed by V. Kostov described in K22/K24} However, there remains a small chance that, although the two eclipsing binaries (EBs) are at near-identical coordinates on the sky, they are separated by substantial distance and therefore unrelated.  Mechanisms for providing indisputable proof of the bound nature of the system, elevating the candidates to confirmed status, are through either radial velocities (RVs) or eclipse timing variations (ETVs).  In this work, through both RVs and ETVs, we present the confirmation of TIC 285853156 and TIC 392229331 as genuine 2+2 hierarchical systems -- the latest additions to the small number of proven compact, tight quadruples.

\begin{table}[htbp]
\centering
\caption{Top 10 Shortest 2+2 Quadruple Outer Periods$^{a}$}
\begin{tabular}{@{}l@{\hspace{4pt}}l@{\hspace{4pt}}l@{\hspace{5pt}}l@{}}
\hline
\# & ID & $P_{\textrm{out}}$ (d) & Ref. \\
\hline
1 & BU CMi$^{b}$ & $121.79\pm0.04$ & \citealt{2023MNRAS.524.4220P} \\
2 & \textbf{TIC 392229331}$^{c}$ & $144.80\pm0.16$ & This work\\
3 & \textbf{TIC 285853156}$^{d}$ & $151.70\pm0.11$ & This work \\
4 & TIC 219006972 & $168.19\pm0.07$& \citealt{2023MNRAS.522...90K} \\
5 & VW LMi & $355.02\pm0.17$ & \citealt{2008MNRAS.390..798P}\\
6 & TIC 454140642 & $432.1\pm0.5$ & \citealt{2021ApJ...917...93K} \\
7 & BG Ind & $720.9\pm3.4$ & \citealt{2021MNRAS.503.3759B} \\
8/9 & TIC 305635022$^{e,f}$ & $844\pm22$ & \citealt{zasche23} \\
8/9 & TIC 278956474$^{g}$ & $858\pm7$ & \citealt{2020AJ....160...76R}\\
10 & KIC 5255552$^{h}$ & 	$878.47\pm0.01$ & \citealt{2023Univ....9..505O} \\
\hline
\end{tabular}
    \begin{tablenotes}
      \item {\em Notes: 
      \item {\em (a)} We do not include EPIC 220204960, originally identified by \citet{2017MNRAS.467.2160R}, due to the uncertainty of the outer period, which the authors suggest to be in the range of 300-500 days, but possibly as long as 4 years. \citet {2023Univ....9..505O} later found a similarly uncertain period of $957^{+717}_{-362}$ days for the same system.
      
      (b-e) First identified as a 2+2 quadruple candidate by: {\em (b)} \citet{2021ARep...65..826V}, {\em (c)} \citet{2022ApJS..259...66K}, {\em (d)} \citet{2024MNRAS.527.3995K}, {\em (e)} \citet{2022A&A...664A..96Z}
      \item {\em (f)} \citet{Czavalinga2023} first found the outer period in Gaia DR3 and identified the system as a 2+2 quadruple candidate; \citet{zasche23} identified the system as WISE J210230.8+610816.
      \item {\em (g)} Our unpublished analysis of recent observations of this system suggests that the period is longer than indicated in the original published work, but we include it here for completeness.
      \item {\em (h)} \citet{2015MNRAS.448..946B} first discovered additional eclipses and found the outer period using ETVs; further analyzed by \citet{2016MNRAS.455.4136B} and \citet{2018A&A...610A..72Z}; first suggested to be a 2+2 quadruple by \citet{2020MNRAS.498.4356G}.
      }
    \end{tablenotes}
\label{tbl:topten}
\end{table}

This paper is organized as follows. In Section \ref{sec:detection} we provide an overview of the systems, outline the initial detections and discuss new data. In Section \ref{sec:photodynamics}, we discuss our modeling methods.  In Section \ref{sec:results}, we present comprehensive photodynamical solutions for the systems. We highlight interesting properties of the two systems in Section \ref{sec:discussion} and summarize our results in Section \ref{sec:summary}. 

\section{Detection and Preliminary Analysis}
\label{sec:detection}
Although the discoveries of TIC 285853156 and TIC 392229331 were first reported in K24 and K22, respectively, we will briefly describe the process by which these systems were first identified. Upon the release of {\em TESS} data from each sector, we build all the light curves brighter than 15th magnitude from the Full Frame Images (FFI).  We then process the light curves through a neural network trained to find eclipses in the light curves (described further in \citealt{2021AJ....161..162P}).  All high-probability candidates for light curves containing eclipses are then manually reviewed by our citizen scientist collaborators in the Visual Survey Group \citep{2022PASP..134g4401K}.  They identify light curves containing multiple sets of eclipses or `extra' eclipses and then we thoroughly vet the candidates.  TIC 285853156 was found through this process by T. Jacobs, while TIC 392229331 was initially identified by B. Powell during development and training of the neural network.  See \citet{2022MNRAS.513.4341R}, \citet{2024A&A...686A..27R}, or K22/K24 for additional details.

\begin{figure*}
   \centering
    \includegraphics[width=0.49\linewidth]{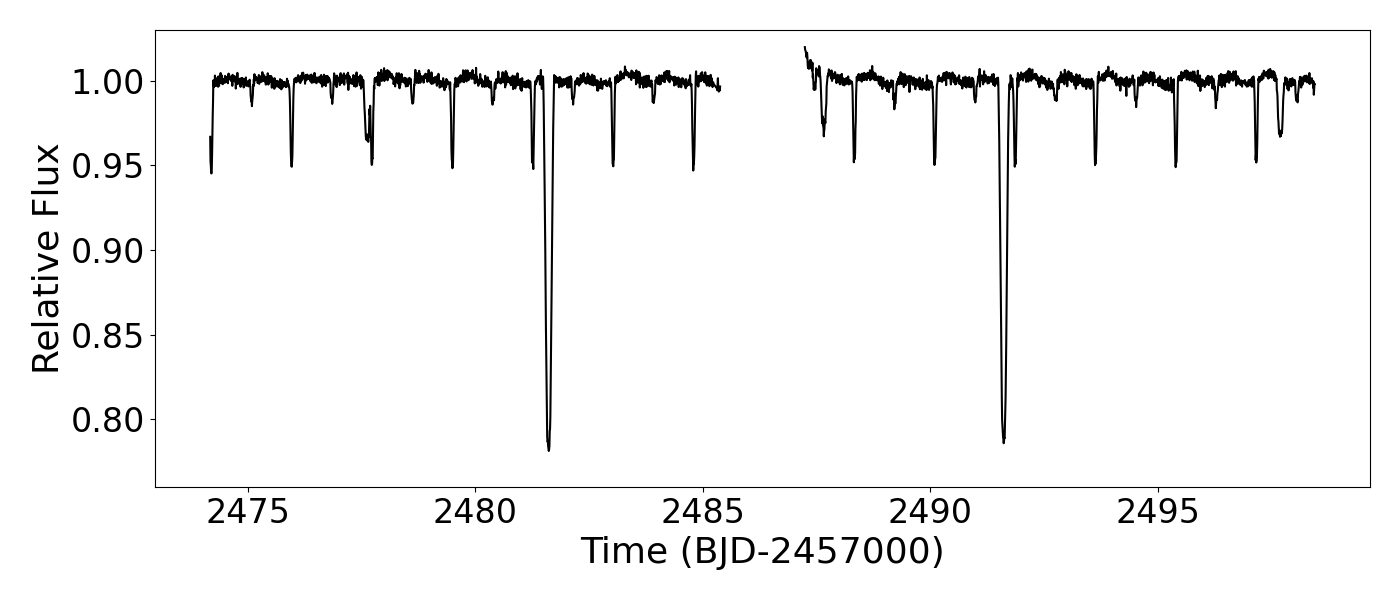}
    \includegraphics[width=0.49\linewidth]{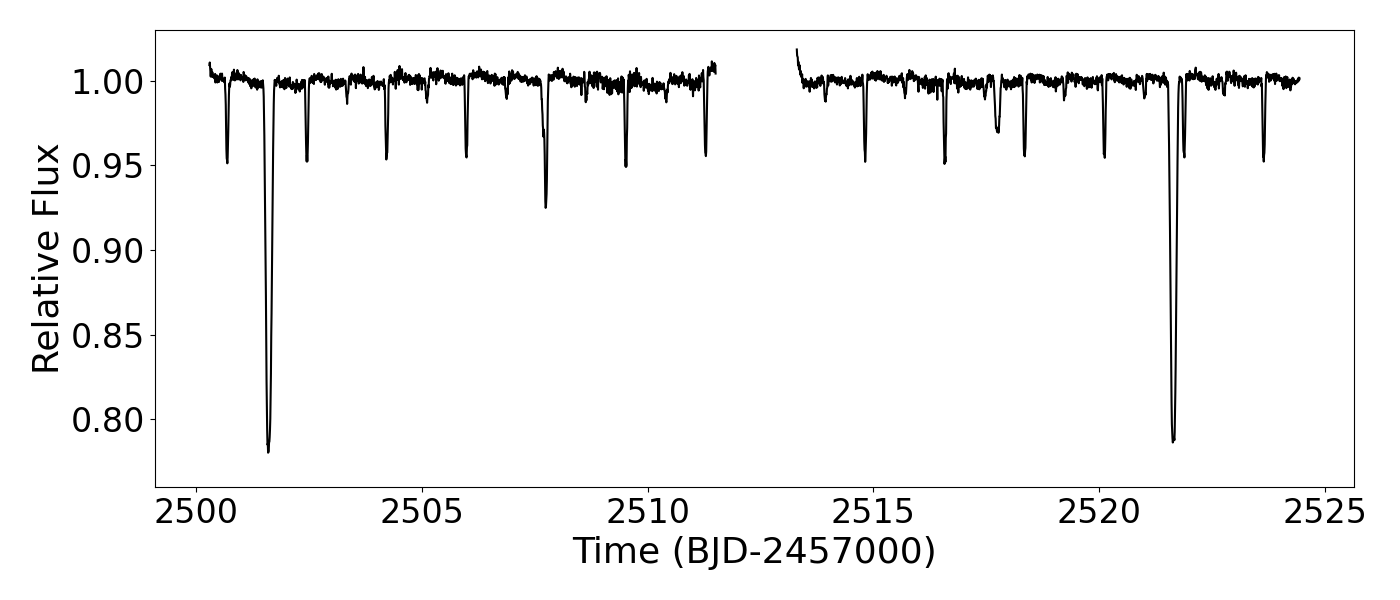}
    \includegraphics[width=0.49\linewidth]{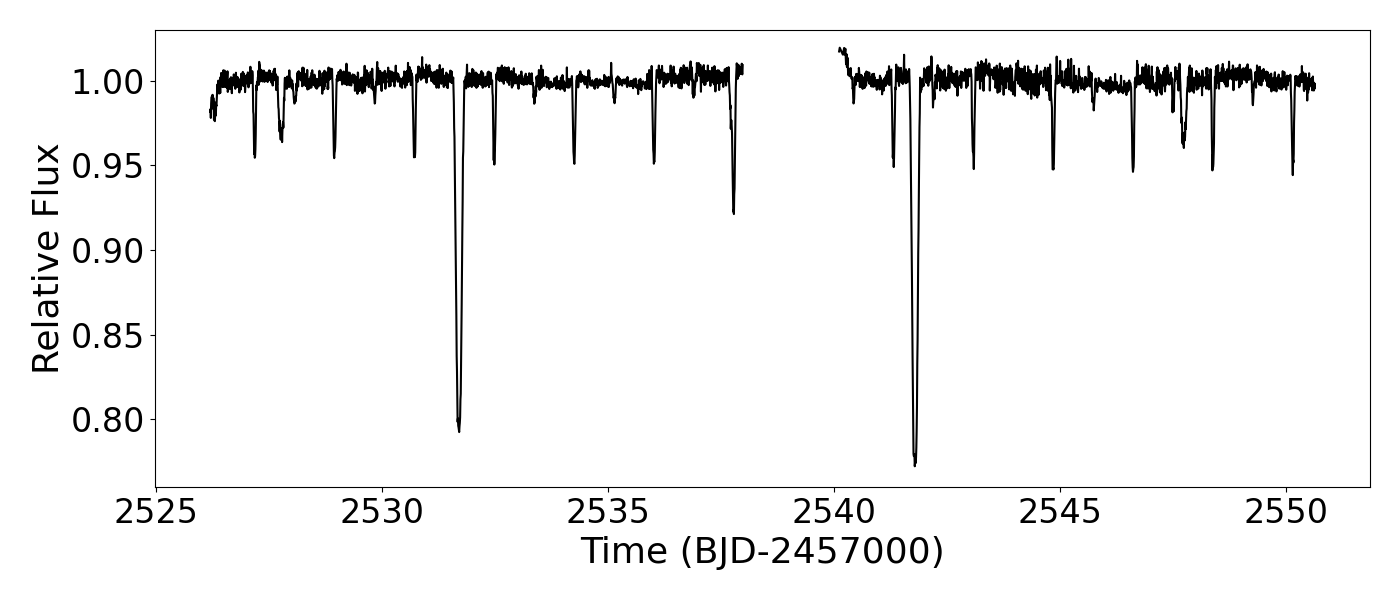}
    \includegraphics[width=0.49\linewidth]{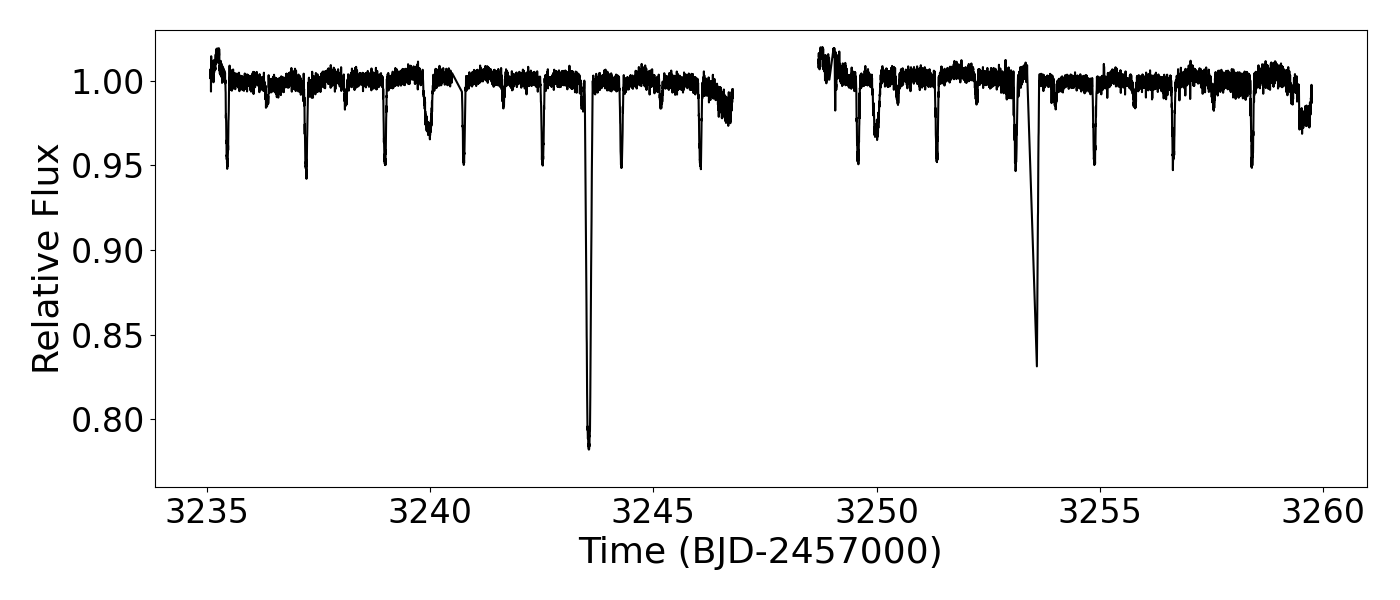}
    \caption{TESS Full-Frame Image {\sc eleanor}\citep{eleanor} light curves for TIC 285853156 from Sectors 43 ({\em upper left}), 44 ({\em upper right}), 45 ({\em lower left}), and 71 ({\em lower right}).}
    \label{fig:tess_lc_285853156}
\end{figure*}

\begin{figure*}
   \centering
    \includegraphics[width=0.49\linewidth]{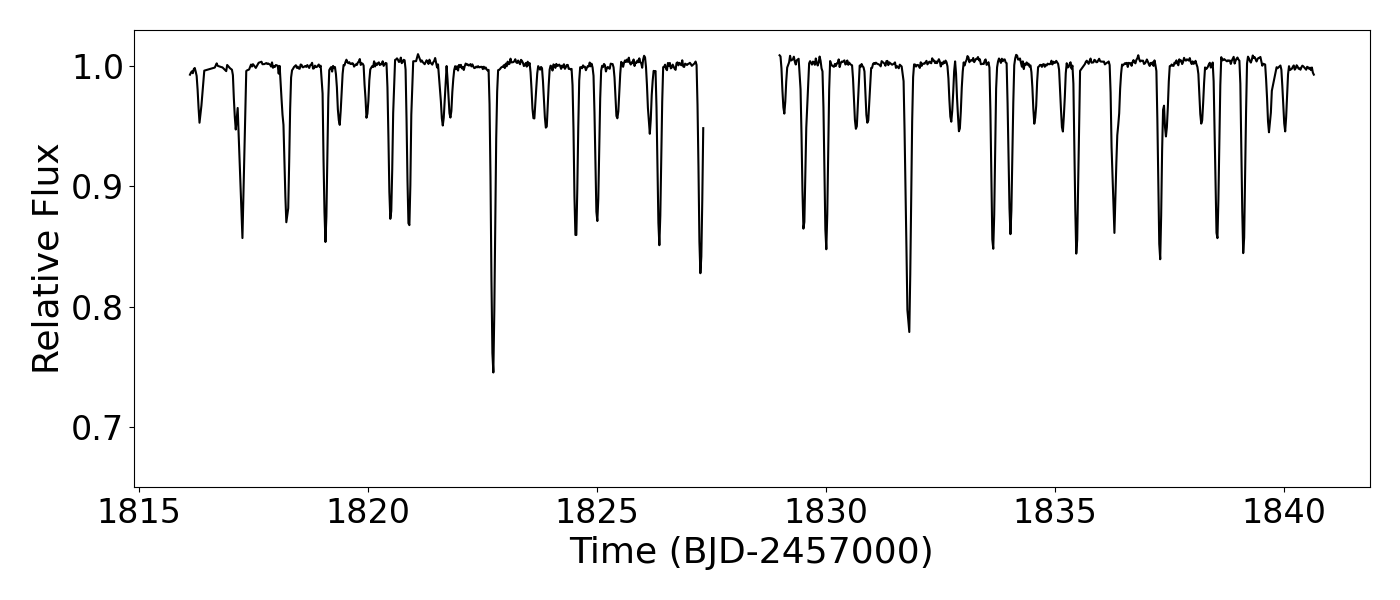}
    \includegraphics[width=0.49\linewidth]{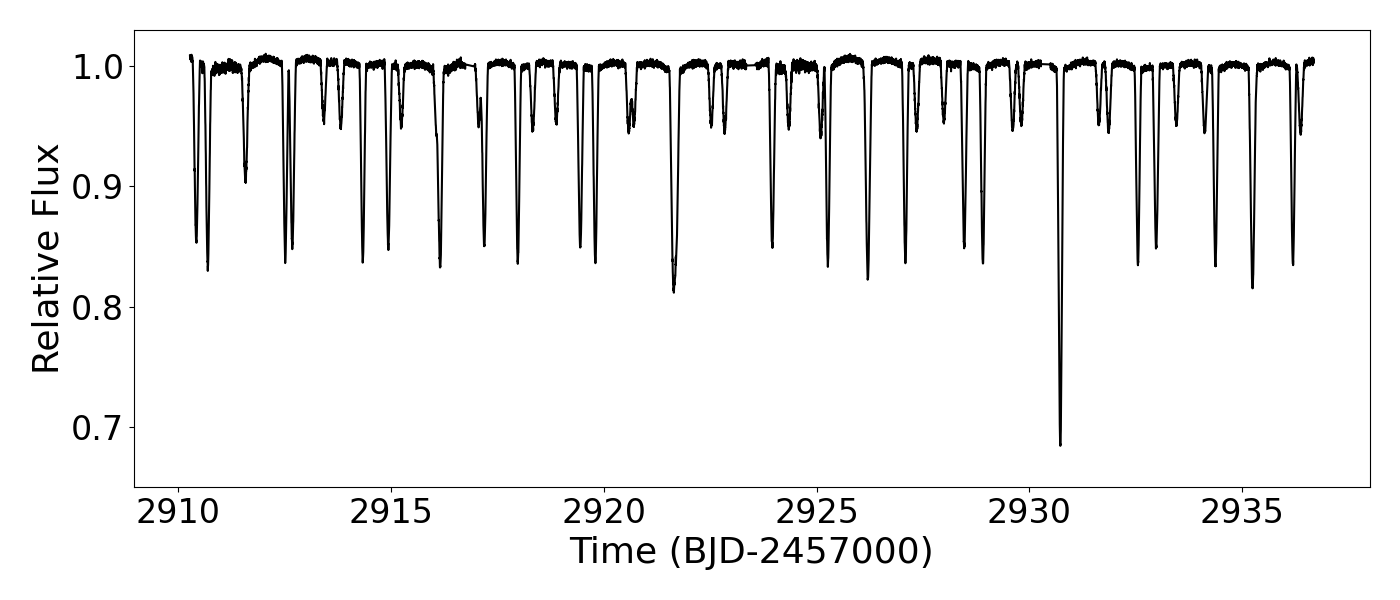}
    \includegraphics[width=0.49\linewidth]{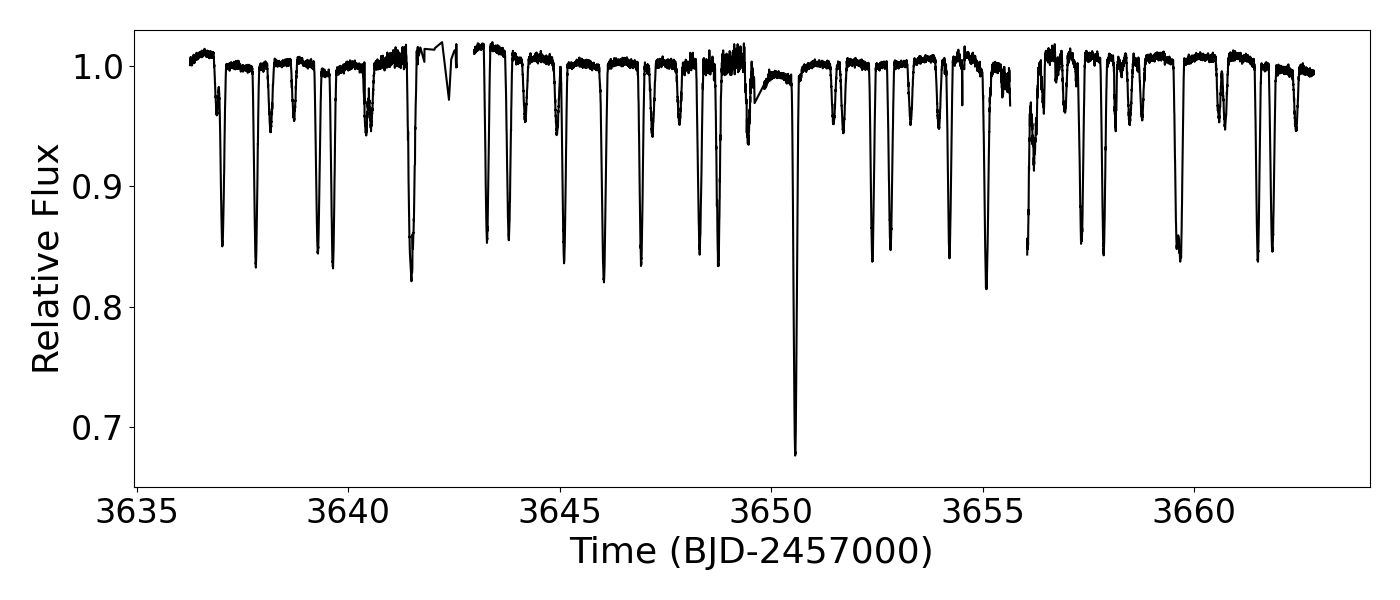}
    \caption{{\em TESS} Full-Frame Image {\sc eleanor}\citep{eleanor} light curves for TIC 392229331 from Sectors 19 ({\em upper left}), 59 ({\em upper right}), and 86 ({\em bottom}).}
    \label{fig:tess_lc_392229331}
\end{figure*}

The initial analysis of the two systems was conducted by K22/K24. TIC 285853156 was first reported in K24 with the periods of the EBs as 1.77 days and 10.03 days.  The authors first noted dramatic, anti-correlated ETVs on the 10-day binary.  {\em TESS} light curves of TIC 285853156 for sectors 43, 44, 45, and 71 are shown in Figure \ref{fig:tess_lc_285853156}.\footnote{MAST FFIs are available from the following DOIs: Sector 43 \dataset[(\url{https://doi.org/10.17909/vanq-c119})]{https://doi.org/10.17909/vanq-c119}, Sector 44 \dataset[(\url{https://doi.org/10.17909/wncz-ja61})]{https://doi.org/10.17909/wncz-ja61}, Sector 45 \dataset[(\url{https://doi.org/10.17909/ncs9-a589})]{https://doi.org/10.17909/ncs9-a589}, Sector 71 \dataset[(\url{https://doi.org/10.17909/e2dc-8076})]{https://doi.org/10.17909/e2dc-8076}}
TIC 392229331 was first reported in K22 with EB periods of 2.26 days and 1.82 days.  {\em TESS} light curves of TIC 392229331 for sectors 19, 59, and 86 are shown in Figure \ref{fig:tess_lc_392229331}.\footnote{MAST FFIs are available from the following DOIs:  Sector 19 \dataset[(\url{https://doi.org/10.17909/msxy-d755})]{https://doi.org/10.17909/msxy-d755}, Sector 59 \dataset[(\url{https://doi.org/10.17909/b7zs-6675})]{https://doi.org/10.17909/b7zs-6675}, Sector 86 \dataset[(\url{https://doi.org/10.17909/b7zs-6675})]{https://doi.org/10.17909/b7zs-6675}}  Table 4 of K22 notes this system as an ideal candidate for follow up based on $T<12$ and eclipse depths $>1$\%, criteria which are also met by TIC 285853156. The basic parameters of the two systems presented here are listed in Table \ref{tab:EBparameters}.

\begin{deluxetable*}{l r r r}[!ht]
\tabletypesize{\scriptsize}
\tablecaption{Basic parameters for TIC 285853156 and TIC 392229331\label{tab:EBparameters}}
\tablewidth{0pt}
\tablehead{
\colhead{Parameter} & \colhead{Value} & \colhead{Value} &\colhead{Source}
}
\startdata
\multicolumn{4}{l}{\bf Identifying Information} \\
\hline
TIC ID d & 285853156 & 392229331 & TIC\\
Gaia DR3 ID & 3445472758271399936 & 486430957815961344 & Gaia DR3\\
$\alpha$ (J2000, degrees)  & 81.8928 & 54.7679 & TIC\\
$\delta$ (J2000, degrees) & 28.5528 & 61.0642 & TIC\\
\\
\multicolumn{4}{l}{\bf Gaia Measurements} \\
\hline
$\mu_{\alpha}$ (mas~yr$^{-1}$) & -10.5307 & 1.0264 & Gaia DR3\\
$\mu_{\delta}$ (mas~yr$^{-1}$) & -0.2348 & 1.1657 & Gaia DR3\\
$\varpi$ (mas)& 3.4830 & 1.5517 & Gaia DR3\\
RUWE & 3.18 & 0.99 & Gaia DR3\\
\tt{astrometric\_excess\_noise} & 0.38 & 0.10 & Gaia DR3\\
\tt{astrometric\_excess\_noise\_sig} & 181.99 & 14.99 & Gaia DR3\\
${\rm T_{eff}~(K)}$ & 5675 & -- & Gaia DR3\\
\\
\multicolumn{4}{l}{\bf Photometric Properties} \\
\hline
$T$ (mag) & 11.4714 & 10.3359 & TIC \\ 
$G$ (mag) & 12.0708 & 10.5856 & Gaia DR3\\
$B$ (mag) & 13.2540 & 11.0610 & TIC\\
$V$ (mag) & 12.3380 & 10.6160 & TIC\\
$J$ (mag) & 10.5280 & 9.8480 & TIC\\
$H$ (mag) & 10.0960 & 9.7490 & TIC\\
$K$ (mag) & 9.9600 & 9.6900 & TIC\\
$W1$ (mag) & 9.8700 & 9.6640 & TIC\\
$W2$ (mag) & 9.8780 & 9.6770 & TIC\\
$W3$ (mag) & 9.7890 & 9.6720 & TIC\\
$W4$ (mag) & 8.8420 & 9.4650 & TIC\\
\hline
\enddata

\end{deluxetable*}

\subsection{Spectroscopic Observations}
\label{section:spectroscopy}

TIC~285853156 and TIC~392229331 were observed spectroscopically with the 1.5m Tillinghast reflector at the Fred L.\ Whipple Observatory (Mount Hopkins, AZ), using the Tillingast Reflector Echelle Spectrograph \citep[TRES;][]{Furesz:2008, Szentgyorgyi:2007}. This instrument records 51 echelle orders between 3800\,\AA\ and 9100\,\AA, at a resolving power of $R \approx 44,\!000$. We gathered 36 spectra for TIC~285853156 with signal-to-noise ratios between 26 and 50 per resolution element of 6.8~\kms, and 47 spectra of TIC~392229331 with signal-to-noise ratios of 40--74. In both cases, visual inspection of the cross-correlation functions revealed two sets of lines, corresponding to the primary of each binary subsystem. For reference hereafter we define TIC~285853156 with binary A (primary Aa, secondary Ab) having a 10-day period and binary B (primary Ba, secondary Bb) having a 1.77-day period.  The quadruple TIC~392229331 has binary A (primary Aa, secondary Ab) with a 1.82-day period and binary B (primary Ba, secondary Bb) with a 2.26-day period.  In TIC~285853156, the stronger lines correspond to star Aa. The second set of lines is much weaker. Of the two sets of lines seen in TIC~392229331, the sharper ones correspond to star Ba, while the star with the broader lines is Aa.

Radial velocities (RVs) were measured using TODCOR \citep{Zucker:1994}, a two-dimensional cross-correlation technique. Templates were taken from a large library of synthetic spectra that are based on model atmospheres by R.\ L.\ Kurucz \citep[see][]{Nordstrom:1994, Latham:2002}, and a line list manually tuned by J. Morse to improve the match to real stars. The templates are centered on the region of the \ion{Mg}{1}\,b triplet ($\sim$5187~\AA), and velocity determinations used the spectral order near the center of this region, which captures most of the velocity information. The template parameters, of which the effective temperature and rotational broadening are the most important, were determined by running extensive grids of cross-correlations, following \cite{Torres:2002}. For the brighter star in TIC~285853156 (star Aa), we determined $T_{\rm eff} = 6280 \pm 100$~K and $v \sin i = 7 \pm 2~\kms$. The faintness of the other star (Ba) prevented us from determining its properties reliably. We therefore adopted educated guesses of $T_{\rm
eff} = 5000$~K and $v \sin i = 20~\kms$, under the assumption of spin-orbit alignment in its 1.77-day orbit, and that it rotates synchronously with the orbital motion.
For TIC~392229331~Aa, the properties we inferred are 
$T_{\rm eff} = 9220 \pm 200$~K and $v \sin i = 22 \pm 3~\kms$,
and for star Ba they are
$9070 \pm 150$~K and $8 \pm 3~\kms$.

\begin{table*}[htbp]
\centering
\caption{Stellar Parameters deduced from Spectroscopy}
\begin{tabular}{lcccc}
\hline
Parameter & Aa & Ab & Ba & Bb \\
\hline
& & TIC 285853156 & & \\
$T_{\rm eff}$ [K] & $6280 \pm 100$ & ... & ... & ... \\
$v \sin i$ [km s$^{-1}$] & $7 \pm 2$  &... & ...&... \\
Fractional luminosities$^a$ & 0.928 &0.015 & 0.057 & ... \\
\hline
\hline
& & TIC 392229331 & & \\
$T_{\rm eff}$ [K]  & $9220 \pm 200$ & ... & $9070 \pm 150$& ... \\
$v \sin i$ [km s$^{-1}$] & $22 \pm 3$  &... & $8 \pm 3$ & ...\\
Fractional luminosities$^a$ & 0.494 &0.013 & 0.474 & 0.019 \\
\hline
\end{tabular}

\label{tbl:spectroscopy_params}
{Note: (a) Values correspond to the mean wavelength of the spectroscopic
observations (5187~\AA).}
\end{table*}

Preliminary single-lined spectroscopic orbital solutions for each visible star gave a much larger RV scatter than expected. For TIC~285853156, we eventually discovered that the velocity residuals for the two visible stars (Aa and Ba) showed the same periodicity of about 151.5 days, representing the outer period of the quadruple. Accounting for this motion greatly improved the residuals of both inner orbits. The situation for TIC~392229331 was analogous: once the outer period of about 144.7 days was found in the residuals from initial orbits for both primary stars, the quality of the fits for the inner orbits improved dramatically.

With these improved solutions and educated guesses at the properties of the secondary components in each inner binary, we applied an extension of TODCOR to three dimensions \citep[TRICOR]{Zucker:1995} to look for additional components.  We were able to detect the lines of star Ab, the M dwarf companion in binary A of TIC~285853156, and measure its radial velocities. Star Bb was not detected, and must be an even fainter, lower main-sequence object. A similar exercise for TIC~392229331 revealed the lines of star Ab in that system, and hints of the fourth star. With an extension of TODCOR to four dimensions \citep[QUADCOR;][]{Torres:2007}, we were then able to measure the velocities of star Bb in most of the spectra.  The RVs for the two quadruple systems are presented in Table~\ref{tab:T0285853156.rvs} and Table~\ref{tab:T0392229331.rvs}. We also measured the flux ratios relative to the brightest object in both systems at the mean wavelength of our observations ($\sim$5187~\AA), which are as follows: for TIC~285853156,
$\ell_{\rm Ab}/\ell_{\rm Aa} = 0.016$ and
$\ell_{\rm Ba}/\ell_{\rm Aa} = 0.062$;
for TIC~392229331,
$\ell_{\rm Ab}/\ell_{\rm Aa} = 0.027$,
$\ell_{\rm Ba}/\ell_{\rm Aa} = 0.959$, and
$\ell_{\rm Bb}/\ell_{\rm Aa} = 0.039$.

\begin{table*}[]
    \centering
    \begin{tabular}{c|cc|cc|cc}
    \hline
    \hline
    Time & RV(Aa) & $\sigma_{RV}$(Aa) & RV(Ab) & $\sigma_{RV}$(Ab) & RV(Ba) & $\sigma_{RV}$(Ba) \\
    {BJD-2,400,000} & \kms & \kms & \kms & \kms & \kms & \kms \\
    \hline
    59979.8660 & 60.76 & 3.36 & -23.44 & 12.09 & -84.81 & 4.93 \\
    60001.6750 & 53.33 & 3.08 & -75.70 & 11.10 & 63.11 & 4.53 \\
    60007.7648 & -20.25 & 3.04 & 52.78 & 10.93 & -9.79 & 4.45 \\
    60028.6322 & -25.58 & 3.78 & 5.40 & 13.59 & 86.59 & 5.54 \\
    60037.6371 & -33.67 & 2.61 & 46.09 & 9.37 & 36.16 & 3.82 \\
    59883.8584 & -30.16 & 2.52 & 26.93 & 9.07 & 69.86 & 3.70 \\
    59902.8640 & 2.37 & 3.28 & -14.63 & 11.82 & 90.93 & 4.82 \\
    59913.8333 & -9.22 & 2.98 & 38.83 & 10.71 & 53.19 & 4.37 \\
    59924.7939 & -13.25 & 2.62 & 89.62 & 9.44 & -40.69 & 3.85 \\
    59928.7897 & 16.32 & 3.92 & 10.37 & 14.13 & -49.36 & 5.76 \\
    59937.9150 & 2.90 & 2.86 & 44.53 & 10.30 & 18.44 & 4.20 \\
    59954.7595 & 1.28 & 2.68 & 78.33 & 9.64 & -35.91 & 3.93 \\
    60191.9499 & 39.89 & 2.49 & -90.55 & 8.96 & -10.40 & 3.65 \\
    60208.0039 & -19.94 & 2.80 & 66.83 & 10.10 & 10.23 & 4.12 \\
    60227.9469 & -9.26 & 2.22 & 59.55 & 8.02 & 75.26 & 3.27 \\
    60236.9528 & -11.12 & 2.76 & 77.91 & 9.92 & 51.44 & 4.05 \\
    60245.9224 & -7.47 & 2.21 & 85.43 & 7.97 & 25.53 & 3.25 \\
    60253.7804 & 27.40 & 2.40 & 20.37 & 8.60 & -61.64 & 3.51 \\
    60263.9580 & 26.29 & 3.47 & 22.12 & 12.46 & -62.93 & 5.08 \\
    60272.7940 & 66.43 & 2.40 & -42.31 & 8.64 & -69.96 & 3.52 \\
    60282.8397 & 66.44 & 2.24 & -41.03 & 8.06 & 38.11 & 3.29 \\
    60284.8778 & 9.57 & 2.72 & 72.93 & 9.79 & -7.78 & 4.00 \\
    60291.8005 & 74.73 & 2.51 & -72.75 & 9.03 & 20.65 & 3.69 \\
    60295.8181 & -5.45 & 3.38 & 89.41 & 12.20 & -69.67 & 4.97 \\
    60302.7970 & 55.95 & 3.01 & -74.47 & 10.83 & -38.43 & 4.41 \\
    60307.8837 & -19.74 & 2.80 & 66.03 & 10.10 & 10.39 & 4.12 \\
    60327.6780 & -44.66 & 3.12 & 43.41 & 11.20 & -9.46 & 4.57 \\
    60338.7193 & -29.56 & 2.27 & 39.70 & 8.15 & 24.09 & 3.32 \\
    60344.8010 & -24.29 & 2.43 & 5.72 & 8.74 & 69.20 & 3.56 \\
    60353.6400 & 21.10 & 2.40 & -47.67 & 8.64 & 64.22 & 3.52 \\
    60356.6360 & -30.91 & 2.61 & 71.75 & 9.38 & 64.45 & 3.83 \\
    60363.6776 & 26.93 & 2.29 & -35.33 & 8.23 & 62.53 & 3.35 \\
    60370.7316 & 25.37 & 2.40 & -20.44 & 8.60 & 49.84 & 3.51 \\
    60411.6232 & 63.65 & 4.25 & -48.63 & 15.25 & 45.53 & 6.22 \\
    60418.6434 & 10.03 & 2.26 & 71.05 & 8.09 & 50.43 & 3.29 \\
    \hline
    \end{tabular}
    \caption{TRES radial velocity measurements for TIC~285853156. Aa and Ab are the primary and secondary stars of the 10-day binary, Ba is the primary star of the 1.77-day binary}
    \label{tab:T0285853156.rvs}
\end{table*}

\begin{table*}[]
\centering
\begin{tabular}{c|cc|cc|cc|cc}
\hline
\hline
Time & RV(Aa) & $\sigma_{RV}$(Aa) & RV(Ab) & $\sigma_{RV}$(Ab) & RV(Ba) & $\sigma_{RV}$(Ba) & RV(Bb) &  $\sigma_{RV}$(Bb)\\
{BJD-2,400,000} & \kms & \kms & \kms & \kms & \kms & \kms & \kms & \kms\\
\hline
59478.0083  &  -9.13  &  2.10  &  -138.53  & 8.08 &  -40.96  &  2.11  &  122.17  &  7.00  \\
59486.8580  &  -73.55  &  1.57  &  24.04  & 6.02 &  -72.31  &  1.57  &  ...  &   ...  \\
59493.9392  &  -109.66  &  1.26  &  13.75  & 6.56 &  -33.89  &  1.26  &  57.27  &  4.18  \\
59497.8583  &  -41.69  &  1.71  &  ...  & ... &  -74.82  &  1.71  &  123.21  &  5.69  \\
59502.9676  &  -108.77  &  1.56  &  145.03  & 6.00 &  -41.39  &  1.56  &  67.98  &  5.20  \\
59507.7799  &  29.12  &  2.12  &  -95.88  & 8.13 &  27.69  &  2.12  &  -66.69  &  7.05  \\
59517.8859  &  -30.94  &  1.63  &  ...  & ... &  -40.53  &  1.63  &  30.54  &  5.42  \\
59520.8497  &  -54.85  &  1.20  &  80.53  & 4.59 &  -85.45  &  1.20  &  95.61  &  3.98  \\
59523.8717  &  80.61  &  1.81  &  -176.95  & 6.96 &  64.36  &  1.82  &  -160.74  &  6.03  \\
59525.7085  &  83.76  &  1.68  &  ...  & ... &  -13.18  &  1.68  &  -46.77  &  5.58  \\
59527.8088  &  46.19  &  1.45  &  -114.47  & 5.59 &  -52.60  &  1.46  &  26.46  &  4.84  \\
59531.7577  &  -41.37  &  1.55  &  ...  & ... &   -105.06  &  1.56  &  121.00  &  5.18  \\
59534.8558  &  88.80  &  1.60  &  -163.42  & 6.15 &  10.75  &  1.60  &  -80.28  &  5.32  \\
59547.7351  &  81.21  &  1.72  &  ...  & ... &   -117.82  &  1.73  &  109.87  &  5.74  \\
59567.7015  &  100.54  &  1.75  &  -145.46  & 6.73 &   -103.11  &  1.76  &  68.78  &  5.83  \\
59584.7696  &  -57.65  &  1.50  &  174.06  & 5.77 &  38.91  &  1.50  &  -181.60  &  5.00  \\
59596.5942  &  60.52  &  1.74  &  -135.48  & 6.69 &  19.44  &  1.74  &  -76.35  &  5.79  \\
59605.7724  &  13.02  &  1.54  &  -224.90  & 5.92 &  42.90  &  1.54  &  22.43  &  5.13  \\
59620.7043  &  9.83  &  1.65  &  -176.28  & 6.36 &  82.35  &  1.66  &  -63.35  &  5.51  \\
59623.6571  &  -50.36  &  1.26  &  ...  & ... &  63.55  &  1.26  &  -43.40  &  4.20  \\
59632.6259  &  -76.16  &  1.83  &  37.08  & 7.04 &  63.72  &  1.84  &  -88.40  &  6.10  \\
59650.6441  &  -101.63  &  2.03  &  134.52  & 7.81 &  52.38  &  2.04  &  -104.29  &  6.77  \\
59657.6555  &  -87.04  &  1.63  &  ...  & ... &  -4.60  &  1.63  &  -26.79  &  5.42  \\
59934.8229  &  -112.77  &  1.26  &  153.69  & 4.84 &  76.78  &  1.26  &  -133.99  &  4.19  \\
59969.6798  &  -61.55  &  1.60  &  103.69  & 6.14 &   -112.46  &  1.60  &  127.59  &  5.32  \\
59980.6834  &  -36.34  &  1.76  &  105.20  & 6.76 &  -98.29  &  1.76  &  89.04  &  5.86  \\
60008.6616  &  106.36  &  1.66  &  ...  & ... &  -20.72  &  1.67  &  -92.17  &  5.54  \\
59912.8271  &  -128.34  &  1.88  &  91.57  & 7.24 &  -12.03  &  1.89  &  84.48  &  6.27  \\
59923.7433  &  -110.50  &  1.62  &  106.85  & 6.24 &  56.72  &  1.63  &  -75.74  &  5.41  \\
59928.7592  &  16.08  &  1.87  &  -120.28  & 7.20 &  -58.43  &  1.88  &  106.71  &  6.24  \\
59954.6351  &  -68.70  &  1.68  &  104.05  & 6.45 &  31.45  &  1.68  &  -97.31  &  5.58  \\
60191.9345  &  -143.90  &  1.65  &  ...  & ... &  127.52  &  1.66  &  -116.93  &  5.51  \\
60207.9749  &  -106.34  &  1.57  &  82.89  & 6.05 &  79.52  &  1.58  &  -95.22  &  5.24  \\
60222.9195  &  -98.71  &  1.13  &  109.99  & 4.35 &  -11.69  &  1.13  &  20.17  &  3.76  \\
60227.9264  &  -63.80  &  1.56  &  63.46  & 6.00 &  77.99  &  1.56  &  -148.89  &  5.20  \\
60235.9479  &  -19.96  &  1.51  &  1.99  & 5.81 &  -98.98  &  1.52  &  145.15  &  5.04  \\
60243.9884  &  45.54  &  1.53  &  -119.51  & 5.89 &  60.47  &  1.54  &  -152.09  &  5.10  \\
60252.7088  &  76.82  &  1.42  &  -154.57  & 5.46 &  59.40  &  1.42  &  -167.77  &  4.73  \\
60271.7361  &  -55.93  &  1.37  &  133.17  & 5.27 &  -92.23  &  1.37  &  76.14  &  4.56  \\
60284.6853  &  -75.45  &  1.89  &  175.39  & 7.28 &  26.89  &  1.90  &  -149.30  &  6.31  \\
60293.8079  &  -72.24  &  1.17  &  179.60  & 4.51 &  5.48  &  1.18  &  -129.82  &  3.91  \\
60302.7587  &  -53.43  &  1.25  &  163.82  & 4.80 &  17.43  &  1.25  &  -155.08  &  4.16  \\
60305.7551  &  104.98  &  1.18  &  -132.10  & 4.54 &   -124.96  &  1.18  &  99.83  &  3.93  \\
60316.6460  &  92.82  &  1.79  &  -142.79  & 6.88 &  -45.02  &  1.79  &  -19.46  &  5.96  \\
60339.6900  &  -107.40  &  1.33  &  35.44  & 5.09 &  -51.11  &  1.33  &  172.54  &  4.41  \\
60353.6178  &  -66.93  &  1.49  &  21.05  & 5.72 &  -37.74  &  1.49  &  113.29  &  4.96  \\
60370.6641  &  -66.42  &  1.68  &  50.47  & 6.47 &  16.09  &  1.69  &  -51.03  &  5.61  \\
\hline
\end{tabular}
\caption{TRES radial velocity measurements for TIC~392229331. Aa and Ab are the primary and secondary stars of the 1.82-day binary, Ba and Bb are the primary and secondary stars of the 2.26-day binary.}
\label{tab:T0392229331.rvs}
\end{table*}


The A-type stars in TIC~392229331 have unexpectedly slow rotations. For typical radii of $\sim$1.7~$R_{\odot}$ each, spin-orbit alignment and synchronization would predict equatorial rotational velocities of about 40~\kms\ for Ba and 50~\kms\ for Aa. The measured values are several times smaller, particularly for Ba. This would imply that the stars are not spinning synchronously with the orbit, or that their spin axes are tilted, or a combination of both. We note also that abnormally slow rotation in stars of spectral type A (and early F) is often seen in the class of chemically peculiar ``metallic-line'' stars, which are overwhelmingly found in binaries \citep[see, e.g.,][and references therein]{Abt:2000}.  While a detailed chemical analysis of TIC~392229331 is beyond the scope of this paper, it would be of considerable interest in order to confirm this possibility, challenging as it may be, given the composite nature of the spectra.

\subsubsection{Archival Photometry}

In addition to the TESS photometry, we also examined the publicly-available photometric data from ground-based surveys such as ASAS-SN \citep{2017PASP..129j4502K}, ATLAS \citep{2018AJ....156..241H}, DASCH \citep{2012IAUS..285...29G}, and SuperWASP \citep{2006PASP..118.1407P}. The detection of eclipses in archival photometry confirms that the corresponding orbital inclinations have not changed dramatically during the observations, which span almost 137 years for the case of TIC 392229331. A summary of the archival photometry is provided in Table \ref{tbl:archival} .

\begin{table}[h]
\centering
\caption{Archival Data$^a$}
\begin{tabular}{lccc}
\hline
\hline
Source & ASAS-SN & DASCH & ATLAS  \\
\hline

 \hline
   TIC 285853156$^b$ & 3722 points  & ... & 2720 points \\ 
   10.02 d & 10.026068 d &... & not detected \\
   1.76 d  & not detected & ... & not detected \\

 \hline
   TIC 392229331$^c$ & 2343 points & 4151 points & 2580 points \\  
 1.82  d & 1.822387 d  & 1.822375 d & too bright  \\
 2.25  d & 2.256562 d  & 2.256540 d & too bright \\
 \hline
\hline
\label{tbl:archival}
\end{tabular}

{Notes. (a) Line one under each source lists the number of archival data points. Line two is for the A binary, and line three is for the B binary.  Only data sets with more than 1000 archival points were considered, otherwise the entries are marked with an ellipsis. Entries marked as ``too bright'' indicate that saturation effects have set in.  In all cases, the phases of the archival eclipses are consistent, within the uncertainties, with those found with {\it TESS}. (b) G = 12.1. (c) G = 10.6.}

\end{table}


\section{Photodynamical Modeling of the Systems}
\label{sec:photodynamics}


{\tt Lightcurvefactory} \citep{2013MNRAS.428.1656B,2017MNRAS.467.2160R,2018MNRAS.478.5135B} is a code developed over the past decade to emulate (i) photometric eclipsing lightcurves, (ii) eclipse timing variations, and (iii) radial velocity variations in a multi-stellar system, and fit the results to the existing data for a given source (see also \citealt{Borkovits2019,2019MNRAS.487.4631B,Borkovits2020,2020MNRAS.496.4624B,2021MNRAS.503.3759B}).  The code uses a numerical (small-)N-body integrator (a seventh-order Runge-Kutta-Nystr\"om algorithm) for stellar systems with 2, 2+1, 2+2, 2+1+1 and 2+2+2 configurations.  (iv) The spectral energy distribution (SED) is also fit within {\tt Lightcurvefactory} given that the trial values of the stellar parameters and distance are available.

The code can also (optionally) make use of the {\tt PARSEC} stellar evolution tracks and isochrones \citep{2012MNRAS.427..127B}.  These relate the stellar mass, age, and chemical composition to  the stellar radius and effective temperature, $T_{\rm eff}$. The SED fitting further ties the stellar parameters to typically readily available archival photometric observations.  The use of these evolutionary tracks requires the assumption that the evolution of all the stars in the system have evolved in a coeval manner (in particular, that there has been no prior episode of mass transfer).

Especially with good RV and ETV data, it is possible in quadruple star systems to fit accurate values for 12 stellar parameters ($R$, $T_{\rm eff}$, $M$ for each of four stars), as well as 15 orbital parameters (5 for each of two binary orbits and one outer orbit): orbital period, $P$, eccentricity, $e$, argument of periastron, $\omega$, time of periastron passage, $\tau$, and observational orbital inclination angle, $i$. Extra light from each binary can also optionally be set, but in this case was unused. In addition, the code fits for the following 4 system parameters: distance, age, metallicity, and $E(B-V)$. When RV data are available, the systemic RV, $\gamma$, must also be fitted.  This yields a total of 
$$12 + 17 + 4 + 1 = 34 ~~{\rm fitted ~parameters}. $$  The code uses a Markov Chain Monte Carlo (MCMC)-based search routine to find the best-fit system parameters and their statistical uncertainties.  This part of the code uses our own implementation of the generic Metropolis-Hastings algorithm \citep[see, e.g., ][]{2005AJ....129.1706F}.  We report the median of the distributions of posteriors and the rms of these distributions as the corresponding uncertainties. 

Aside from the more familiar inferences of parameter determinations that are possible in binary systems with photometric and RV data, we note a few special effects that we utilize here for the two current quadruple systems (as well as for triples in other works). \\
\noindent
(i). In addition to the usual mass-function information from RVs, the contribution to the ETVs from light travel time effects (LTTE) can either further enhance the RVs (if the motion of the same object is being measured) or add independent supplementary information if the ETVs are of the reflex-partner to the object whose RVs have been measured. \\
(ii). Furthermore, there is a “dynamical delay” component to the ETVs which has no counterpart in RV measurements.  This is basically due to the dynamical effects (including an increase in the orbital period of the binary) from the presence of a third body (or another binary).  Among other things, this yields information on the ratio of the mass of the “other object" to the mass of the composite system. \\
(iii). The SED fits yield relations among such quantities as the sum of $R^2T_{\rm eff}$ and the flux at long wavelengths (i.e., in the Rayleigh-Jeans tail), and $R^2T_{\rm eff}^4$ (when integrated over the entire spectrum).\\
(iv). The photometric shapes/profiles of any third (and fourth) body transits (not applicable in the two current systems) can provide crucial information about the mutual inclination angles. \\
(v). The precession of the orbital planes or lines of the apsidal nodes provides further dynamical information on  $P_{\rm out}^2/P_{\rm in}$ timescales about the system.

Finally, for the more in-depth, comprehensive analysis of the systems via {\tt Lightcurvefactory}, we reprocessed the \textit{TESS} full frame image (FFI) data using the convolution-based differential image analysis methods of the \texttt{fitsh} package \citep{2012MNRAS.421.1825P}.

\section{System Parameter Results}
\label{sec:results}

In the next two subsections we describe the results of the photodynamic analysis of first TIC 285853156, and then TIC 392229331.  For each quadruple system we present five types of results.  These are (i) the RV measurements vs.~time; (ii) the RVs as a function of orbital phase for each of the three binary subsystems (the two EBs and the outer orbit); (iii) the ETV curves; (iv) illustrative photodynamical fits to segments of the \textit{TESS} lightcurve; and (v) the composite spectral energy distribution (SED) for the quadruple.  In all five presentations we show both the measured values and the model curves generated by {\tt Lightcurvefactory}.  For each quadruple, we also present a table summarizing all the system parameters with uncertainties. 

\subsection{The Quadruple TIC 285853156}

For TIC 285853156 we summarize the system parameters and uncertainties in Table \ref{tbl:simlightcurve285853156}, as determined by {\tt Lightcurvefactory}.  TIC 285853156 is now the third most compact known quadruple, with an outer period of only 151.7 days.  A particularly interesting feature of TIC 285853156 is the period and eccentricity of the orbit of binary A (10.0 days and 0.23, respectively).  While this orbit is unexciting for a isolated binary, or even as part of a wide multiple system, its existence as part of a very tight quadruple system with a particularly short-period outer orbit is rather remarkable.  While we will discuss the stability of the system further in Section \ref{sec:stability}, we note here that neither the period ratio stability criterion of Equation \ref{eqn:p} nor the semi-major axis ratio stability criterion of Equation \ref{eqn:a} are strictly met by TIC 285853156.  However, we demonstrate the long-term stability of the system through simulation, again discussed in section \ref{sec:stability}.  Further validation of the long-term stability of the system is provided by its age at 3.8 Gyr.

The system flux is dominated by the G-type primary star of binary A, with a mass of 1.072 $M_\odot$, that contributes $\sim$85.5\% of the system light. The primary of binary B has a mass of 0.731 $M_\odot$ and contributes $\sim$10.9\% of the system light.  The secondaries of each binary are substantially smaller M-type stars, with $M_{Ab}/M_{Aa}=0.495$ and $M_{Bb}/M_{Ba}=0.566$.  The secondary of binary B, with a mass of 0.414 $M_\odot$, is not reliably detectable in the spectra, as we noted in Section \ref{section:spectroscopy}.  As such, we were only able to obtain the RVs from the primary of binary B.

\begin{figure}
   \centering
   \includegraphics[width=.99\columnwidth]{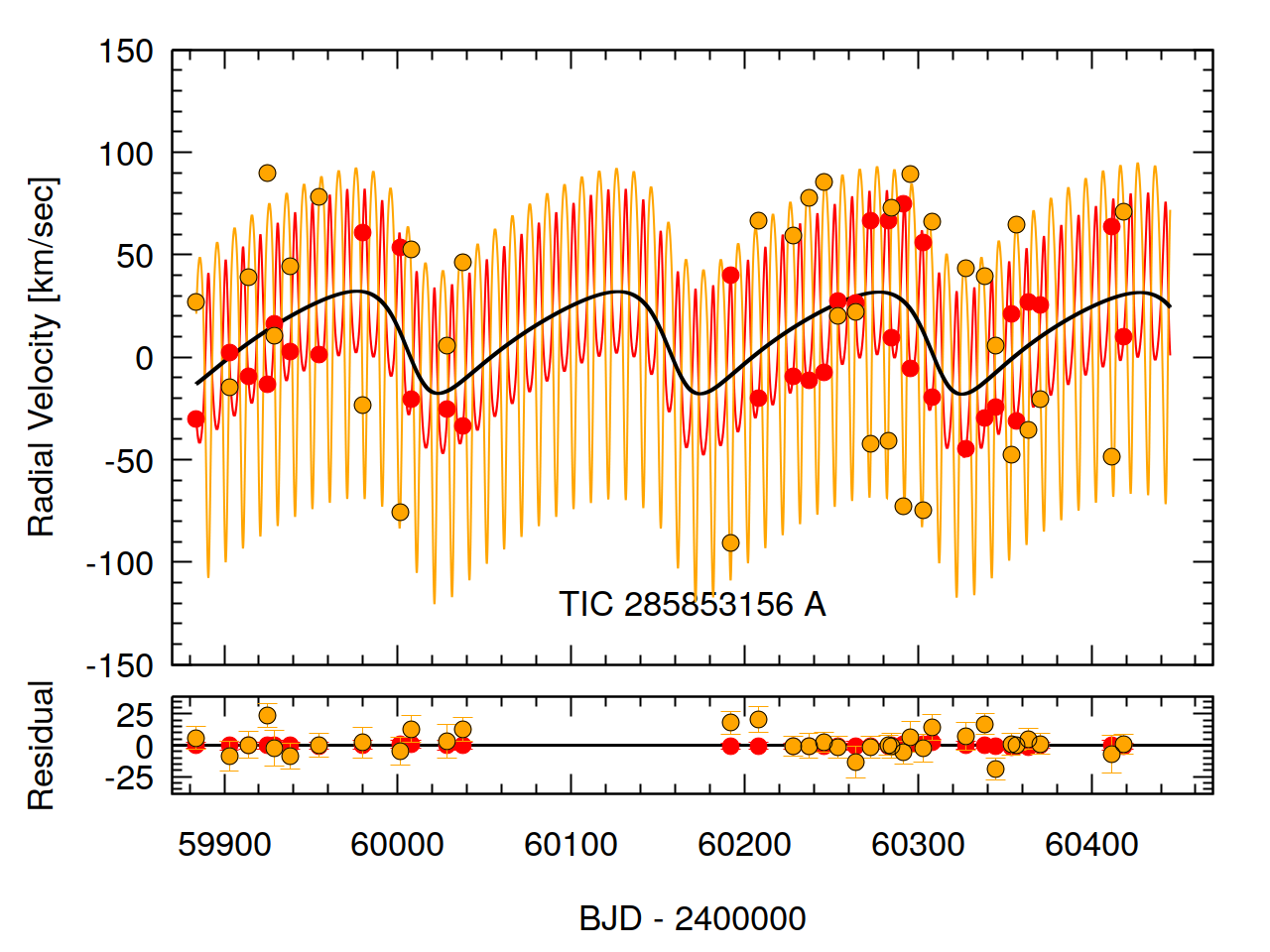}
   \includegraphics[width=.99\columnwidth]{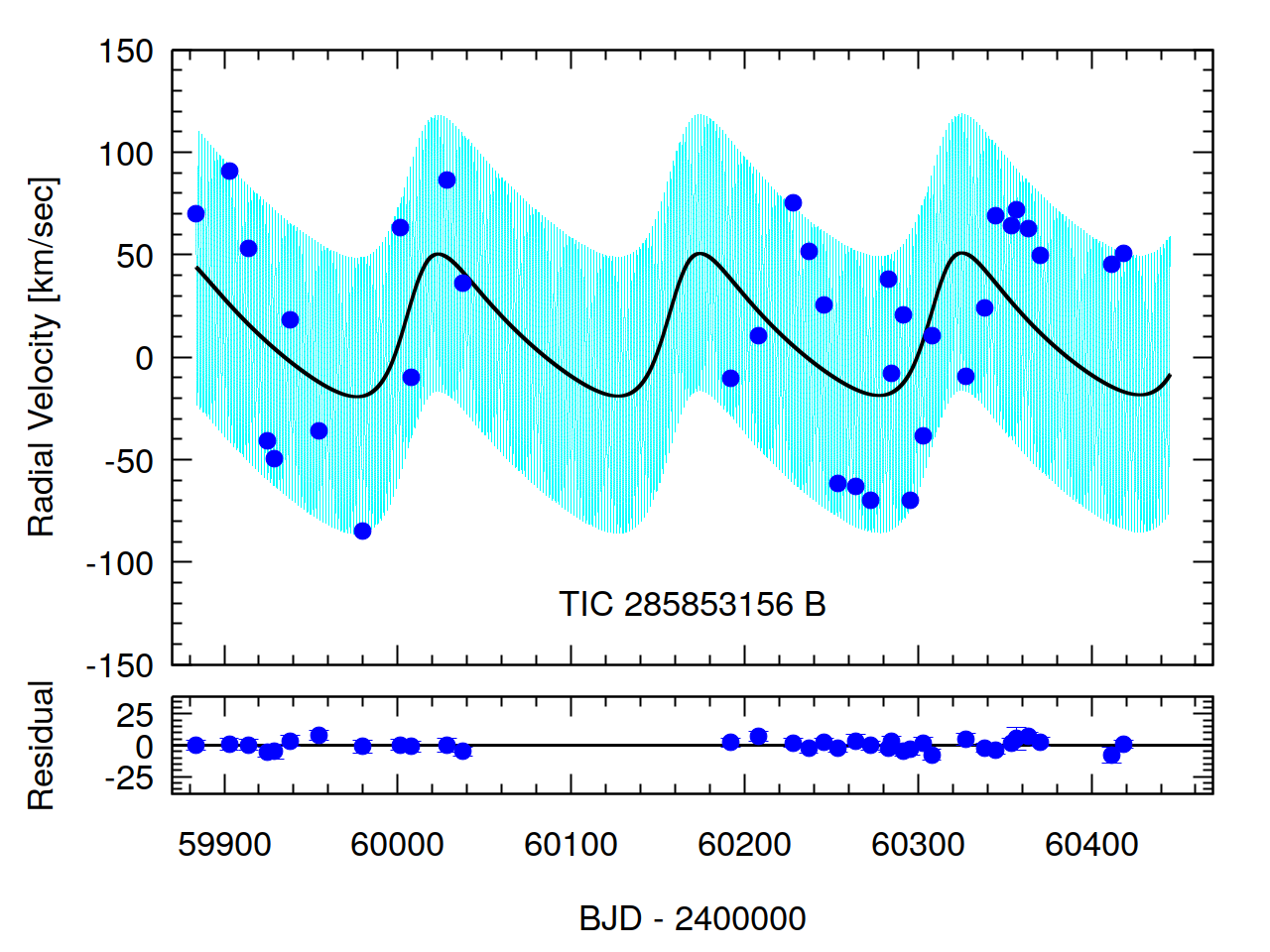}
   \caption{Radial velocity data vs.~time and model fits for TIC 285853156 binaries A ({\em upper panel}) and B ({\em lower panel}) shown over the full duration of the collected spectra (see Table \ref{tab:T0285853156.rvs}).  Points are measured RVs for star Aa (red) and Ab (gold) in the top panel, and blue points are for star Ba in the bottom panel. Smooth curves are the model fits.  The black curves represents the motion of binaries A and B around the system center of mass in the top and bottom panels, respectively.}
   \label{fig:285853156_rvt}
\end{figure}

In Figure \ref{fig:285853156_rvt}, we show the 35 measured RVs and model fits for both binaries of TIC 285853156 as a function of time.  The orange and gold curves in the top panel are the model curves for the primary and secondary of binary A, respectively, while the blue points and curve in the bottom panel are for the primary of binary B. The black curves show the radial velocity motion of binary A (top panel) and binary B (bottom panel) around the system center of mass.  

\begin{figure}
   \centering
   \includegraphics[width=.99\columnwidth]{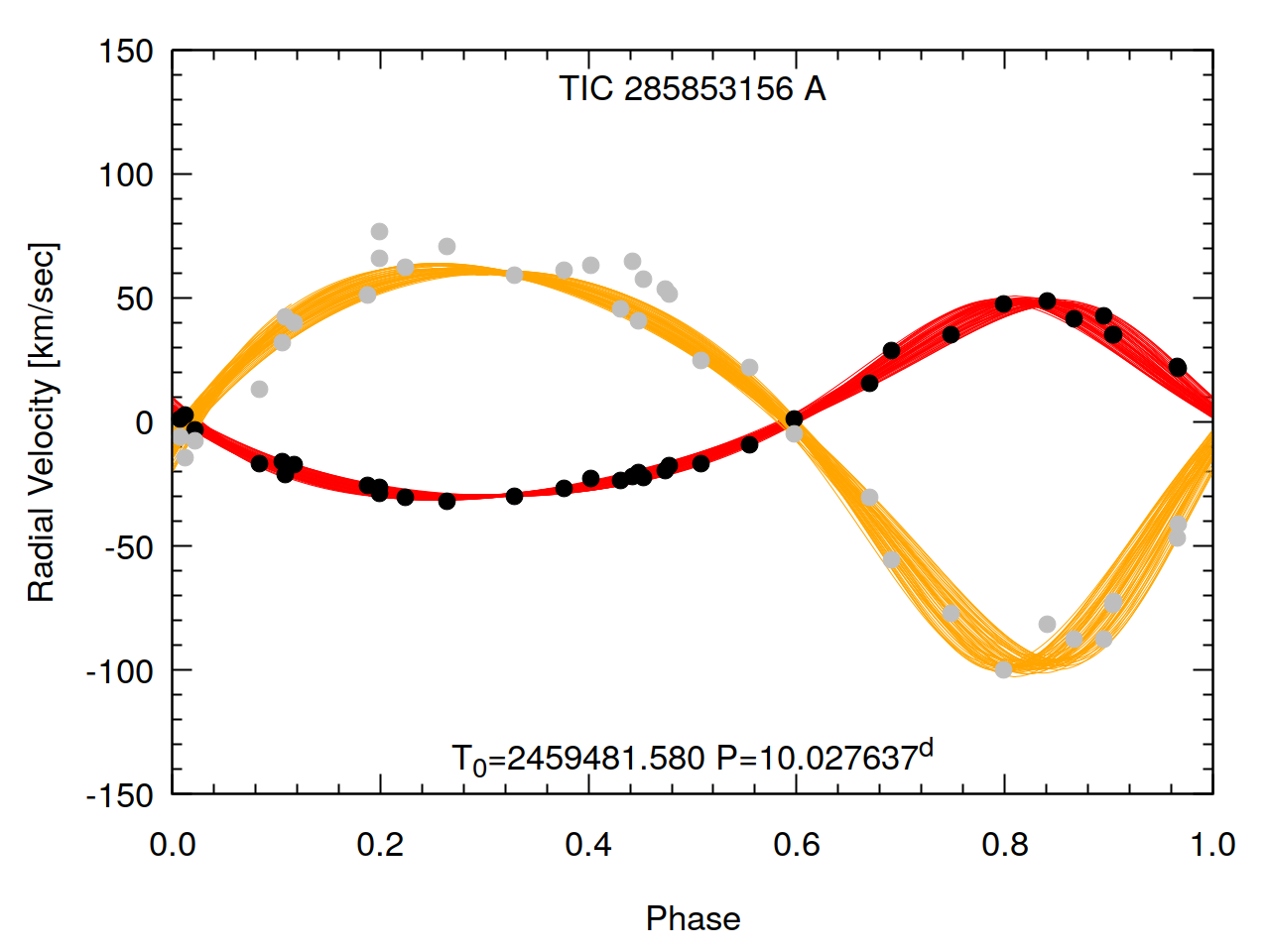}
   \includegraphics[width=.99\columnwidth]{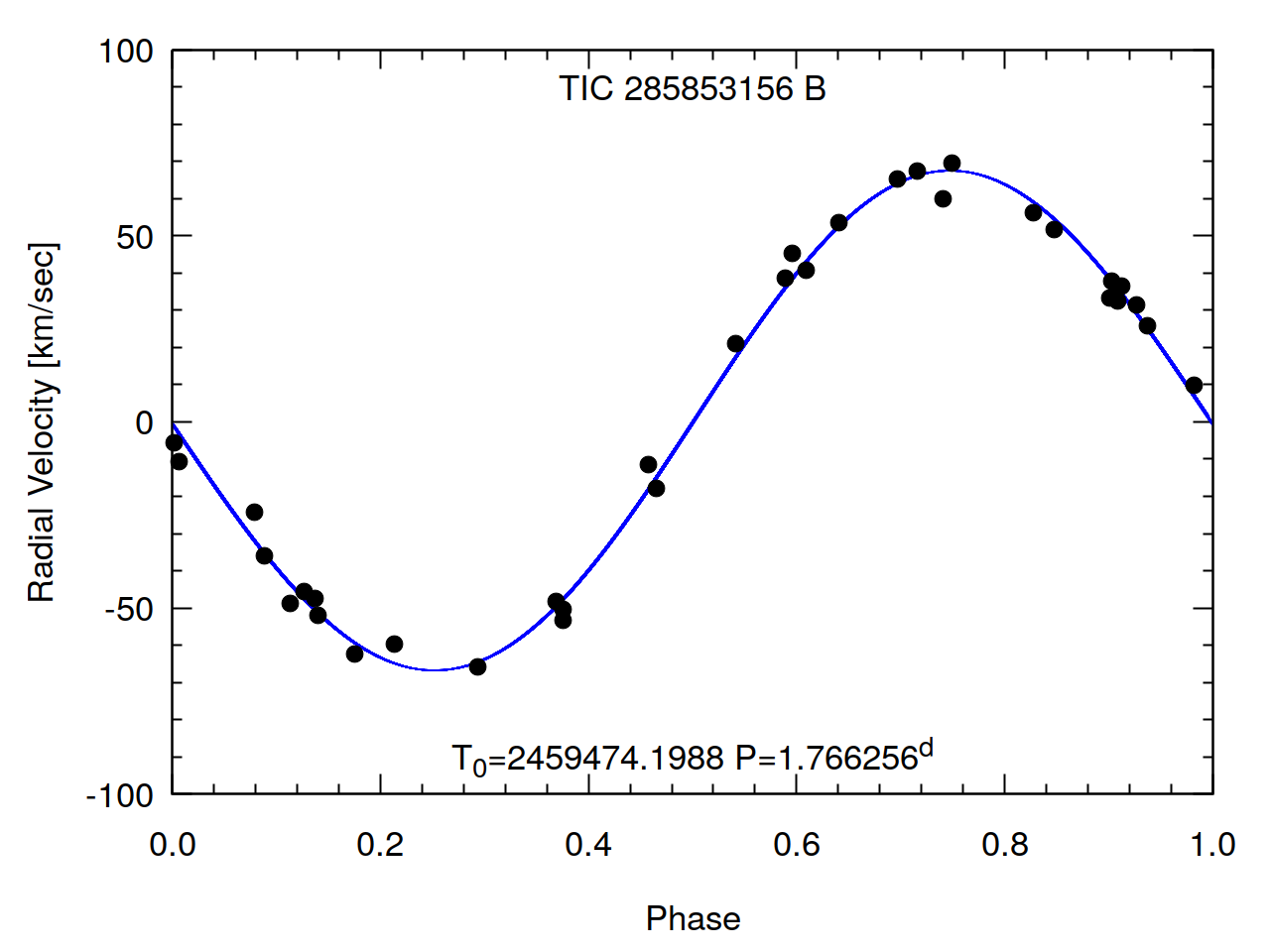}
   \includegraphics[width=.99\columnwidth]{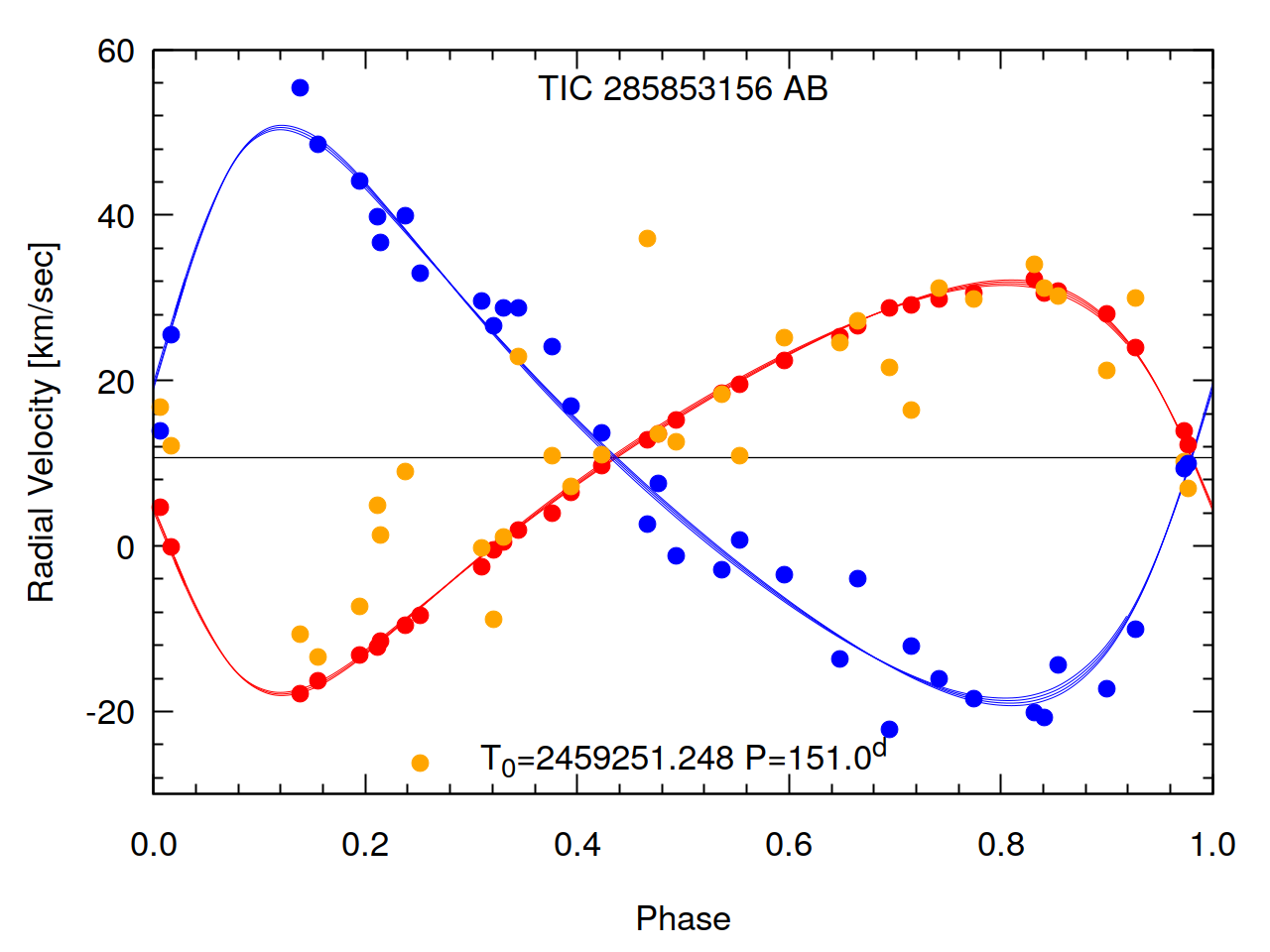}
   \caption{Radial velocity data vs.~orbital phase and model fits for TIC 285853156.  The RV data and the corresponding model fits are shown in the center of mass reference frame of Binary A ({\em top panel}), of binary B ({\em middle panel}), and of the quadruple AB ({\em bottom panel}). Points in the bottom panel correspond to those described in Figure \ref{fig:285853156_rvt}. Note that binary B is single-lined due to the relatively low luminosity of the secondary, Bb.  Driven apsidal motion can be seen in the model curves for Binary A. }
   \label{fig:285853156_rvp}
\end{figure}

Figure \ref{fig:285853156_rvp} shows the same RV data but for Aa and Ab in the center of mass of binary A (top panel), for Ba in the center of mass of binary B (middle panel), and for binary A orbiting binary B in the system center of mass (bottom panel).  The model curves show the dynamically forced apsidal motion in binary A as well as in the outer orbit.

\begin{figure}
   \centering
   \includegraphics[width=.99\columnwidth]{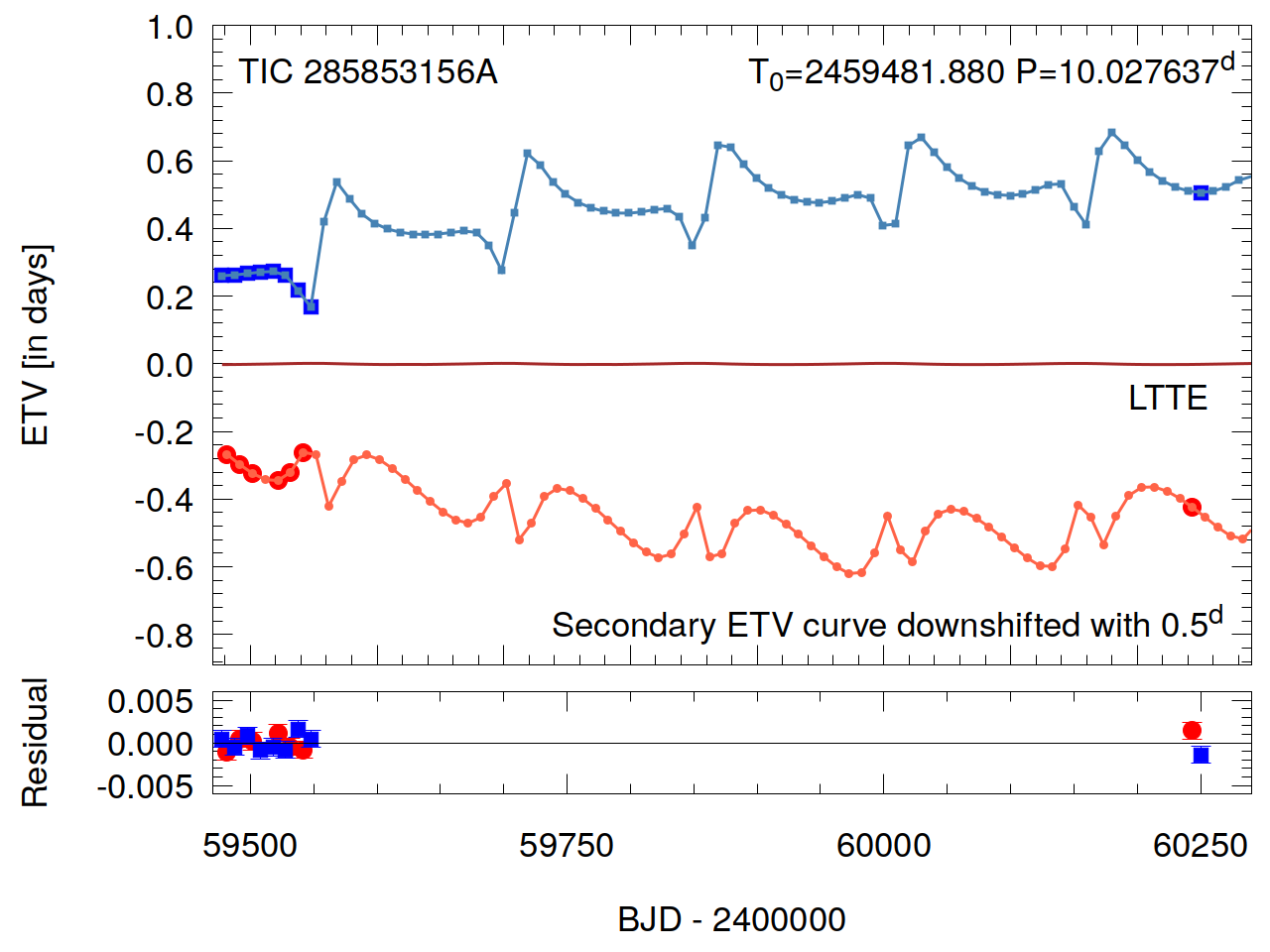}
   \includegraphics[width=.99\columnwidth]{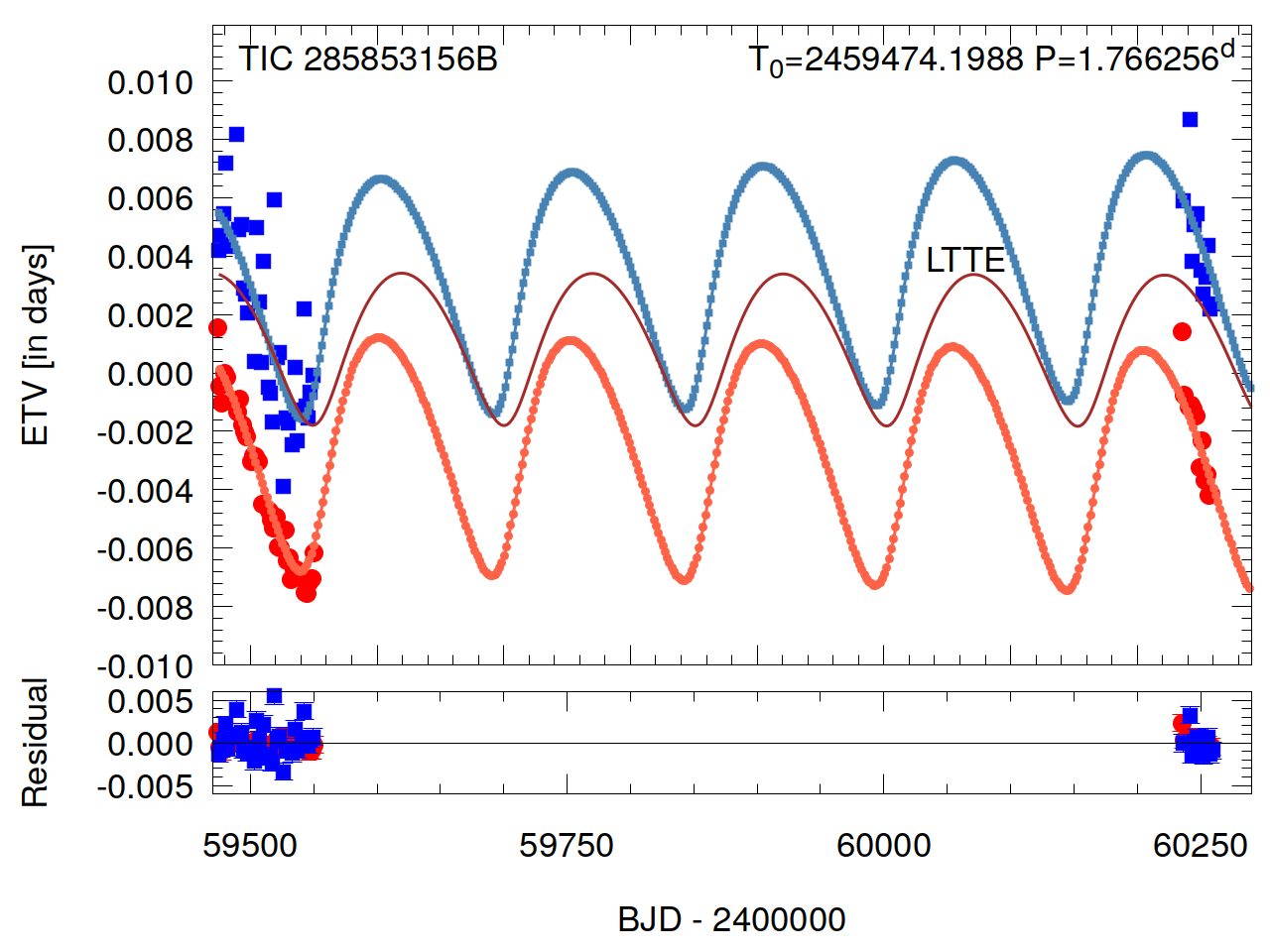}
   \caption{Measured ETV points and the corresponding model fits for TIC 285853156.  The results for binary A and binary B are displayed in the top panel and bottom panel, respectively.  The ETVs and for the primary and secondary eclipses are shown with red circles and blue squares, respectively. The smooth model ETV curves for the primary and secondary eclipses are shown in the same corresponding colors.  The A binary is dominated by dynamical delays, while for binary B, the contribution of the LTTE and DE to the ETVs are more comparable. The thin brown curve shows the LTTE contribution to the ETVs by itself. Residuals from the fit are shown in the bottom section of each panel.}
   \label{fig:285853156_etv}
\end{figure}

In Figure \ref{fig:285853156_etv}, we show the measured ETVs for binary A (top panel) and binary B (bottom panel).  The  ETV data are sparse with a large gap between the two \textit{TESS} observations of about two years.  Nonetheless, the outer model for the 152-d outer orbit is secure because of all the other information being fit simultaneously with {\sc Lightcurvefactory}. The binary A ETV curve is dominated by dynamical effects (DE) due to the long period of binary A and the relatively short outer period, $P_{\rm out}$\footnote{Equation (6) of \citet{2022Galax..10....9B}, shows that the amplitude of the DE is directly proportional to $P_{\text{in}}^{2}/P_{\text{out}}$}.  In contrast, the ETV curve for binary B is a combination of the Light Travel Time Effect (LTTE) and dynamical delays.  (For a thorough discussion of ETVs, as well as the separate LTTE and DE contributions to the ETVs, see \citealt{2022Galax..10....9B}.) The LTTE contribution to the delays is shown as a thin brown curve labeled ``LTTE.''  We also note the diverging ETV curves for the primary and secondary eclipses, which indicates apsidal motion in that binary.  

\begin{figure}
   \centering
   \includegraphics[width=.99\columnwidth]{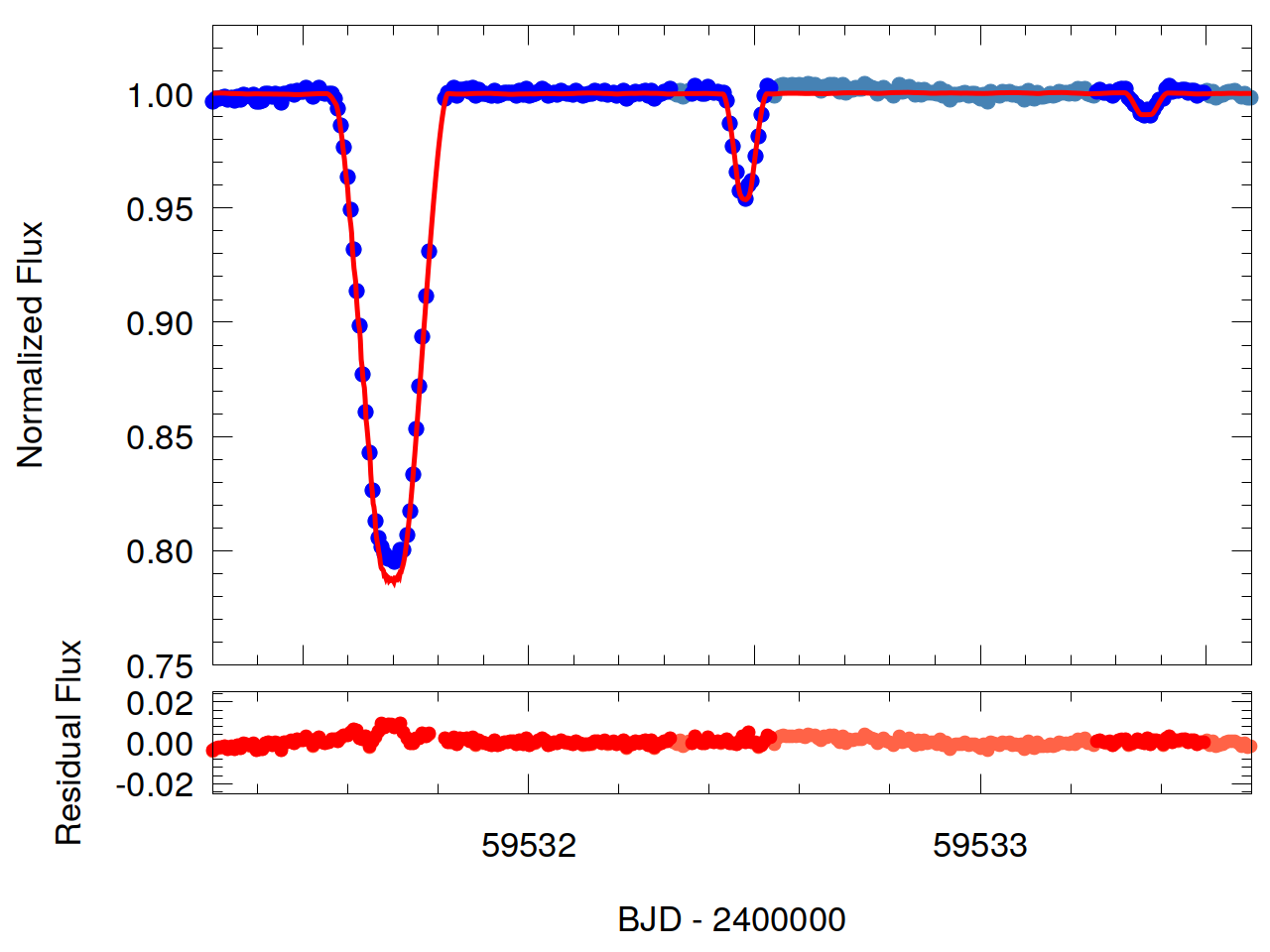}
   \includegraphics[width=.99\columnwidth]{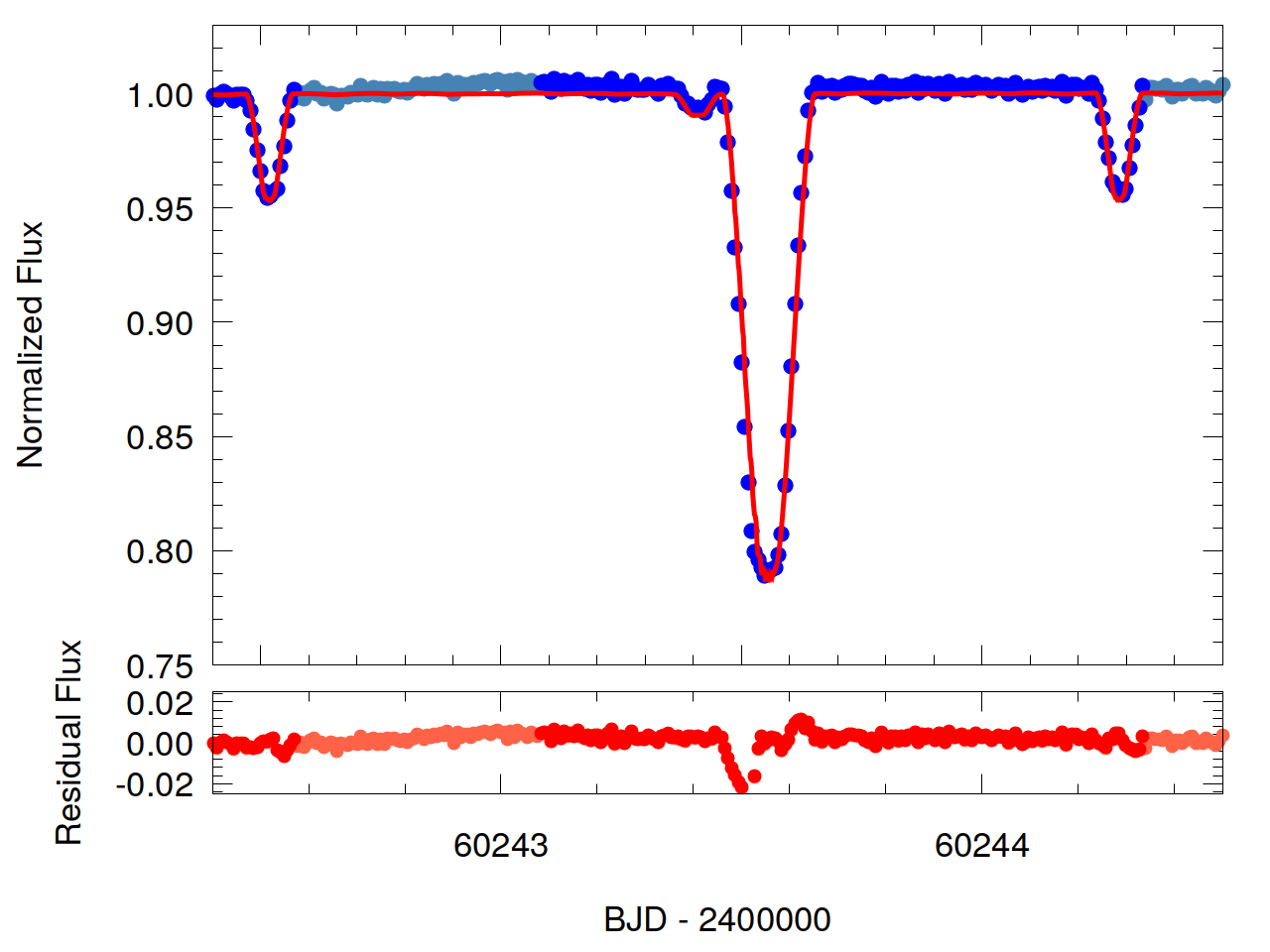}
   \caption{Lightcurve segments for TIC 285853156 from sector 45 ({\em top panel}) and sector 71 ({\em bottom panel}).  Lightcurve data points used for the fit (dark blue circles) and lightcurve data points not used for the fit (light blue circles) are plotted with the model light curve (red line) superposed.  Residuals of the data from the fit are shown in the bottom section of each panel.}
   \label{fig:285853156_lc}
\end{figure}

Two segments of the \textit{TESS} lightcurve from sectors 45 and 71 are shown in Figure \ref{fig:285853156_lc}.  The deeper eclipses in this figure are, naturally, caused by occultations of star Aa, which produces $\sim$85\% of the system light, by star Ab.  The others are the primary and secondary eclipses of binary B.  Superposed on the lightcurve is the photometric model from the global fit to all the data by {\sc Lightcurvefactory}.  

\begin{figure}
   \centering
   \includegraphics[width=.99\columnwidth]{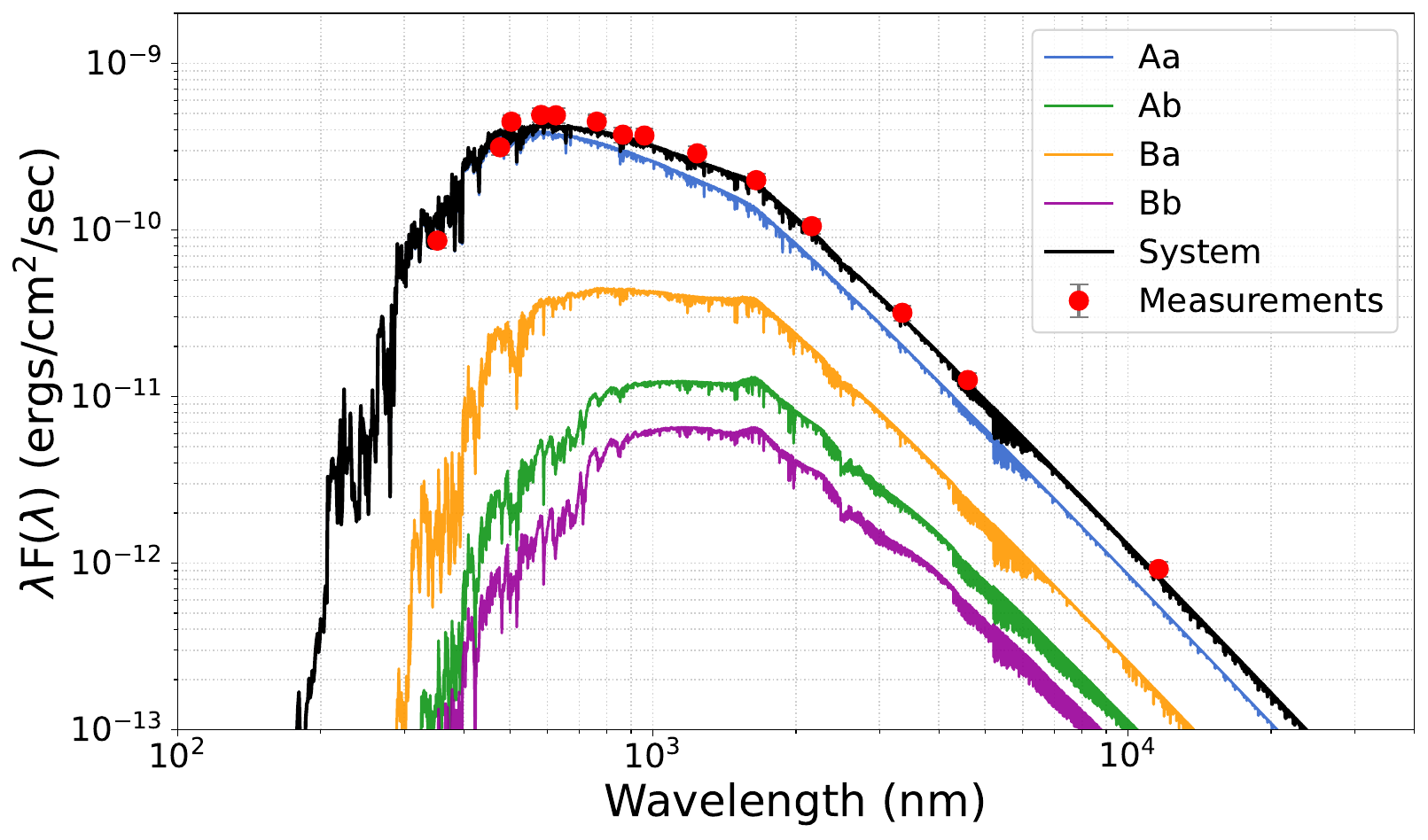}
   \caption{SED data points and model fit for TIC 285853156.  The SED measurements are shown as red circles, while the model contributions from stars Aa (blue), Ab (green), Ba (orange), Bb (purple), and the total system (black) are shown superposed on the data. The system age is 3.8 Gyr.}
   \label{fig:285853156_sed}
\end{figure}

Finally, the model SED fit is shown in Figure \ref{fig:285853156_sed}.  We were able to retrieve 19 SED measurements spanning 0.43 to 11.6 microns on VizieR \citep{vizier2000}.  The figure shows the separate (model) contributions to the flux from all four stars, as well as the total system flux.  By itself, there is insufficient information in SED data like these to find the parameters of four constituent stars without additional information.  That is because, even under the assumption that all the stars in the system are coeval, there are still at least six-seven free parameters to fit, including four masses, one system age, the interstellar extinction, $A_V$, and possibly the distance. However, {\sc Lightcurvefactory} is able to work with many other input parameters; thus, the SED fit is just one of the ingredients contributing to the overall solution.

\begin{table*}
\centering
\caption{Median values of the parameters of TIC 285853156 from the double EB simultaneous lightcurve,  double ETV, joint SED and \texttt{PARSEC} evolutionary track solution from {\sc Lightcurvefactory}.}
\begin{tabular}{lccccc}
\hline
\multicolumn{6}{c}{Orbital elements$^a$} \\
\hline
   & \multicolumn{3}{c}{subsystem}  \\
   & \multicolumn{2}{c}{A} & \multicolumn{2}{c}{B} & A--B \\
  \hline
$P_\mathrm{a}$ [days]            & \multicolumn{2}{c}{$9.99873_{-0.00042}^{+0.00042}$} & \multicolumn{2}{c}{$1.7660006_{-0.0000021}^{+0.0000022}$}       & $151.70_{-0.11}^{+0.11}$ \\
semimajor axis  [$R_\odot$]      & \multicolumn{2}{c}{$22.87_{-0.15}^{+0.10}$}           & \multicolumn{2}{c}{$6.436_{-0.029}^{+0.020}$}                & $167.74_{-0.98}^{+0.63}$ \\  
$i$ [deg]                        & \multicolumn{2}{c}{$88.93_{-0.06}^{+0.08}$}          & \multicolumn{2}{c}{$89.90_{-0.56}^{+0.45}$}                & $89.07_{-0.88}^{+0.45}$  \\
$e$                              & \multicolumn{2}{c}{$0.2313_{-0.0004}^{+0.0005}$}     & \multicolumn{2}{c}{$0.0111_{-0.0033}^{+0.0032}$}           & $0.3251_{-0.0030}^{+0.0030}$ \\  
$\omega$ [deg]                   & \multicolumn{2}{c}{$132.77_{-0.13}^{+0.10}$}             & \multicolumn{2}{c}{$115.3_{-7.2}^{+13.7}$}               & $288.55_{-0.39}^{+0.40}$ \\
$\tau^b$ [BJD]     & \multicolumn{2}{c}{$2\,459\,478.383_{-0.003}^{+0.003}$}         & \multicolumn{2}{c}{$2\,459\,473.439_{-0.035}^{+0.067}$} & $2\,459\,251.09_{-0.34}^{+0.33}$\\
$\Omega$ [deg]                   & \multicolumn{2}{c}{$0.0$}                           & \multicolumn{2}{c}{$-0.63_{-0.69}^{+0.54}$}                    & $0.49_{-0.47}^{+0.24}$ \\
$(i_\mathrm{m})_{A-...}^c$ [deg]   & \multicolumn{2}{c}{$0.0$}                         & \multicolumn{2}{c}{$1.15_{-0.41}^{+0.76}$}                    & $0.72_{-0.24}^{+0.26}$ \\
$(i_\mathrm{m})_{B-...}$ [deg]   & \multicolumn{2}{c}{$1.15_{-0.41}^{+0.76}$}          & \multicolumn{2}{c}{$0.0$}                                  & $1.49_{-0.54}^{+0.75}$ \\
$\varpi_\mathrm{dyn}^d$ [deg]    & \multicolumn{2}{c}{$312.76_{-0.12}^{+0.10}$}        & \multicolumn{2}{c}{$295.3_{-7.2}^{+13.7}$}                 & $108.55_{-0.39}^{+0.40}$ \\
$i_\mathrm{dyn}^d$ [deg]         & \multicolumn{2}{c}{$0.59_{-0.23}^{+0.23}$}          & \multicolumn{2}{c}{$1.37_{-0.51}^{+0.72}$}                 & $0.15_{-0.04}^{+0.04}$ \\
$\Omega_\mathrm{dyn}^d$ [deg]    & \multicolumn{2}{c}{$69_{-21}^{+112}$}              & \multicolumn{2}{c}{$134_{21}^{+19}$}                        & $282_{-29}^{+64}$ \\
$i_\mathrm{inv}^e$ [deg]         & \multicolumn{5}{c}{$89.09_{-0.75}^{+0.38}$} \\
$\Omega_\mathrm{inv}^e$ [deg]    & \multicolumn{5}{c}{$0.38_{-0.38}^{+0.20}$} \\
mass ratio $[q=m_\mathrm{sec}/m_\mathrm{pri}]$ & \multicolumn{2}{c}{$0.495_{-0.006}^{+0.006}$} & \multicolumn{2}{c}{$0.566_{-0.013}^{+0.013}$} & $0.715_{-0.005}^{+0.006}$ \\
$K_\mathrm{pri}$ [km\,s$^{-1}$] & \multicolumn{2}{c}{$39.4_{-0.4}^{+0.4}$} & \multicolumn{2}{c}{$66.6_{-1.2}^{+1.0}$} & $24.66_{-0.09}^{+0.08}$ \\ 
$K_\mathrm{sec}$ [km\,s$^{-1}$] & \multicolumn{2}{c}{$79.6_{-0.5}^{+0.4}$} & \multicolumn{2}{c}{$117.8_{-0.9}^{+0.9}$} & $34.49_{-0.29}^{+0.22}$ \\ 
  \hline  
\multicolumn{6}{c}{\textbf{Apsidal and nodal motion related parameters$^f$}} \\  
\hline  
$P_\mathrm{apse}$ [year] & \multicolumn{2}{c}{$17.52_{-0.09}^{+0.09}$} & \multicolumn{2}{c}{$62.2_{-0.3}^{+0.3}$} & $91.0_{-0.5}^{+0.6}$\\
$P_\mathrm{apse}^\mathrm{dyn}$ [year] & \multicolumn{2}{c}{$7.2_{-0.1}^{+0.5}$} & \multicolumn{2}{c}{$31.9_{-0.1}^{+0.2}$} & $38.1_{-0.2}^{+0.2}$\\
$P_\mathrm{node}^\mathrm{dyn}$ [year] & \multicolumn{2}{c}{$12.3_{-1.4}^{+0.4}$} & \multicolumn{2}{c}{$65.6_{-0.3}^{+0.3}$} & \\
$\Delta\omega_\mathrm{3b}^\mathrm{dyn}$ [arcsec/cycle] & \multicolumn{2}{c}{$4918_{-303}^{+92}$} & \multicolumn{2}{c}{$186.4_{-0.9}^{+0.8}$} & $14125_{-71}^{+70}$\\
$\Delta\omega_\mathrm{GR}$ [arcsec/cycle] & \multicolumn{2}{c}{$0.612_{-0.008}^{+0.005}$} & \multicolumn{2}{c}{$1.47_{-0.01}^{+0.01}$} & $0.151_{-0.002}^{+0.001}$\\
$\Delta\omega_\mathrm{tide}$ [arcsec/cycle] & \multicolumn{2}{c}{$0.21_{-0.01}^{+0.01}$} & \multicolumn{2}{c}{$8.45_{-0.18}^{+0.15}$} & $-$\\
\hline
\multicolumn{6}{c}{Stellar parameters} \\
\hline
   & Aa & Ab &  Ba & Bb & \\
  \hline
 \multicolumn{6}{c}{Relative quantities} \\
  \hline
fractional radius [$R/a$]               & $0.0475_{-0.0006}^{+0.0005}$ & $0.0230_{-0.0004}^{+0.0003}$  & $0.1054_{-0.0004}^{+0.0006}$ & $0.0636_{-0.0016}^{+0.0011}$ & \\
fractional flux [in \textit{TESS}-band] & $0.855_{-0.007}^{+0.005}$    & $0.027_{-0.002}^{+0.001}$     & $0.109_{-0.005}^{+0.006}$    & $0.009_{-0.001}^{+0.001}$    & \\
 \hline
 \multicolumn{6}{c}{Physical Quantities} \\
  \hline 
 $m$ [M$_\odot$]   & $1.072_{-0.020}^{+0.013}$ & $0.530_{-0.011}^{+0.009}$ & $0.731_{-0.008}^{+0.007}$ & $0.414_{-0.010}^{+0.008}$ & \\
 $R^g$ [R$_\odot$] & $1.086_{-0.021}^{+0.016}$ & $0.525_{-0.012}^{+0.009}$ & $0.678_{-0.003}^{+0.003}$ & $0.409_{-0.012}^{+0.008}$ & \\
 $T_\mathrm{eff}^g$ [K]& $5845_{-51}^{+57}$    & $3507_{-34}^{+28}$        & $4454_{-58}^{+56}$        & $3252_{-38}^{+30}$      & \\
 $L_\mathrm{bol}^g$ [L$_\odot$] & $1.230_{-0.045}^{+0.049}$ & $0.037_{-0.002}^{+0.002}$ & $0.162_{-0.009}^{+0.009}$ & $0.017_{-0.001}^{+0.001}$ &\\
 $M_\mathrm{bol}^g$ & $4.55_{-0.04}^{+0.04}$    & $8.35_{-0.05}^{+0.05}$    & $6.74_{-0.06}^{+0.06}$    & $9.21_{-0.06}^{+0.06}$    &\\
 $M_V^g           $ & $4.58_{-0.05}^{+0.05}$    & $10.14_{-0.10}^{+0.13}$   & $7.40_{-0.11}^{+0.12}$    & $11.46_{-0.12}^{+0.15}$    &\\
 $\log g^g$ [dex]   & $4.395_{-0.009}^{+0.011}$ & $4.721_{-0.009}^{+0.011}$ & $4.638_{-0.003}^{+0.002}$ & $4.829_{-0.010}^{+0.014}$ &\\
 \hline
\multicolumn{6}{c}{Global Quantities} \\
\hline
$\log$(age)$^g$ [dex] &\multicolumn{5}{c}{$9.584_{-0.048}^{+0.043}$} \\
$ [M/H]^g$  [dex]      &\multicolumn{5}{c}{$0.197_{-0.088}^{+0.074}$} \\
$E(B-V)$ [mag]    &\multicolumn{5}{c}{$0.257_{-0.016}^{+0.016}$} \\
$(M_V)_\mathrm{tot}^g$           &\multicolumn{5}{c}{$4.50_{-0.05}^{+0.05}$} \\
distance [pc]                &\multicolumn{5}{c}{$274_{-4}^{+3}$}  \\  
\hline
\end{tabular}
\label{tbl:simlightcurve285853156}

{\em Notes.} (a) Instantaneous, osculating orbital elements at epoch $t_0=2\,459\,474.0$; (b) Time of periastron passsage; (c) Mutual (relative) inclination; (d) Longitude of pericenter ($\varpi_\mathrm{dyn}$) and inclination ($i_\mathrm{dyn}$) with respect to the dynamical (relative) reference frame (see text for details); (e) Inclination ($i_\mathrm{inv}$) and node ($\Omega_\mathrm{inv}$) of the invariable plane to the sky; (f) See Sect~\ref{Sect:apsidal} for a detailed discussion of the tabulated apsidal motion parameters; (g) Interpolated from the \texttt{PARSEC} isochrones;  
\end{table*}

\subsection{The Quadruple TIC 392229331}

We summarize the system parameters and uncertainties for TIC 392229331 in Table  \ref{tbl:simlightcurve392229331}.  As for TIC 285853156, {\tt Lightcurvefactory} was used to model simultaneously all the RV, ETV, photometric, and SED data.  We discuss the results for TIC 392229331 in this section.

With an outer orbital period of 144.8 days, TIC 392229331 now has the second shortest known outer orbit of any quadruple system.  The outer orbital eccentricity of 0.558 makes this even more impressive from the perspective of long-term dynamical stability, which we will discuss in Section \ref{sec:stability}.  The component binaries A and B both have near-circular orbits with periods of 1.82 days and 2.25 days, respectively.  The two binaries have similar constituent masses, with $M_{\text{Aa}} \simeq 2.052 M_\odot$, $M_{\text{Ab}} \simeq 1.097 M_\odot$, $M_{\text{Ba}} \simeq 1.948 M_\odot$, and $M_{\text{Bb}} 
\simeq 1.147 M_\odot$.  Most of the system light in the {\em TESS} band is from star Aa ($\sim$48\%) and star Ba ($\sim$42\%), together accounting for $\sim$90\% of the system light.

\begin{figure}
   \centering
   \includegraphics[width=.99\columnwidth]{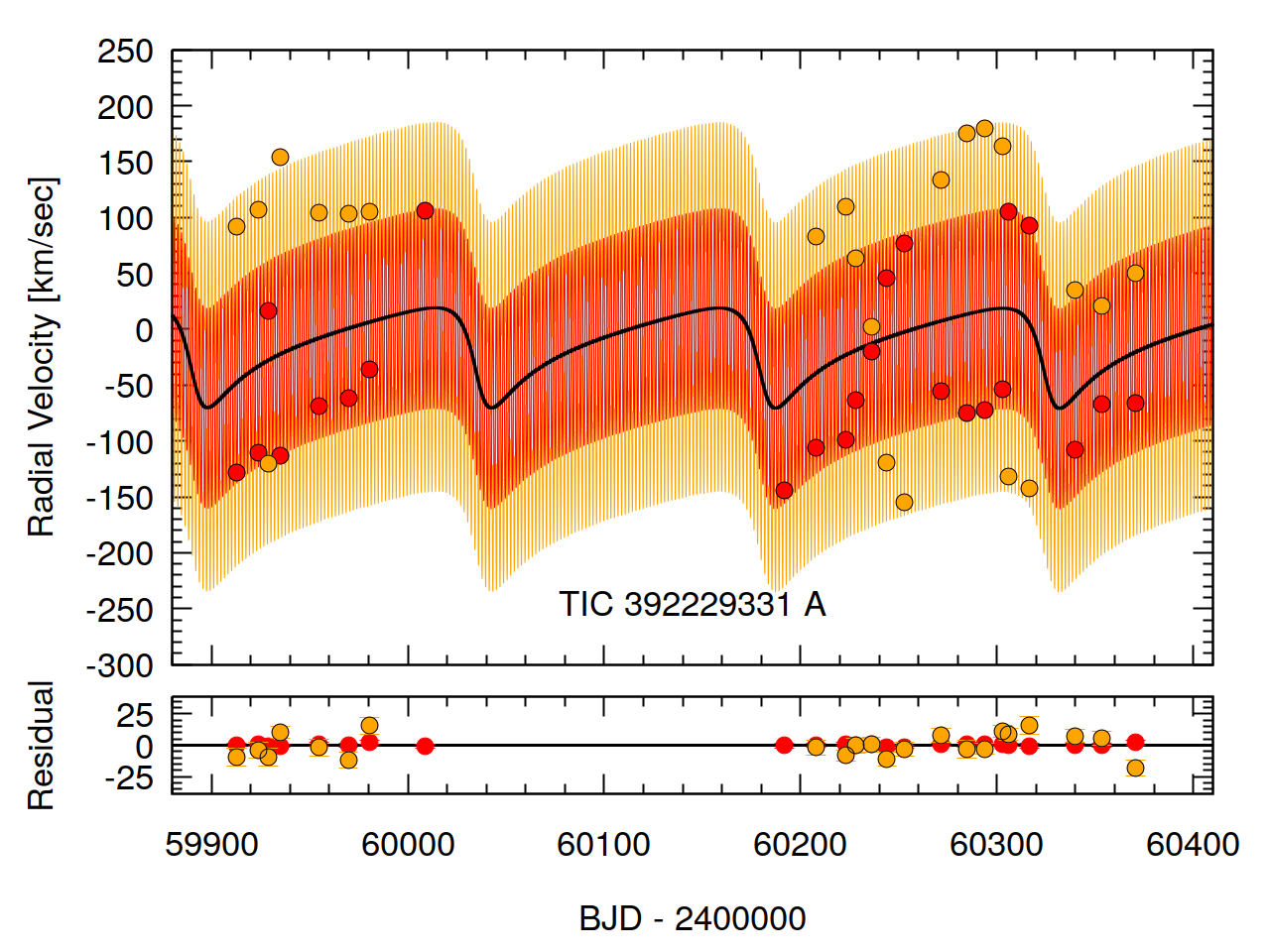}
   \includegraphics[width=.99\columnwidth]{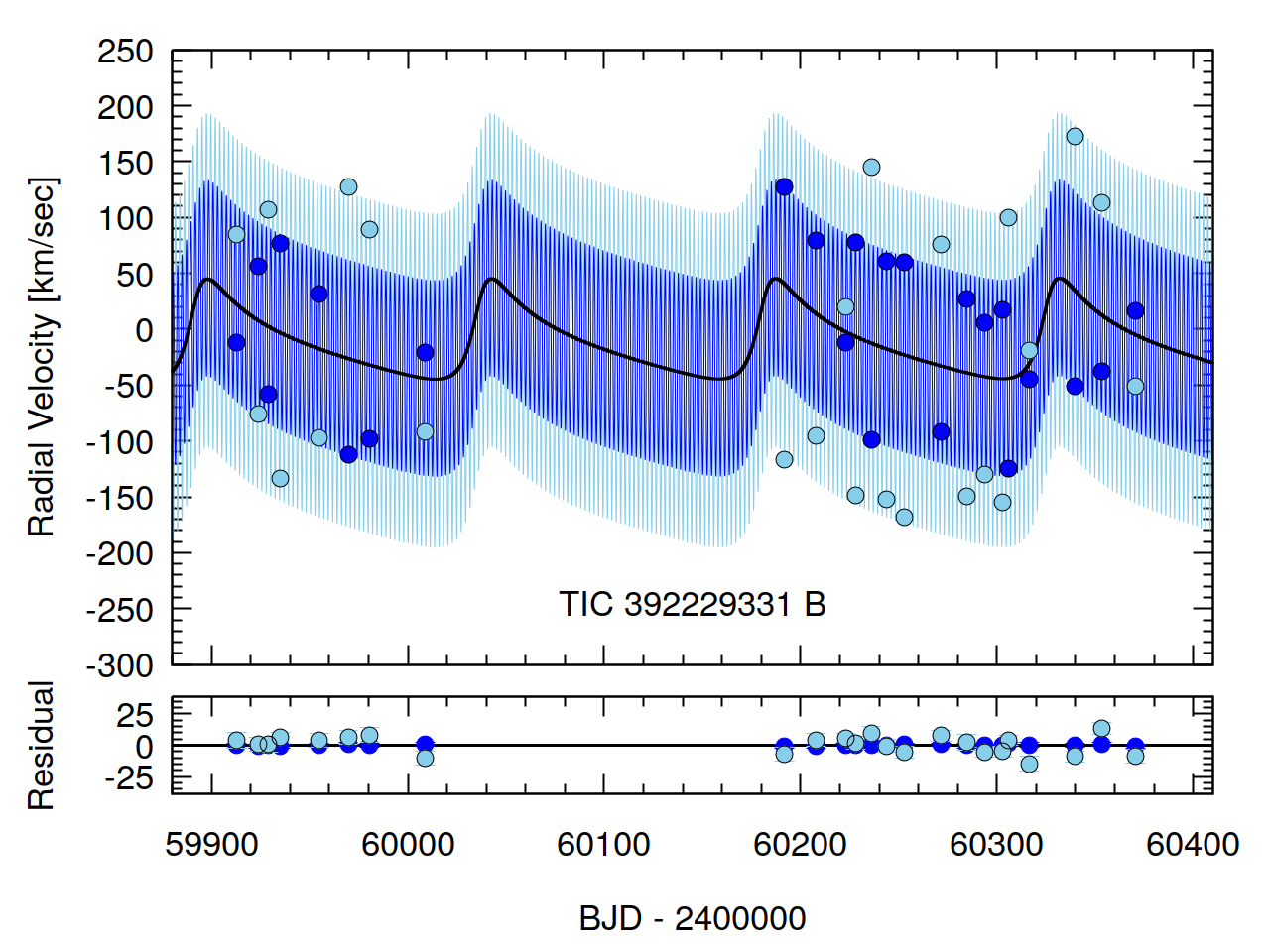}
   \caption{Radial velocity measurements and model fits vs.~time for TIC 392229331.  The upper and lower panels show the RVs for binary A and binary B, respectively. All other descriptors are the same as given in the caption to Figure  \ref{fig:285853156_rvt} for TIC 285853156.}
   \label{fig:392229331_rvt}
\end{figure}

In Figure \ref{fig:392229331_rvt}, we show the 47 measured RVs as a function of time together with the model fits for both binaries of TIC 392229331.  The orange and gold points (top panel) and blue and cyan points (bottom panel) are the RVs for the primary and secondary stars of binary A and binary B, respectively.  The smooth curves of the same color are the corresponding model fits from {\sc Lightcurvefactory}.  The black curves indicate the motion of the center of mass of binary A (B) around the system center of mass in the upper (lower) panel.  

\begin{figure}
   \centering
   \includegraphics[width=.99\columnwidth]{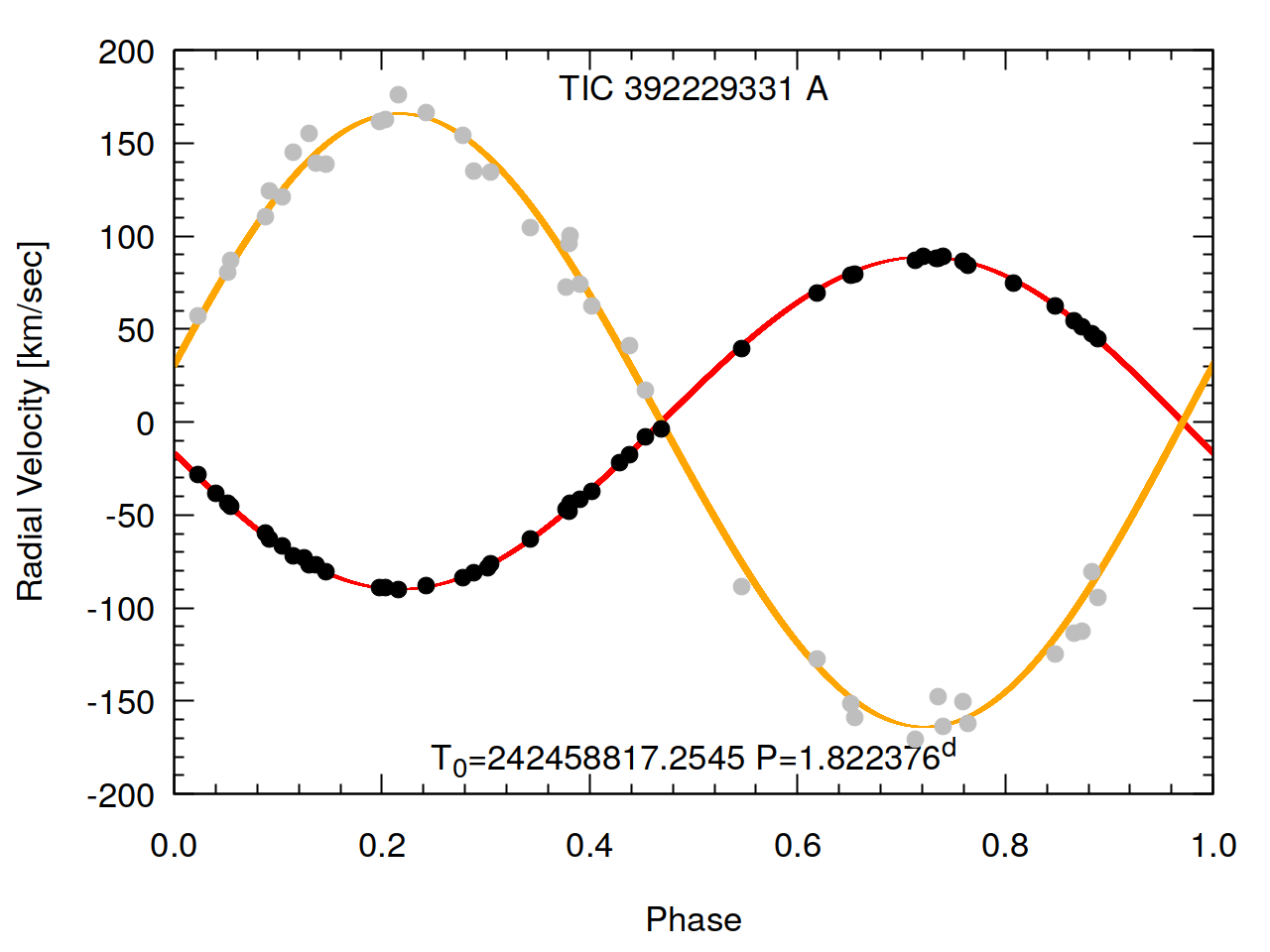}
   \includegraphics[width=.99\columnwidth]{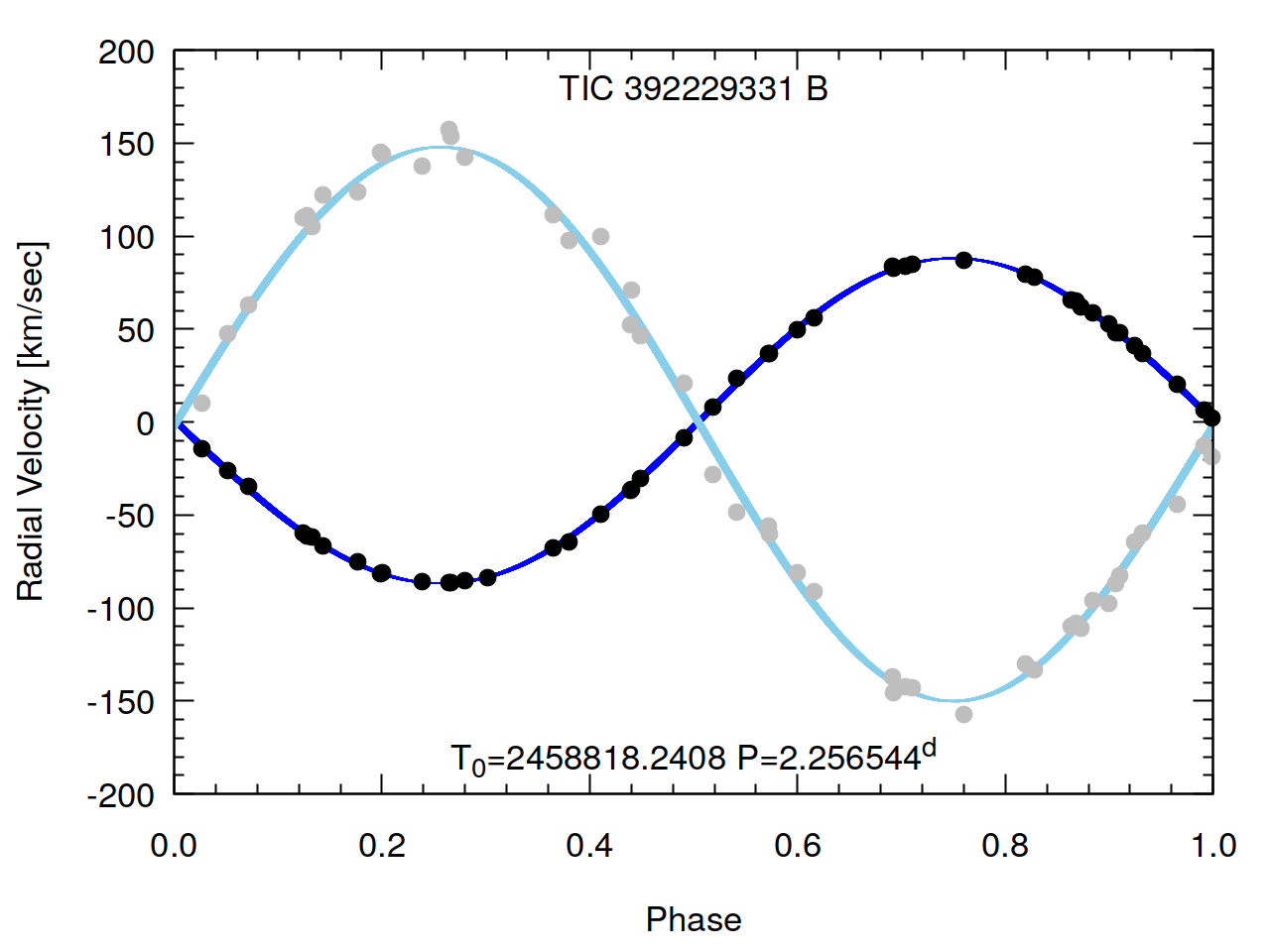}
   \includegraphics[width=.99\columnwidth]{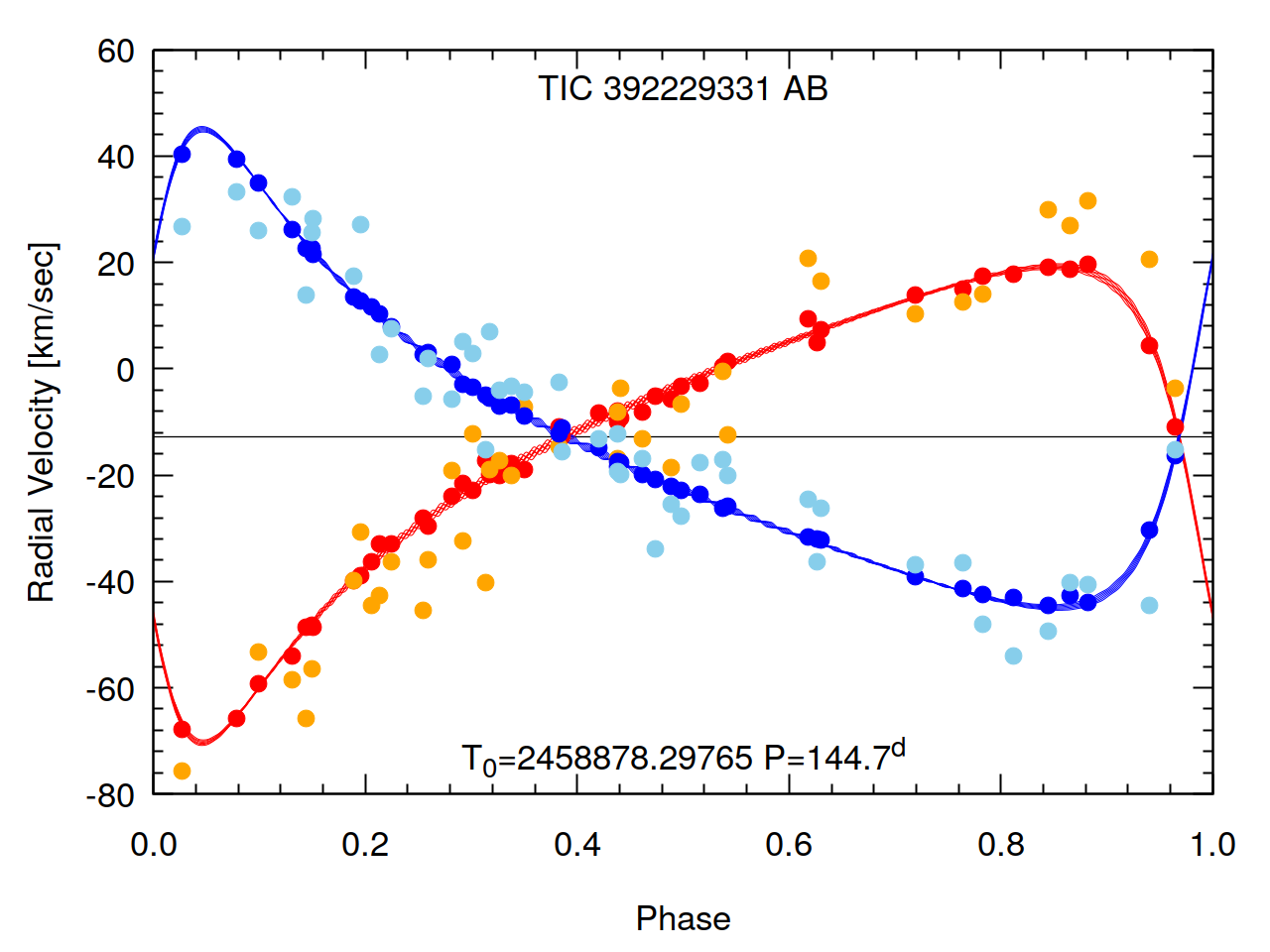}
   \caption{Radial velocity measurements and model fits vs.~orbital phase for the three binary orbits in TIC 392229331. These are the binary A components in A's center of mass (top panel), binary B components in B's center of mass (middle panel), and binaries A and B orbiting the system center of mass (bottom panel). All other descriptors are the same as in the caption to Figure \ref{fig:285853156_rvp} for TIC 285853156.}
   \label{fig:392229331_rvp}
\end{figure}

Figure \ref{fig:392229331_rvp} shows the same RV data, but here as a function of the orbital phase of binary A (top panel), of binary B (middle panel), and of the outer orbit of A orbiting B (bottom panel).  In the top and middle panels, the RV of the respective binaries around the system center of mass have been removed.  In the bottom panel, the RV motions of the four stars with respect to the system center of mass are shown---after removing the internal motions of those stars that are shown in the top and middle panels.  Specifically, this shows how the four stars trace the outer orbit.  As with the masses of the stars, here again we find a remarkable similarity between the orbits of the two binaries (i.e., compare the top and middle panels of Figure \ref{fig:392229331_rvp}).  The bottom panel of the same figure shows the substantial eccentricity of the outer orbit.

\begin{figure}
   \centering
   \includegraphics[width=.99\columnwidth]{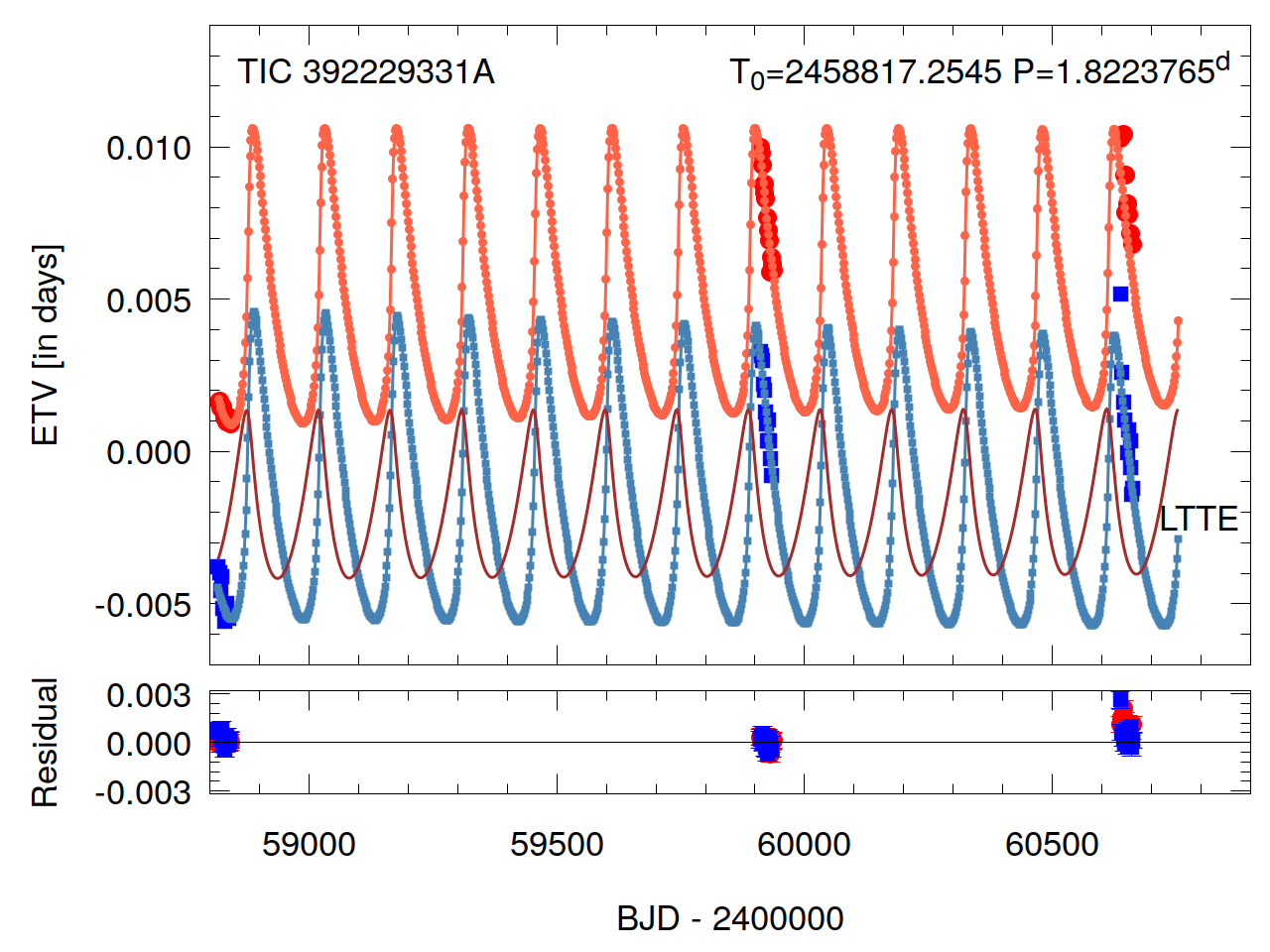}
   \includegraphics[width=.99\columnwidth]{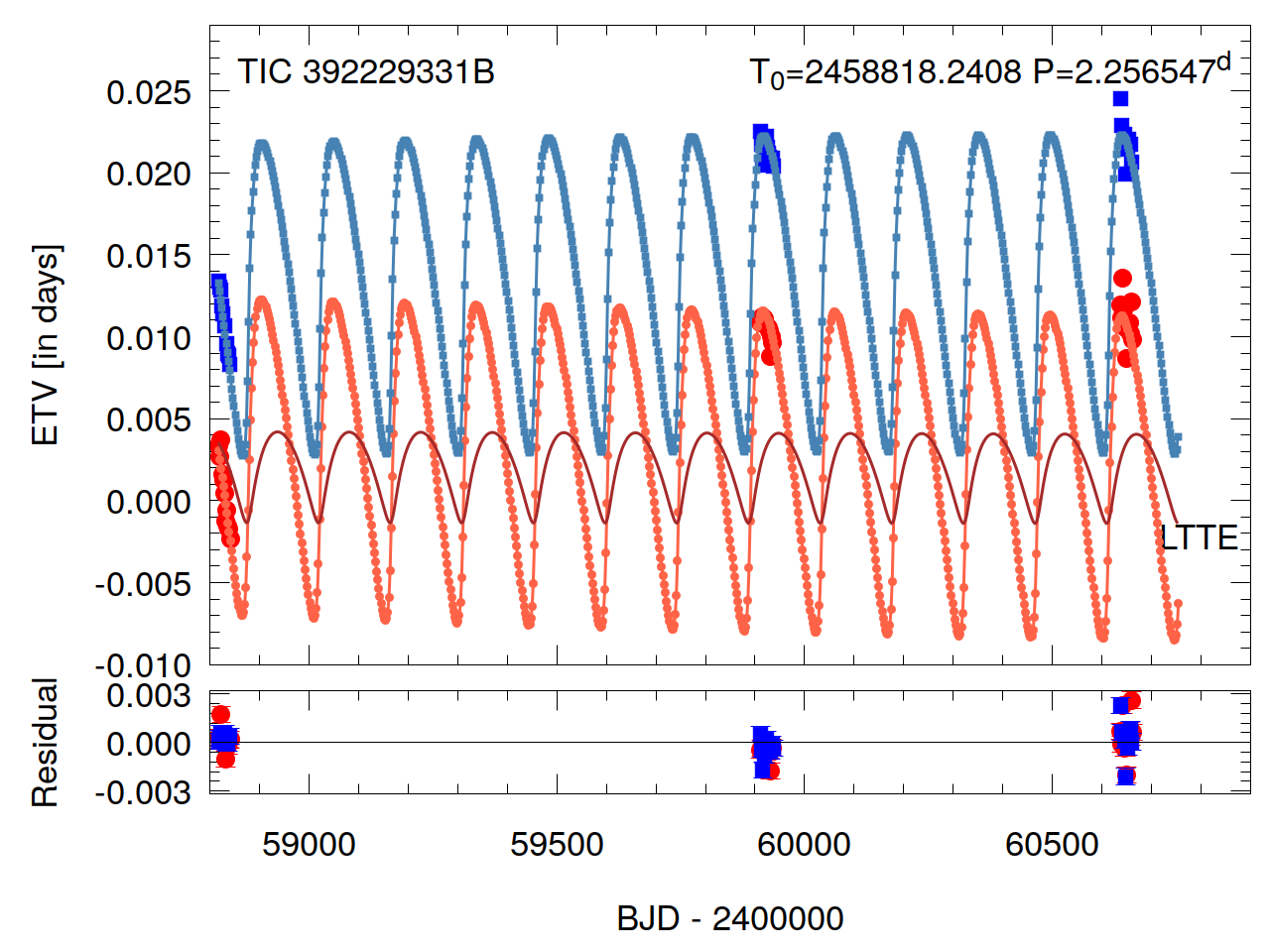}
   \caption{ETV points and model curves for TIC 392229331.  The top and bottom panels are for binary A and binary B, respectively, with the  red circles indicating the primary eclipses and blue squares the secondary eclipses.  The smooth curves of the corresponding colors are the model fits.  For both binaries, we show the contribution of the LTTE to the ETVs (thin brown curve), while the remainder of the ETVs are accounted for by dynamical effects. Residuals of the measured points from the fit are shown in the bottom section of each panel.}
   \label{fig:392229331_etv}
\end{figure}

The ETV data and model fits for TIC 392229331 are shown in Figure \ref{fig:392229331_etv}. These ETV data are similar to those for TIC 285853156 in that there is also a long (in this case, $\sim$3-year) data gap between between the first two {\textit TESS} observations.  However, the rest of the data for this source are sufficiently rich, that {\sc Lightcurvefactory} is readily able to incorporate these ETV snippets into the analysis and return a complete and robust solution.  In fact, we show in Figure \ref{fig:392229331_etv} the ETV points based on the most recent {\textit TESS} data from Sector 86, which was not fit with the model. Rather, we demonstrate the robustness of the photdynamical solution with data taken after the model was made. Note also that in each panel, the thin brown curve gives the LTTE contribution to the ETVs, with the remainder being due to the DEs. 

\begin{figure}
   \centering
   \includegraphics[width=.99\columnwidth]{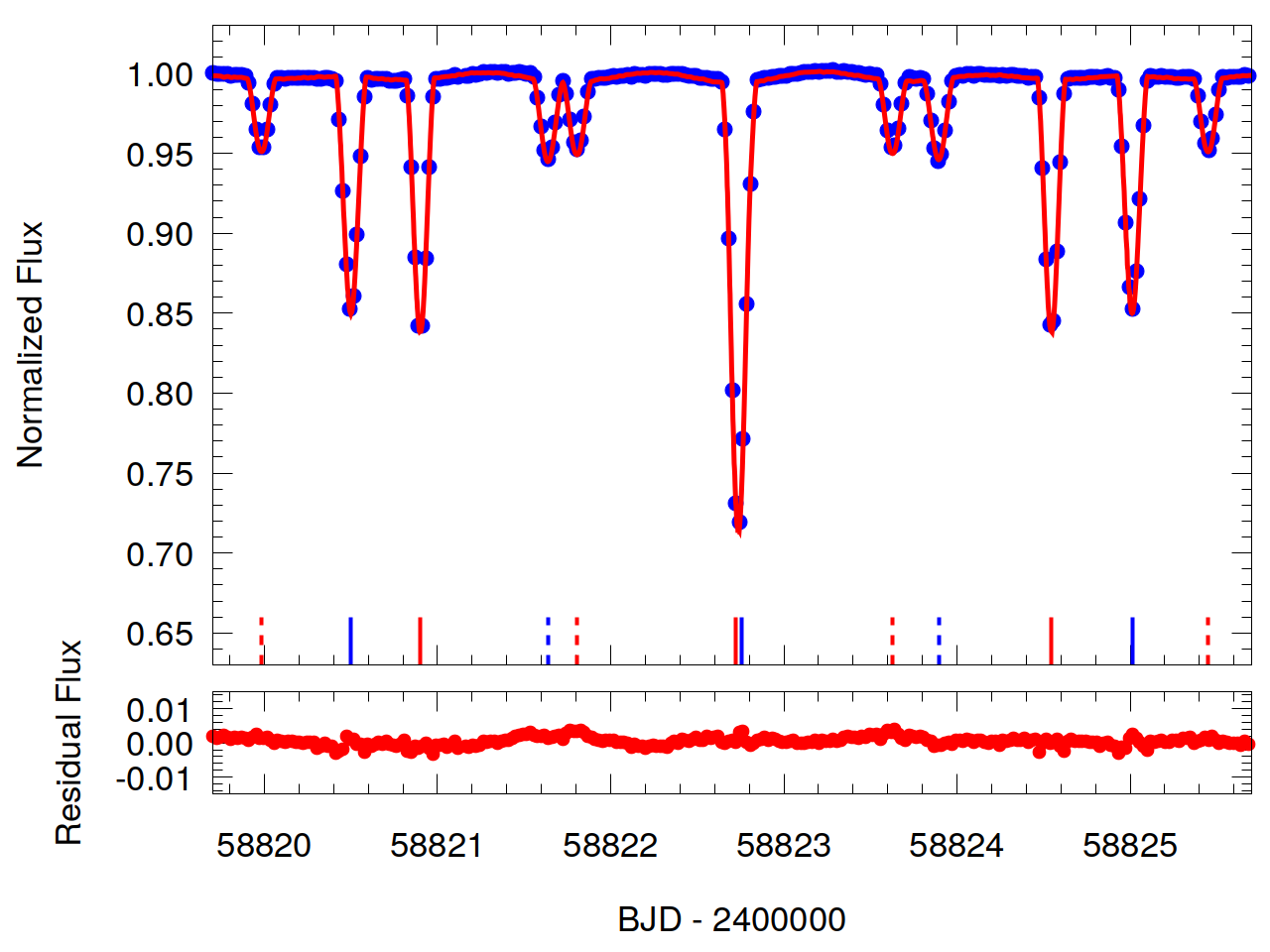}
   \includegraphics[width=.99\columnwidth]{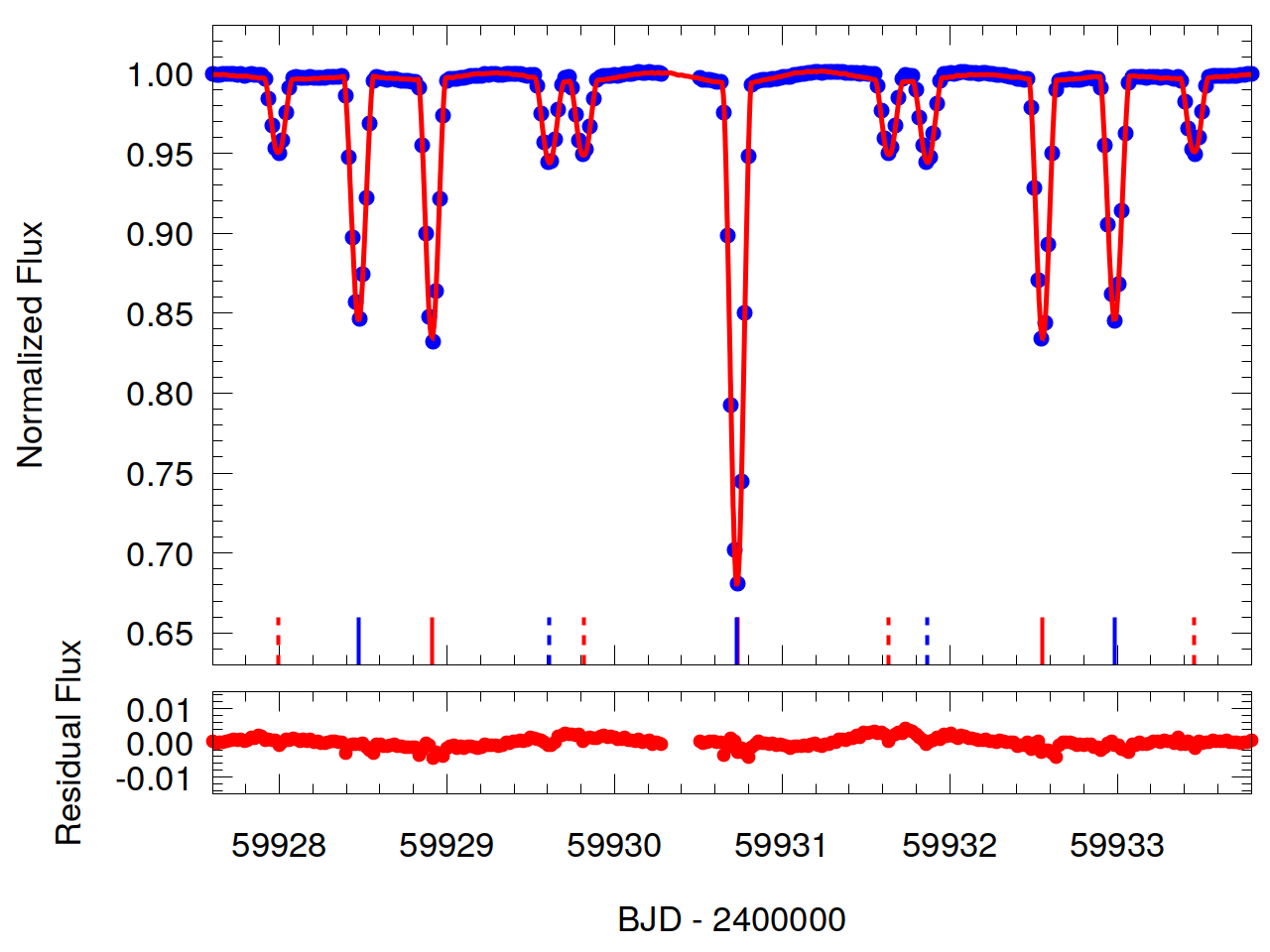}
   \includegraphics[width=.99\columnwidth]{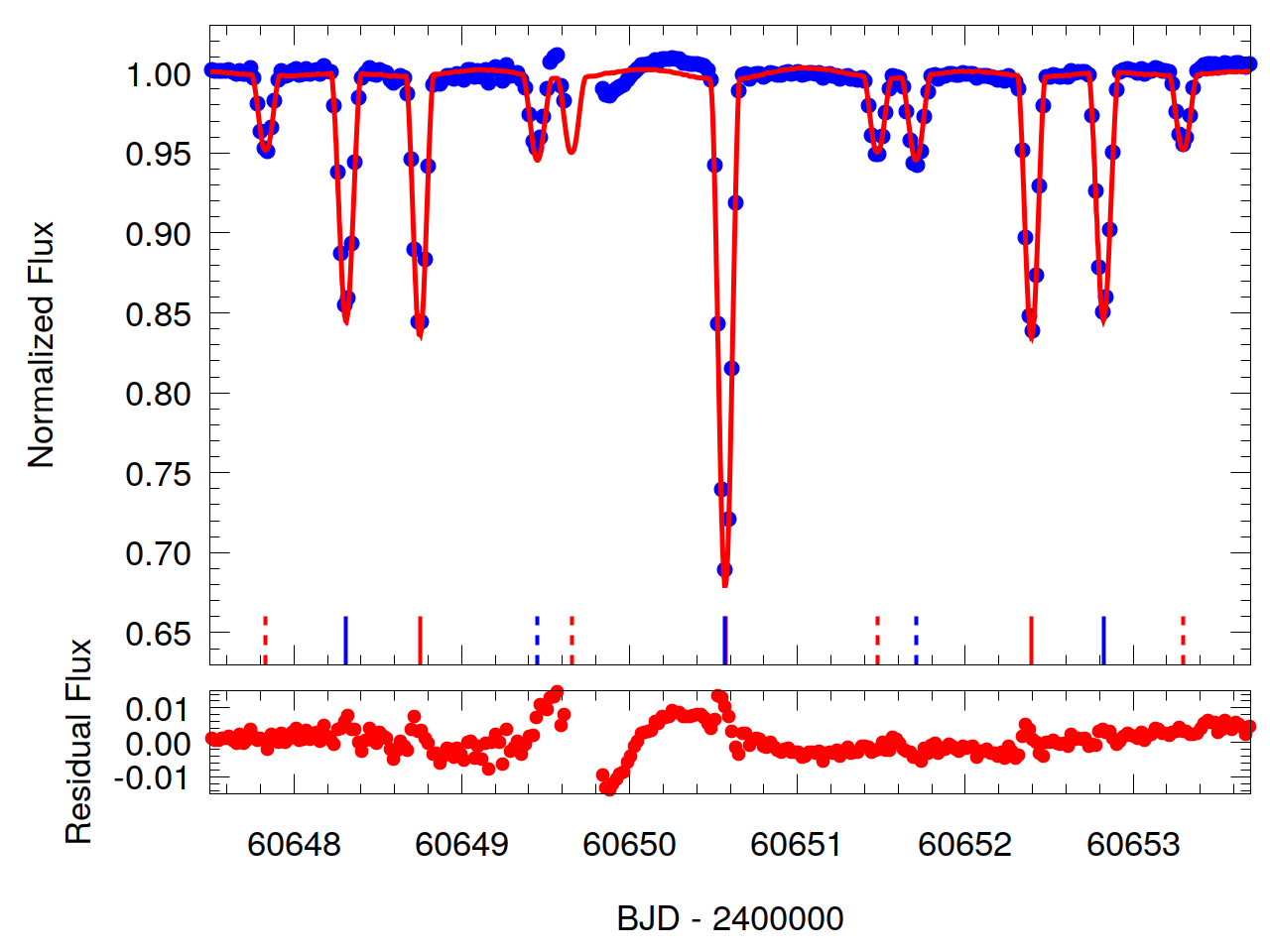}
   \caption{Lightcurve segments from {\em TESS} sectors 19 ({\em top}), sector 59 ({\em middle}), and sector 86 ({\em bottom}) for TIC 392229331. Lightcurve data points (blue circles) are compared to the model lightcurve (red line).  Residuals of the data points from the model fit are shown in the bottom section of each panel.  As with Figure \ref{fig:392229331_etv}, the sector 86 data were not fit with the model, but are shown here to demonstrate the robustness of the photodynamical solution.}
   \label{fig:392229331_lc}
\end{figure}

We show segments of the {\it TESS} photometry from sectors 19, 59, and 86 in Figure \ref{fig:392229331_lc}, along with their model fits.  Note that, although these light curve segments look identical at first glance, there are slight differences.  The relationship of $P_B/P_A = 26/21 \simeq 5/4$ 
causes frequent in-phase repeating patterns in the light curve, which can be readily seen upon closer inspection of the full {\em TESS} light curve in Figure \ref{fig:tess_lc_392229331}.  

\begin{figure}
   \centering
   \includegraphics[width=.99\columnwidth]{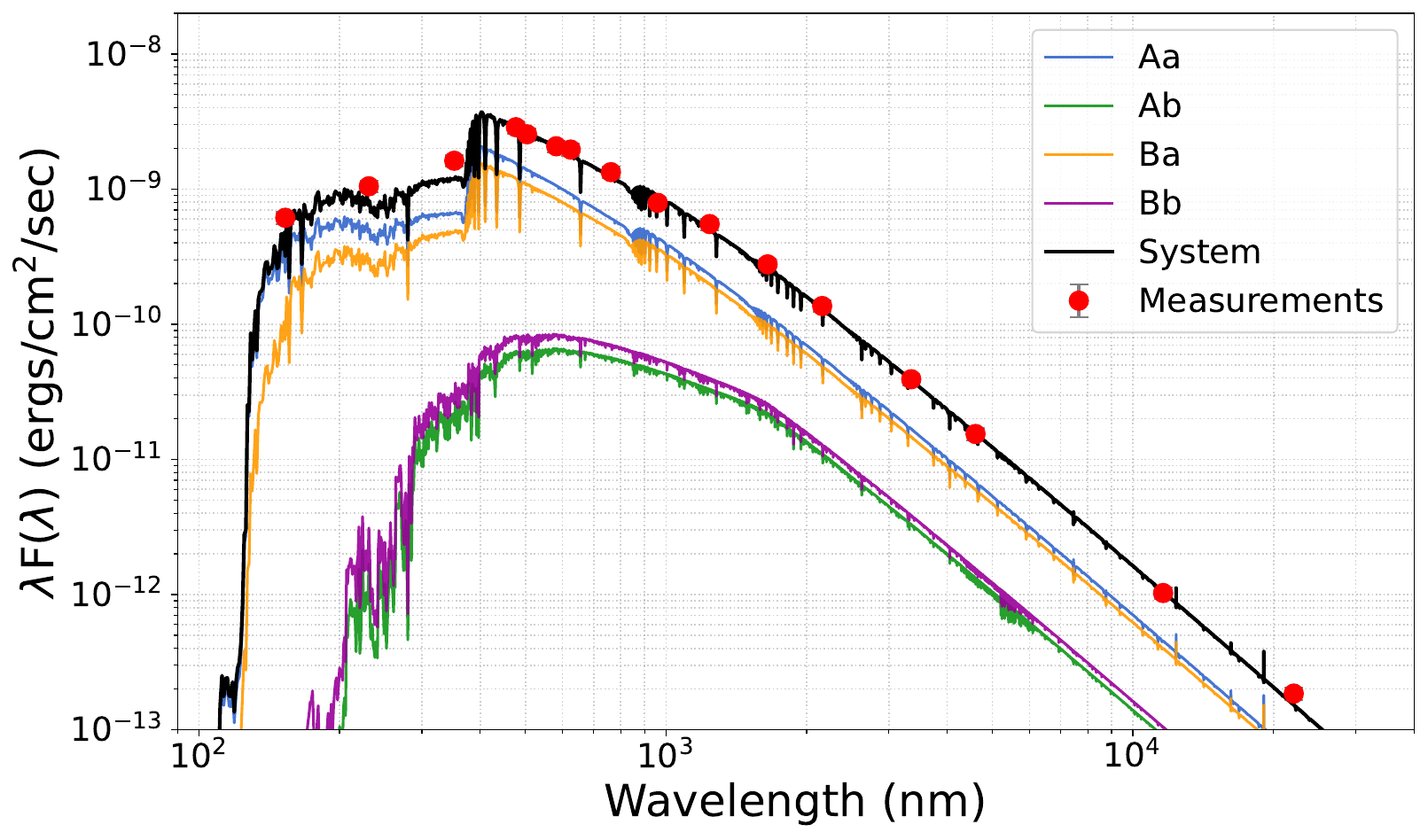}
   \caption{SED data points and model fit for TIC 392229331.  The SED measurements are shown as red circles, while the model contributions from stars Aa (blue), Ab (green), Ba (orange), Bb (purple), and the total system (black) are shown superposed on the data. The system age is 103 Myr.}
   \label{fig:392229331_sed}
\end{figure}

Finally, in Figure \ref{fig:392229331_sed}, we show the SED data and model fit for TIC 392229331.  We retrieved 22 flux measurements from VizieR \citep{vizier2000}, including the Galex NUV and FUV fluxes, and these are all plotted as orange circles.  We show the model flux curves vs.~wavelength for each of the four stars in the system as well as the composite model spectrum.  The fit is quite decent and indicates two similar primaries and two similar secondaries.  As was the case for TIC 285853156, we remind the reader that some 20 SED points over the wavelength range 0.15 to 11.6 microns are not adequate, by themselves, to infer all the parameters  for the four stars in the system.  However, when coupled with supplementary information, e.g., some temperature ratios between secondary and primary stars, mass estimates, stellar evolution tracks, etc., this becomes entirely possible.  Thus, the SED is just one input, among several others, that allow for a total solution for the system.  

\begin{table*}
\centering
\caption{Median values of the parameters of TIC 392229331 from the double EB simultaneous lightcurve,  double ETV, joint SED and \texttt{PARSEC} evolutionary track solution from {\sc Lightcurvefactory}.}
\begin{tabular}{lccccc}
\hline
\multicolumn{6}{c}{Orbital elements$^a$} \\
\hline
   & \multicolumn{3}{c}{subsystem}  \\
   & \multicolumn{2}{c}{A} & \multicolumn{2}{c}{B} & A--B \\
  \hline
$P_\mathrm{a}$ [days]            & \multicolumn{2}{c}{$1.8220567_{-0.0000064}^{+0.0000067}$} & \multicolumn{2}{c}{$2.256093_{-0.000015}^{+0.000015}$}       & $144.80_{-0.14}^{+0.16}$ \\
semimajor axis  [$R_\odot$]      & \multicolumn{2}{c}{$9.206_{-0.064}^{+0.049}$}           & \multicolumn{2}{c}{$10.554_{-0.057}^{+0.060}$}                & $213.8_{-1.3}^{+0.9}$ \\  
$i$ [deg]                        & \multicolumn{2}{c}{$84.88_{-0.35}^{+0.42}$}          & \multicolumn{2}{c}{$85.13_{-0.34}^{+0.26}$}                & $84.46_{-1.05}^{+0.78}$  \\
$e$                              & \multicolumn{2}{c}{$0.0113_{-0.0005}^{+0.0006}$}     & \multicolumn{2}{c}{$0.0135_{-0.0005}^{+0.0005}$}           & $0.558_{-0.010}^{+0.010}$ \\  
$\omega$ [deg]                   & \multicolumn{2}{c}{$298.0_{-2.3}^{+2.4}$}             & \multicolumn{2}{c}{$120.5_{-1.7}^{+1.8}$}               & $296.2_{-2.0}^{+2.4}$ \\
$\tau^b$ [BJD]     & \multicolumn{2}{c}{$2\,458\,815.577_{-0.012}^{+0.012}$}         & \multicolumn{2}{c}{$2\,458\,817.309_{-0.011}^{+0.012}$} & $2\,458\,877.95_{-1.12}^{+1.34}$\\
$\Omega$ [deg]                   & \multicolumn{2}{c}{$0.0$}                           & \multicolumn{2}{c}{$0.02_{-0.49}^{+0.35}$}                    & $-0.92_{-0.45}^{+0.35}$ \\
$(i_\mathrm{m})_{A-...}^c$ [deg]   & \multicolumn{2}{c}{$0.0$}                         & \multicolumn{2}{c}{$0.62_{-0.31}^{+0.37}$}                    & $1.25_{-0.55}^{+0.72}$ \\
$(i_\mathrm{m})_{B-...}$ [deg]   & \multicolumn{2}{c}{$0.62_{-0.31}^{+0.37}$}          & \multicolumn{2}{c}{$0.0$}                                  & $1.15_{-0.46}^{+0.88}$ \\
$\varpi_\mathrm{dyn}^d$ [deg]    & \multicolumn{2}{c}{$118.1_{-2.3}^{+2.4}$}        & \multicolumn{2}{c}{$300.5_{-1.7}^{+1.8}$}                 & $116.2_{-2.0}^{+2.4}$ \\
$i_\mathrm{dyn}^d$ [deg]         & \multicolumn{2}{c}{$1.07_{-0.49}^{+0.65}$}          & \multicolumn{2}{c}{$1.00_{-0.41}^{+0.76}$}                 & $0.17_{-0.06}^{+0.12}$ \\
$\Omega_\mathrm{dyn}^d$ [deg]    & \multicolumn{2}{c}{$247_{-32}^{+53}$}              & \multicolumn{2}{c}{$233_{20}^{+52}$}                        & $61_{-24}^{+48}$ \\
$i_\mathrm{inv}^e$ [deg]         & \multicolumn{5}{c}{$84.53_{-0.90}^{+0.68}$} \\
$\Omega_\mathrm{inv}^e$ [deg]    & \multicolumn{5}{c}{$-0.79_{-0.41}^{+0.32}$} \\
mass ratio $[q=m_\mathrm{sec}/m_\mathrm{pri}]$ & \multicolumn{2}{c}{$0.536_{-0.006}^{+0.008}$} & \multicolumn{2}{c}{$0.590_{-0.008}^{+0.007}$} & $0.983_{-0.012}^{+0.017}$ \\
$K_\mathrm{pri}$ [km\,s$^{-1}$] & \multicolumn{2}{c}{$88.9_{-0.6}^{+0.6}$} & \multicolumn{2}{c}{$87.5_{-0.7}^{+0.7}$} & $44.4_{-0.5}^{+0.6}$ \\ 
$K_\mathrm{sec}$ [km\,s$^{-1}$] & \multicolumn{2}{c}{$165.9_{-1.8}^{+1.2}$} & \multicolumn{2}{c}{$148.4_{-1.2}^{+1.1}$} & $45.1_{-0.6}^{+0.6}$ \\ 
  \hline  
\multicolumn{6}{c}{\textbf{Apsidal and nodal motion related parameters$^f$}} \\  
\hline  
$P_\mathrm{apse}$ [year] & \multicolumn{2}{c}{$30.2_{-0.6}^{+0.6}$} & \multicolumn{2}{c}{$32.0_{-0.7}^{+0.7}$} & $252.2_{-8.4}^{+8.2}$\\
$P_\mathrm{apse}^\mathrm{dyn}$ [year] & \multicolumn{2}{c}{$18.0_{-0.4}^{+0.4}$} & \multicolumn{2}{c}{$16.8_{-0.4}^{+0.4}$} & $31.1_{-0.8}^{+0.8}$\\
$P_\mathrm{node}^\mathrm{dyn}$ [year] & \multicolumn{2}{c}{$44.8_{-1.2}^{+1.2}$} & \multicolumn{2}{c}{$35.4_{-0.9}^{+0.9}$} & \\
$\Delta\omega_\mathrm{3b}^\mathrm{dyn}$ [arcsec/cycle] & \multicolumn{2}{c}{$277.5_{-7.0}^{+7.4}$} & \multicolumn{2}{c}{$433_{-11}^{+11}$} & $16544_{-430}^{+456}$\\
$\Delta\omega_\mathrm{GR}$ [arcsec/cycle] & \multicolumn{2}{c}{$2.83_{-0.04}^{+0.03}$} & \multicolumn{2}{c}{$2.43_{-0.03}^{+0.03}$} & $0.351_{-0.007}^{+0.007}$\\
$\Delta\omega_\mathrm{tide}$ [arcsec/cycle] & \multicolumn{2}{c}{$77.8_{-2.5}^{+2.9}$} & \multicolumn{2}{c}{$40.1_{-1.2}^{+1.4}$} & $-$\\
\hline
\multicolumn{6}{c}{Stellar parameters} \\
\hline
   & Aa & Ab &  Ba & Bb & \\
  \hline
 \multicolumn{6}{c}{Relative quantities} \\
  \hline
fractional radius [$R/a$]               & $0.1857_{-0.0014}^{+0.0016}$ & $0.1068_{-0.0012}^{+0.0014}$  & $0.1573_{-0.0012}^{+0.0011}$ & $0.0988_{-0.0016}^{+0.0018}$ & \\
fractional flux [in \textit{TESS}-band] & $0.481_{-0.015}^{+0.010}$    & $0.046_{-0.001}^{+0.001}$     & $0.417_{-0.010}^{+0.012}$    & $0.056_{-0.003}^{+0.003}$    & \\
 \hline
 \multicolumn{6}{c}{Physical Quantities} \\
  \hline 
 $m$ [M$_\odot$]   & $2.052_{-0.053}^{+0.036}$ & $1.097_{-0.014}^{+0.017}$ & $1.948_{-0.038}^{+0.035}$ & $1.147_{-0.017}^{+0.020}$ & \\
 $R^g$ [R$_\odot$] & $1.709_{-0.022}^{+0.021}$ & $0.983_{-0.016}^{+0.017}$ & $1.659_{-0.014}^{+0.017}$ & $1.043_{-0.022}^{+0.024}$ & \\
 $T_\mathrm{eff}^g$ [K]& $9038_{-197}^{+210}$    & $5845_{-83}^{+96}$        & $8696_{-219}^{+240}$        & $5984_{-98}^{+126}$      & \\
 $L_\mathrm{bol}^g$ [L$_\odot$] & $17.55_{-1.75}^{+1.64}$ & $1.019_{-0.084}^{+0.077}$ & $14.22_{-1.51}^{+1.51}$ & $1.264_{-0.134}^{+0.129}$ &\\
 $M_\mathrm{bol}^g$ & $1.66_{-0.10}^{+0.11}$    & $4.75_{-0.08}^{+0.09}$    & $1.89_{-0.11}^{+0.12}$    & $4.52_{-0.11}^{+0.12}$    &\\
 $M_V^g           $ & $1.70_{-0.06}^{+0.08}$    & $4.79_{-0.09}^{+0.11}$   & $1.89_{-0.08}^{+0.10}$    & $4.53_{-0.11}^{+0.13}$    &\\
 $\log g^g$ [dex]   & $4.282_{-0.006}^{+0.007}$ & $4.492_{-0.009}^{+0.009}$ & $4.286_{-0.0056}^{+0.008}$ & $4.460_{-0.012}^{+0.013}$ &\\
 \hline
\multicolumn{6}{c}{Global Quantities} \\
\hline
$\log$(age)$^g$ [dex] &\multicolumn{5}{c}{$8.014_{-0.173}^{+0.201}$} \\
$ [M/H]^g$  [dex]      &\multicolumn{5}{c}{$0.183_{-0.093}^{+0.052}$} \\
$E(B-V)$ [mag]    &\multicolumn{5}{c}{$0.301_{-0.020}^{+0.021}$} \\
$(M_V)_\mathrm{tot}^g$           &\multicolumn{5}{c}{$0.96_{-0.06}^{+0.08}$} \\
distance [pc]                &\multicolumn{5}{c}{$607_{-9}^{+9}$}  \\  
\hline
\end{tabular}
\label{tbl:simlightcurve392229331}

{\em Notes.} (a) Instantaneous, osculating orbital elements at epoch $t_0=2\,458\,816.0$; (b) Time of periastron passsage; (c) Mutual (relative) inclination; (d) Longitude of pericenter ($\varpi_\mathrm{dyn}$) and inclination ($i_\mathrm{dyn}$) with respect to the dynamical (relative) reference frame (see text for details); (e) Inclination ($i_\mathrm{inv}$) and node ($\Omega_\mathrm{inv}$) of the invariable plane to the sky; (f) See Sect~\ref{Sect:apsidal} for a detailed discussion of the tabulated apsidal motion parameters; (g) Interpolated from the \texttt{PARSEC} isochrones;  
\end{table*}

\section{Discussion}
\label{sec:discussion}
 
\subsection{Flatness of the Systems}
\label{Sect:flatness}
 
\begin{figure}
   \centering
   \includegraphics[width=0.99\columnwidth]{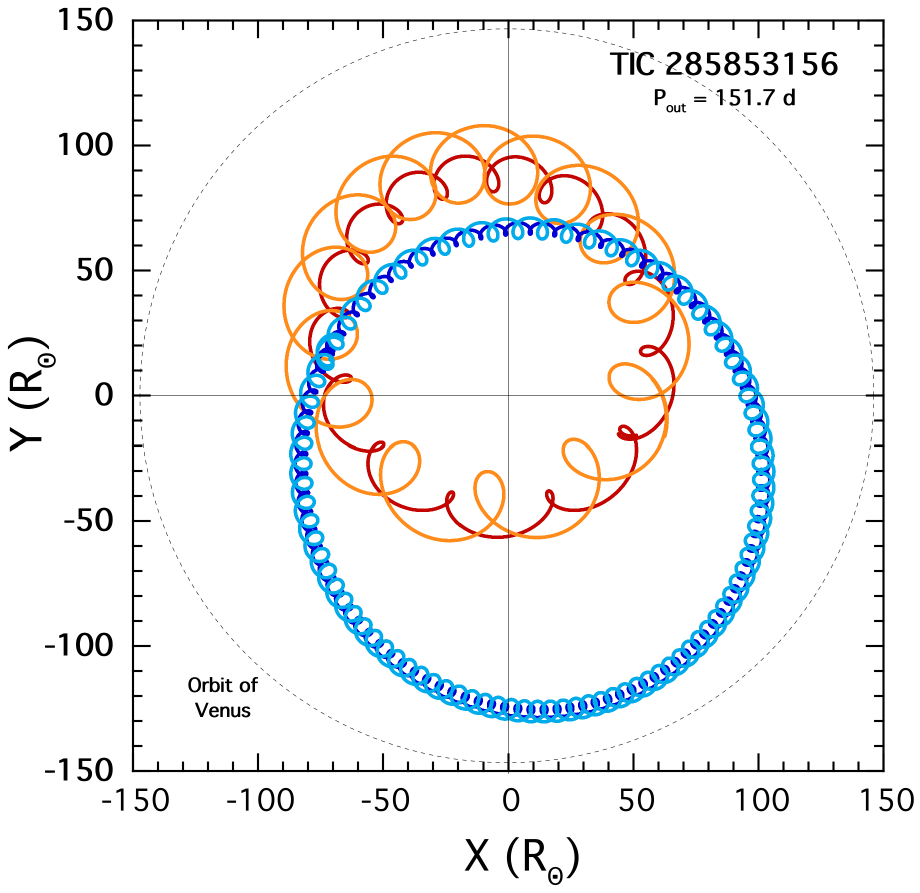} \hglue-0.2cm
    \includegraphics[width=0.97\columnwidth]{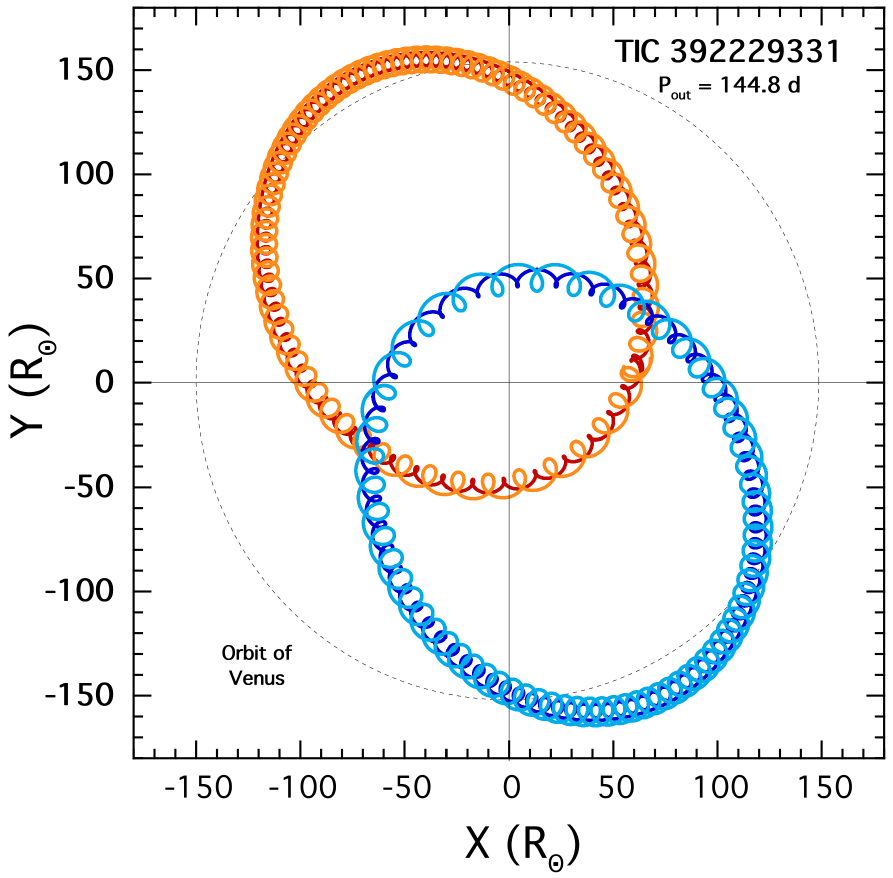}
   \caption{Orbital motions in TIC 285853156 ({\em top panel}) and in TIC 392229331 ({\em bottom panel}), as seen from a vantage point above the orbital plane. The actual observational view direction from the Earth is along the $+y$ axis. A comparison between the sizes of these orbits with that of our planet Venus around the Sun is included.}
   \label{fig:orbits}
\end{figure}

Views of the two quadruple systems studied in this work, TIC 285853156 ($P_{\rm out} = 151.7$ d) and TIC 392229331 ($P_{\rm out} = 144.8$ d), from above the orbital plane are shown in Figure \ref{fig:orbits}.  They are sufficiently compact that TIC 285853156 would fit entirely within the orbit of Venus (300 R$_\odot$ in diameter), while TIC 392229331 would very nearly do the same.  

When seen edge on with respect to the outer orbit, however, TIC 285853156 would be less than 2 R$_\odot$ in thickness, while TIC 392229331 might be a bit thicker at $\sim$5 R$_\odot$.  In other words, the systems are coplanar to within $\lesssim 1^\circ$ in the case of TIC 285853156, and $\lesssim 2^\circ$ for TIC 392229331.  

\begin{figure*}
   \centering
   \includegraphics[width=0.99\textwidth]{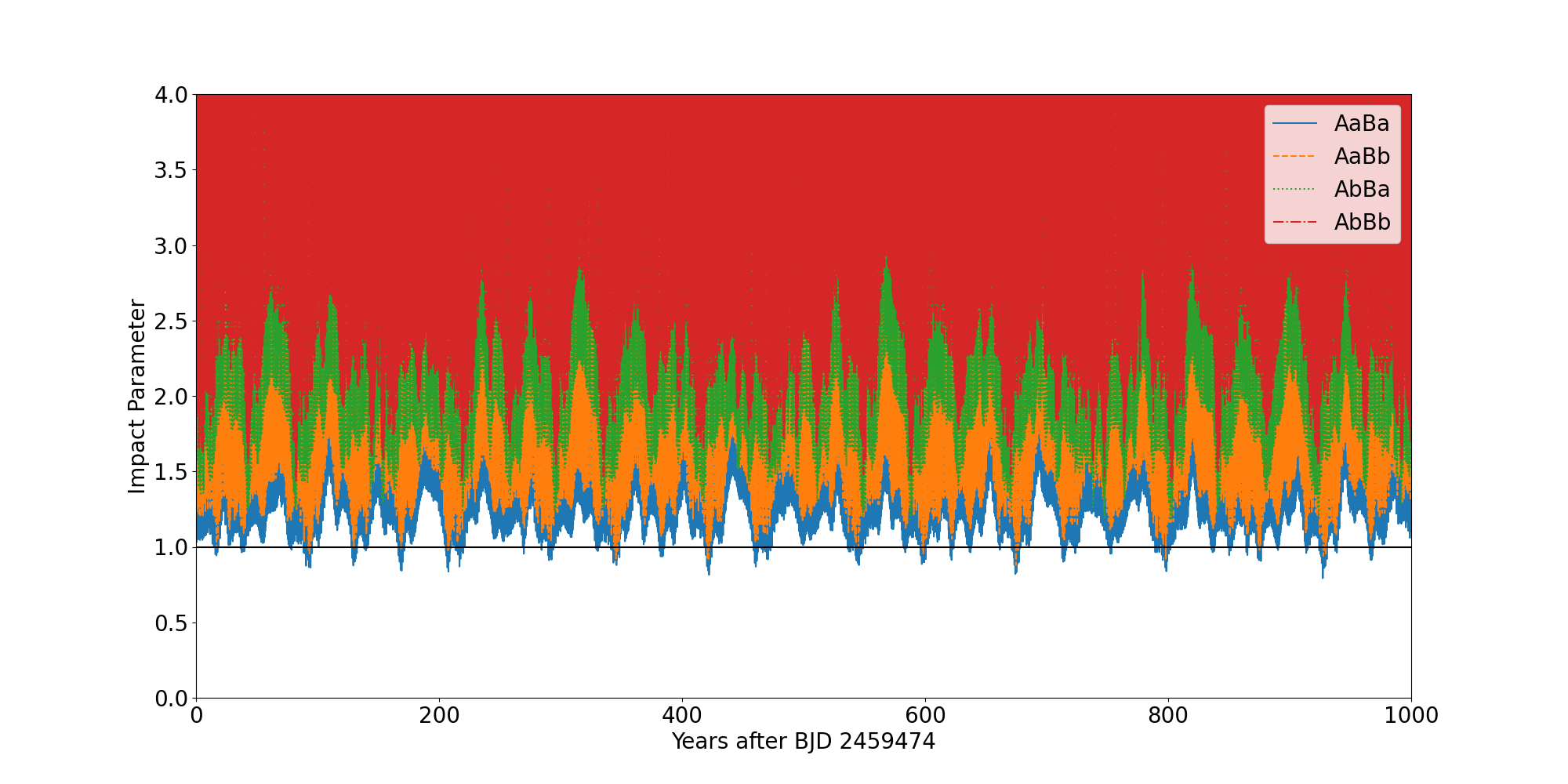}
   \includegraphics[width=0.99\textwidth]{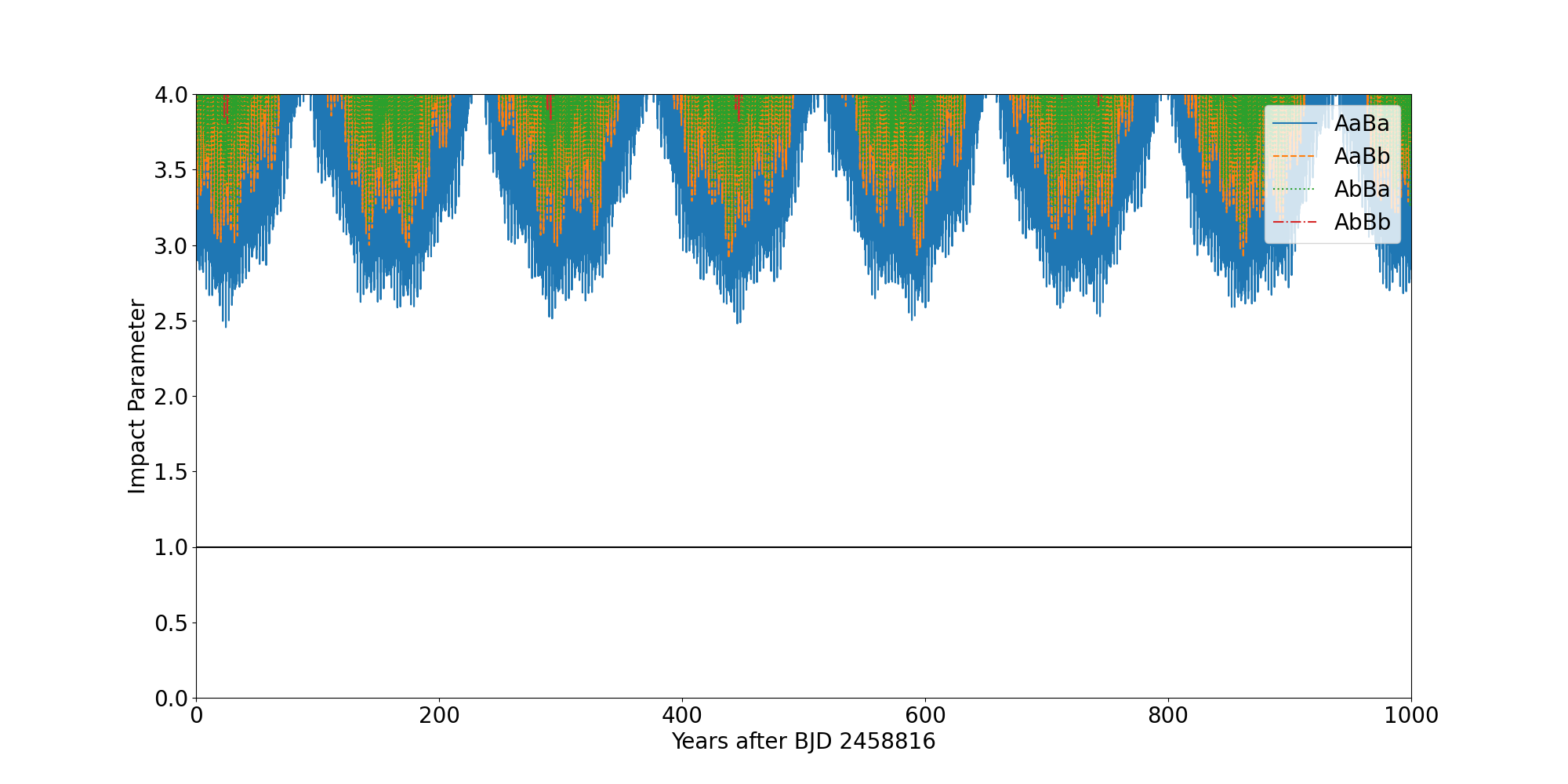}
   \caption{Impact parameter (centroid distance divided by the sum of the radii) of pairs of stars in TIC 285853156 ({\em top panel}) and TIC 392229331 ({\em bottom panel}) for a one thousand year duration. We should expect a pair of stars to produce eclipses on the outer orbit when the impact parameter is below the black line at $y=1$. This will occur for a several year interval once every few decades between pairs Aa/Ba and Aa/Bb of TIC 285853156, while TIC 392229331 will not eclipse on its outer orbit.}
   \label{fig:impact285853156}
\end{figure*}

The geometric flatness of these systems is a remarkable inference given that outer eclipses in these systems have not been observed. In order to determine whether there should be outer eclipses visible from the Earth, we simulated the system for one thousand years using {\sc REBOUND} \citep{Rein12} with the IAS15 integrator \citep{1985ASSL..115..185E,2015MNRAS.446.1424R} and examined the positions of the stars.  Figure \ref{fig:impact285853156} shows the impact parameter between stars for both systems over this time scale.  Due to apsidal motion (see Section \ref{Sect:apsidal}), we expect that TIC 285853156 will produce eclipses on the outer orbit between stars Aa and Ba (and, less frequently, between stars Aa and Bb) for a duration of several years at a time, once every few decades.  In contrast, we do not expect TIC 392229331 to ever produce outer eclipses.  

The ability to determine the mutual inclination angles between the binaries and the outer orbits is made possible by the fact that misalignment between the orbital planes would lead to orbital precession on timescales just a few times longer than the observations (see Tables \ref{tbl:simlightcurve285853156} and \ref{tbl:simlightcurve392229331}).  If the orbital planes of the binaries did precess, then there would be observable eclipse-depth variations which are not seen.  And, quantitative constraints on the motion of the orbital planes can thereby be set. 

\subsection{Apsidal Motion}
\label{Sect:apsidal}

The driven apsidal periods in TIC 285853156 are $\sim$17, 62, and 91 years for binary A, binary B, and the outer orbit respectively.  For TIC 392229331, these apsidal periods are 30, 32, and 250 years, respectively.  We can see just how rapid this precessional motion is in the RVs of binary A of TIC 285853156 from the model curves in the top panel of Fig.~\ref{fig:285853156_rvp}.  Apsidal precession can also be seen, but to a lesser degree, in the outer orbit of this same quadruple in the bottom panel of Fig.~\ref{fig:285853156_rvp}.  Likewise, for TIC 392229331, the model fits to the outer orbit RVs in the bottom panel of Fig.~\ref{fig:392229331_rvp} also exhibit apsidal motion.  By contrast, we don't see obvious precession in these figures from either binary B in TIC 285853156 or either inner binary in TIC 392229331.  This is due only to the small ($\sim$0.01) orbital eccentricities of these binaries, which are indeed undergoing apsidal motion, but just without having a large observational effect.  

A rough estimate of the dynamically forced apsidal precession timescale is given by the proportionality:$$\frac{4}{3}\left(1+1/q_{\rm out}\right)P_{\rm out}^2/P_{\rm in} (1-e_{\rm out})^{3/2}$$ \citep{2022Galax..10....9B}.  For the binaries with periods of $\sim$2 d in systems with outer periods of $\sim$150 d, this timescale is expected to be of order 30 years for both binaries in TIC 392229331 and binary B in TIC 285853156, while for binary A with a period of 10 d, the expected apsidal precession time is distinctly shorter (see Tables \ref{tbl:simlightcurve285853156} and \ref{tbl:simlightcurve392229331} for more precise values).  These are all far more rapid than could be expected from apsidal motion driven by stellar tides.

\subsection{Dynamical Stability} 
\label{sec:stability}

While the outer periods of the two systems are not quite the shortest known, the high outer eccentricities raise the question of whether the systems are long-term dynamically stable. To address this question, we first utilize the formalism of equation 1 from \cite{2022MNRAS.510.1352B}, derived from \citet{1995ApJ...455..640E,2008msah.conf...11M} and reproduced below for completeness, which shows that for the system to be long-term dynamically stable, the semi-major axis ratio and orbital period ratio between the quadruple and each binary, $a_{\rm quad}/a_{\rm bin}$ and $P_{\rm quad}/P_{\rm bin}$, should satisfy the following:

\begin{equation}
    \frac{a_{\rm quad}}{a_{\rm bin}} \gtrsim 2.8\left(\frac{M_{\rm quad}}{M_{\rm bin}}\right)^{2/5}\frac{(1+e_{\rm quad})^{2/5}}{(1-e_{\rm quad})^{6/5}}(1+e_{\rm bin}) 
\label{eqn:a}
\end{equation}

\begin{equation}
    \frac{P_{\rm quad}}{P_{\rm bin}} \gtrsim 4.7 \left(\frac{M_{\rm quad}}{M_{\rm bin}}\right)^{1/10}\frac{(1+e_{\rm quad})^{3/5}}{(1-e_{\rm quad})^{9/5}}(1+e_{\rm bin})^{3/2}
\label{eqn:p}
\end{equation}

We substituted the parameter values from Tables \ref{tbl:simlightcurve285853156} and \ref{tbl:simlightcurve392229331} into Eqns.~(\ref{eqn:a}) and (\ref{eqn:p}) and found that TIC 392229331 has the requisitely large outer period and semimajor axis, by about factors of 2, to be long-term stable.  However, for TIC 285853156 both the outer period and semimajor axis are only at $\sim$95\% of their respective stability criteria.  TIC 285853156 is therefore fairly close to long-term dynamical instability.  On the other hand, Eqns.~(\ref{eqn:a}) and (\ref{eqn:p}) are fitting formulae to the results of numerical experiments representing a wide range of orbital eccentricities, period ratios, and mass ratios, and may not precisely give the exact stability boundary for any given specific system.    

\begin{figure*}
   \centering
   \includegraphics[width=1.1\textwidth]{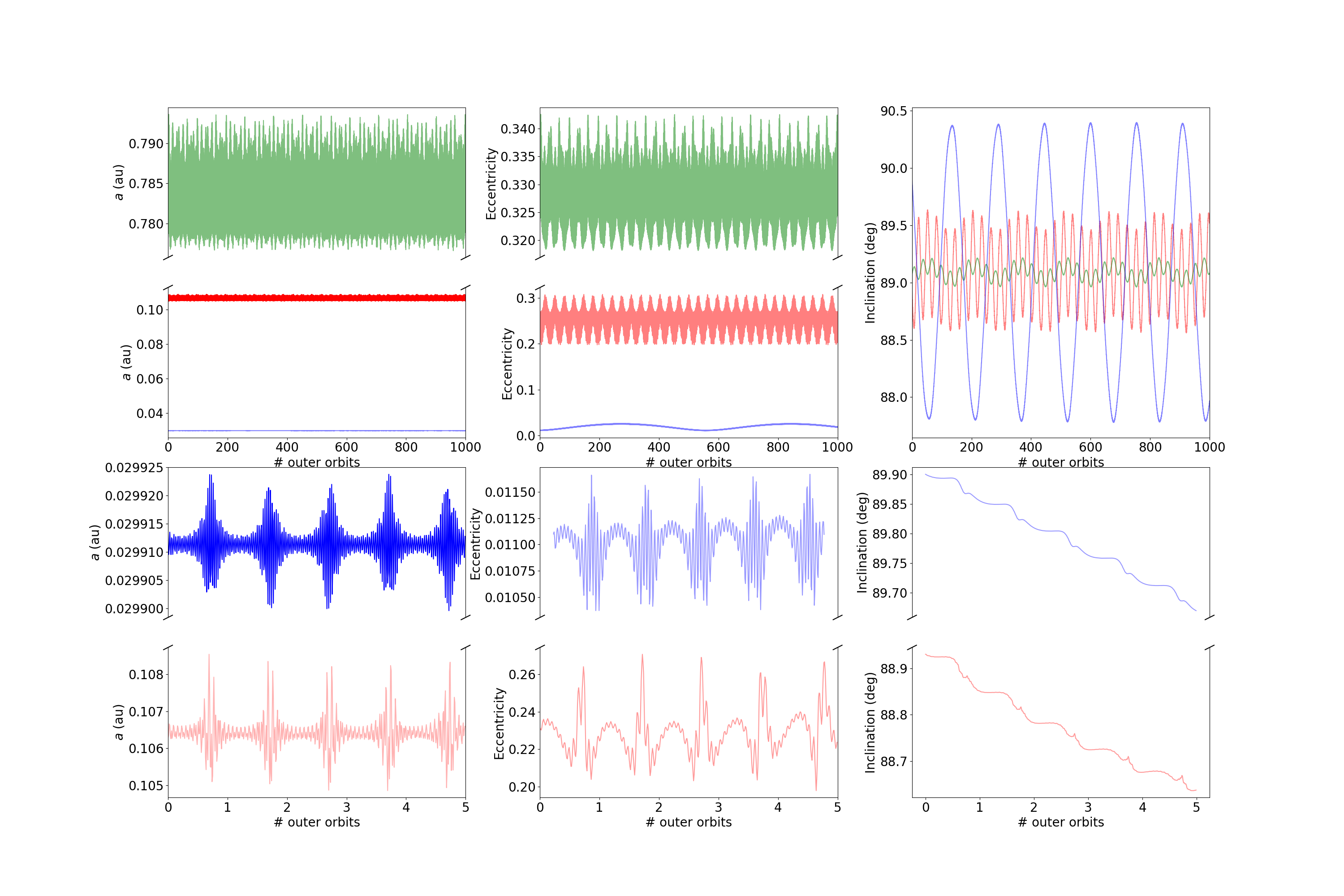}
   \caption{For TIC 285853156, orbital parameters of binary A ({\em red}), binary B ({\em blue}), and quadruple AB ({\em green}).  The top row shows one thousand orbits of the semi-major axes ({\em top left}), the eccentricities ({\em top middle}), and the inclinations ({\em top right}).  The bottom row shows a zoomed in view of five orbits of the binaries for the semi-major axes ({\em bottom left}), the eccentricities ({\em bottom middle}), and the inclinations ({\em bottom right}).}
   \label{fig:285853156_st}
\end{figure*}

\begin{figure*}
   \centering
   \includegraphics[width=1.1\textwidth]{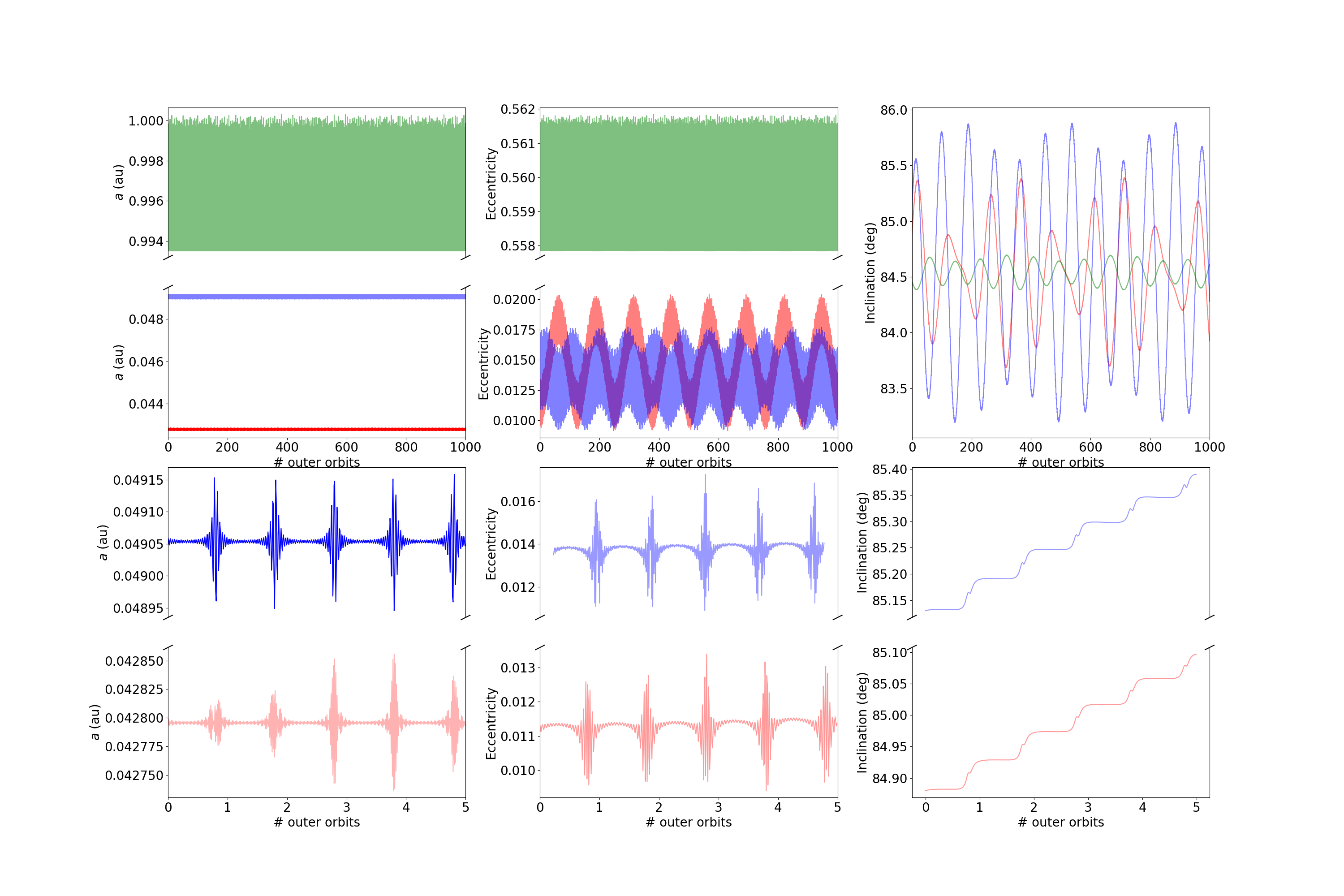}
   \caption{For TIC 392229331, orbital parameters of binary A ({\em red}), binary B ({\em blue}), and quadruple AB ({\em green}).  The top row shows one thousand orbits of the semi-major axes ({\em top left}), the eccentricities ({\em top middle}), and the inclinations ({\em top right}).  The bottom row shows a zoomed in view of five orbits of the binaries for the semi-major axes ({\em bottom left}), the eccentricities ({\em bottom middle}), and the inclinations ({\em bottom right}).}
   \label{fig:39222933_st}
\end{figure*}

Therefore, in order to check the fidelity of the above expressions for the TIC 285853156 system in particular, we carried out a long-timescale numerical simulation using {\sc REBOUND} \citep{Rein12} with the IAS15 integrator \citep{1985ASSL..115..185E,2015MNRAS.446.1424R}, as well as for the TIC 392229331 system (for completeness).  In Figure \ref{fig:285853156_st}, we show the semi-major axes, eccentricities, and inclinations of each of the binaries and the quadruple for the duration of one thousand outer orbits.  In Figure \ref{fig:39222933_st}, we show the same for TIC 392229331.  In both cases, there are no hints of instability. 

We also lengthened the duration of our simulations to test the stability of the systems for one million years.  In each case, there was no significant change in the semi-major axes, eccentricities, or inclinations.  As such, we conclude that both TIC 285853156 and TIC 392229331 are stable for at least millions of years, if not for the lifetime of the Galaxy.  

The inferred age of TIC 285853156 is 3.8 Gyr (Table \ref{tbl:simlightcurve285853156}), and thus, empirically, Nature seems to have determined that this system is indeed long-term stable---in spite of having {\it slightly} failed the stability expressions of Eqns.~({\ref{eqn:a}) and (\ref{eqn:p}).  For TIC 392229331 the observationally determined age is only 103 Myr (Table \ref{tbl:simlightcurve392229331}).  So, this empirical determination of stability still exceeds our calculations by two orders of magnitude.  
 
\subsection{Formation and Evolution of the Quadruples}
\label{sec:formation}
Here we briefly consider how such compact quadruple systems, that are at the same time nearly coplanar but with substantial outer orbital eccentricities, may have formed.  The components of compact quadruples originally formed at much wider separations, likely via an outside-in process \citep{Tokovinin2021}. In this scenario, the original core fragmented on $\sim$1,000 au scales and then each of the components fragmented again, possibly via gravitational disk instabilities on $\sim$100 au scales, resulting in the inner binaries (see \citealt{2023ASPC..534..275O} for a recent review on multiple star formation). Dynamical hardening of the outer orbit down to P$_{\text{out}}$ $\approx$ 150 days would have required substantial circumquadruple accretion, resulting in the nearly coplanar configurations we see today. Moreover, significant circumquadruple accretion would drive the outer binary mass ratio toward unity \citep{2015MNRAS.452.3085Y,2015MNRAS.447L..80F}. Indeed, TIC 392229331 is a near twin with q$_{\text{out}}$ = (M$_{Ba}$+M$_{Bb}$)/(M$_{Aa}$+M$_{Ab}$) = 0.98. 

However, circumbinary accretion does not necessarily dampen the orbital eccentricities. Eccentricity evolution depends inextricably on multiple factors, including the binary mass ratio, initial eccentricity, and the viscosity and sound speed of the accreting gas \citep{2020ApJ...901...25D,2021ApJ...909L..13Z}. The large eccentricities $e_{\rm out}$ $>$ 0.3 of the outer orbits in our two quadruples are relic signatures that they formed at much larger separations and subsequently decayed inward via dynamical friction within the surrounding gas. In contrast, most compact triples with $P_{\rm out}$ $<$ 200 days have $e_{\rm out}$ $<$ 0.3 \citep{2016MNRAS.455.4136B}, suggesting both companions formed through two subsequent episodes of disk fragmentation. Compact 2+2 quadruples {\it cannot} form through disk fragmentation alone. Instead, the outer pair must have originally formed via core fragmentation on substantially larger scales. Hence, for a given final $P_{\rm out}$, compact quadruples will have systematically larger $e_{\rm out}$ compared to their compact triple counterparts. The transition between the disk versus core fragmentation channels appears to be around $e$ = 0.3. For example, the spin-orbit angles of close binaries with $e$ $<$ 0.2 are well-aligned, suggesting they formed out of the same disk, whereas close binaries with $e$ $>$ 0.4 exhibit random spin-orbit orientations, suggesting they dynamically migrated \citep{2024ApJ...975..153S,2024ApJ...975..149M}. It is thus not surprising that for our two compact quadruples, where the outer pairs must have formed via core fragmentation and decayed inward by three orders of magnitude, the outer orbits still retain rather large eccentricities $e_{\rm out}$ $>$ 0.3.

\section{Summary}
\label{sec:summary}

We have presented the discovery of two eclipsing quadruple star systems, TIC 392229331 and TIC 285853156. The systems have a 2+2 hierarchical configuration consisting of two eclipsing binaries. These quadruples have the second and third shortest outer periods of all the known quadruple systems---145 days for TIC 392229331 and 152 days for TIC 285853156.  Both systems have all three orbits (two inner binaries and the outer orbit) aligned in a near coplanar configuration to within about a degree.  Both outer orbits have a substantial eccentricity of 0.558 (for TIC 392229331) and 0.325 (for TIC 285853156). All three orbits of these quadruples would just about fit within the orbit of Venus in our solar system.

The two quadruple systems were analyzed for their full sets of system parameters using a comprehensive photodynamical model, simultaneously fitting the RVs, ETVs, light curves, and the SED.  The fitting code, {\sc Lightcurvefactory}, also utilizes theoretical stellar evolution tracks to interrelate the stellar masses, age, and metallicity with their radii, and $T_{\rm eff}$ (see Sect.~\ref{sec:photodynamics}).

The periods of the binaries in TIC 392229331 are 1.82 days and 2.25 days, both with eccentricities of $\sim$0.01.  The two binaries are rather similar, with mass ratios $q_{\text{in,A}} = 0.51$ and $q_{\text{in,B}} = 0.59$, with the outer mass ratio $q_{\text{out}} = 0.98$, close to unity.  The system is relatively young, with an age of 103 Myr.  Of particular note, the outer period is 144.8 days, the second shortest period among known quadruples, with a substantial eccentricity of 0.558.

The periods of the binaries in TIC 285853156 are 10.0 days and 1.77 days with eccentricities 0.23 and 0.01, respectively.  The system flux is dominated by the primary of the 10.0 day binary.  Like TIC 392229331, mass ratios of the binaries, $q_{\text{in,A}} = 0.49$ and $q_{\text{in,B}} = 0.57$, are close to $0.5$, with $q_{\text{out}} = 0.71$.  The system age is 3.8 Gyr. The outer orbital period is 151.7 days, the third shortest period of known quadruples, with an eccentricity of 0.325.

We note that the basic stellar properties of the two quadruple systems, including mass, radius, $T_{\rm eff}$, and age, are well enough determined (to within better than a couple of percent) so that these should be useful to check against stellar evolution models.  There are very few such multistellar systems, especially 2+2 quadruples, that are available for this purpose.

Our spectroscopic observations of TIC 285853156 and TIC 392229331, following initial identification as quadruples by K22/K24, provide powerful examples of the value of continuing to pursue follow-up observations for interesting multistellar targets.  Additionally, the radial velocity measurements in these systems were particularly important because the ETV data were limited and there are no outer eclipses (third and fourth body eclipses) observed.  
TIC 285853156 and TIC 392229331 are both compact systems, per the definitions of \citet{2021Univ....7..352T} and \citet{2022Galax..10....9B}, and, additionally, TIC 285853156 is a ``tight'' system with $P_{\rm out}/P_{\rm in}$ as low as 15.  They are both rather remarkable, for having such short outer periods, yet with substantial eccentricities. These properties will inform our understanding of the formation and evolution of such multistellar systems (see Section \ref{sec:formation}).

Finally, the short outer periods of these systems raise the question of the shortest physically possible outer period in a quadruple star system (the current record holder is BU CMi with $P_{\rm out}=122$ days; \citealt{2023MNRAS.524.4220P}).  In terms of basic dynamical stability requirements, a coplanar system with non-eccentric orbits, with, e.g., inner binary periods of $\sim$2 days and an outer period of $\sim$25 days should be quite stable.  A simpler analog of such a system is the triple system TIC 290061484 \citep{2024ApJ...974...25K} with a binary period of 1.8 days and an outer orbital period of 24.6 days.  However, the formation of a quadruple system is likely more complex, and it is not clear whether such short outer period quadruples will ever be found.

\clearpage
\begin{acknowledgments}

This paper includes data collected by the {\em TESS} mission, which are publicly available from the Mikulski Archive for Space Telescopes (MAST). Funding for the {\em TESS} mission is provided by NASA's Science Mission directorate. 

Resources supporting this work were provided by the NASA High-End Computing (HEC) Program through the NASA Center for Climate Simulation (NCCS) at Goddard Space Flight Center.

This project has received funding from the HUN-REN Hungarian Research Network.

This work has made use of data from the European Space Agency (ESA) mission {\it Gaia} (\url{https://www.cosmos.esa.int/gaia}), processed by the {\it Gaia} Data Processing and Analysis Consortium (DPAC, \url{https://www.cosmos.esa.int/web/gaia/dpac/consortium}). Funding for the DPAC has been provided by national institutions, in particular the institutions participating in the {\it Gaia} Multilateral Agreement. 

T.\,B. acknowledges the financial support of the Hungarian National Research, Development and Innovation Office -- NKFIH Grant K-147131

V.\,B.\,K. is grateful for financial support from NASA grant 80NSSC21K0631 and NSF grant AST-2206814.
\end{acknowledgments}

\facilities{
\emph{Gaia},
MAST,
TESS,
ASAS-SN,
ATLAS,
NCCS,
}

\software{
{\tt Astropy} \citep{astropy2013,astropy2018}, 
{\tt Eleanor} \citep{eleanor},
{\tt IPython} \citep{ipython},
{\tt Keras} \citep{keras},
{\tt LcTools} \citep{2019arXiv191008034S,2021arXiv210310285S},
{\tt Lightcurvefactory} \citep{Borkovits2019,Borkovits2020},
{\tt Lightkurve} \citep{lightkurve},
{\tt Matplotlib} \citep{matplotlib},
{\tt Mpi4py} \citep{mpi4py2008},
{\tt NumPy} \citep{numpy}, 
{\tt Pandas} \citep{pandas},
{\tt SciPy} \citep{scipy},
{\tt Tensorflow} \citep{tensorflow},
{\tt Tess-point} \citep{tess-point}
}

\bibliography{refs}{}
\bibliographystyle{aasjournal}



\end{document}